%% file: MyThesis.tex
\documentclass[11pt,Chicago]{uuthesis2e1}
\usepackage{acronym}
\usepackage{uuthesis-2011}
\usepackage{amssymb}
\usepackage{graphicx}
\usepackage{amsmath}
\usepackage{flafter}
\usepackage{epstopdf}
\usepackage{afterpage}
\usepackage{color}
\usepackage{enumerate}
\usepackage{url}
\newcommand{\Sun}{\ensuremath{{\odot}}}
\definecolor{darkgreen}{rgb}{0.0, 0.4, 0.0}
\definecolor{darkblue}{rgb}{0.0, 0.0, 0.6}
\definecolor{darkred}{rgb}{0.6, 0.0, 0.0}
\title{AXIONS IN COLD DARK MATTER\protect\\ AND INFLATION MODELS}
\author{Luca Visinelli}
\thesistype{dissertation}
\graduatedean{Charles A. Wight}
\department{Department of Physics and Astronomy}
\degree{Doctor of Philosophy\\ \mbox{} \\ in \\ \mbox{}\\ Physics}
\departmentchair{David Kieda}
\committeechair{Paolo Gondolo}
\firstreader{Yong Shi Wu}
\secondreader{Wayne Springer}
\thirdreader{Oleg Starykh}
\fourthreader{Aaron Bertram}
\chairsigndate{9/26/2011}
\firstsigndate{9/26/2011}
\secondsigndate{9/26/2011}
\thirdsigndate{9/26/2011}
\fourthsigndate{9/26/2011}
\chairtitle{Professor}
\submitdate{December 2011}
\copyrightyear{2011}
\fourlevels
\dedication{A mia moglie Erika}

\begin{document}
\frontmatterformat
\titlepage
\copyrightpage
\committeeapproval
\preface{abstract}{Abstract}
\dedicationpage
\tableofcontents
\listoffigures
\listoftables
\preface{acronyms}{Acronyms used}
\preface{acknowledge}{Acknowledgements}
 \addcontentsline{toc}{groupheader}{\protect\vspace{-30pt}}
\addcontentsline{toc}{groupheader}{CHAPTERS}
\maintext
\include{chap1}
\include{chap2}
\include{chap3}
\include{chap4}

\include{chap5}
\include{chap6}

\include{chap7}

\bibliography{MyThesisRefs1}
\bibliographystyle{acm-plain}

\end{document}

%% file: chap1.tex
 \addcontentsline{toc}{groupheader}{\protect\vspace{-25pt}}
\chapter{Introduction}\label{Chapter_Introduction}

My interest in the invisible axion, a hypothetical pseudo-scalar particle arising in the QCD sector of the Standard Model, addresses the fact that this particle might play the role of the Cold Dark Matter (CDM). This possibility has been intensively studied for more than thirty years; original work on the subject by myself and my advisor prof. Paolo Gondolo is exposed in Chapters~\ref{Revising the axion as the cold dark matter} and~\ref{Dark matter axions in nonstandard cosmologies}, drawn from Refs.~\cite{Visinelli:2009zm} and~\cite{Visinelli:2009kt}, respectively. 

In Chapter~\ref{Revising the axion as the cold dark matter}, I considered the invisible axion as the CDM particle in the standard cosmological scenario, updating the bounds on the axion parameter space in light of the most recent cosmological data, and improving the treatment of anharmonicities in the axion potential and axionic strings. In Chapter~\ref{Dark matter axions in nonstandard cosmologies}, it is shown how the invisible axion can be used as a probe to study the properties of the universe before Big Bang nucleosynthesis (BBN) took place. In fact, we only know that the universe has to be radiation dominated (standard cosmology) at the time of BBN, but a plethora of different possibilities could have taken place before this period. We studied the properties of the axion CDM in two different nonstandard cosmological scenarios, the low temperature reheating (LTR) cosmology and the kination cosmology.

The last chapter of the thesis deals with the role of axion-like particles, a generalization of the invisible axion theory, as the inflaton in the specific scenario of warm inflation, and refers to the paper in Ref.~\cite{Visinelli:2011jy}. Axion-like particles have long been successfully considered as the inflaton particle in the standard inflaton, with the model going under the name of Natural Inflation (NI). One problem of the NI model resides in that the energy scale at which the axion symmetry spontaneously breaks is of the order of the Planck mass, which makes it difficult to embed NI into larger Grand Unification Theories (GUT).
In Chapter~\ref{Natural warm inflation}, I show that this problem can be overcome if we consider the warm inflation scenario in place of standard inflation, thus developing the natural warm inflation model (NWI).
 \addcontentsline{toc}{groupheader}{\protect\vspace{-33pt}}

%% file: chap2.tex
\chapter{Concordance Cosmology} \label{Concordance Cosmology }

A quantitative approach to cosmology began in the early 20th century, after Einstein's General Relativity (GR) was formulated. In fact, cosmology as we know today is firmly established on the basis of the GR theory, as we will explore in detail in Sec.~\ref{General relativity and cosmology}. Before relativity, space and time were considered fixed, an idea that directly followed from the Newtonian concept of an absolute framework in which physical processes would take place. The birth of GR provided scientists with a new concept, the fact that space-time is a dynamic entity whose evolution is described by Einstein equation. Within the GR theory, gravity can be understood as a geometric property of the space-time. Soon after the introduction of GR in its final version, the scientific community began speculating on the origin of the universe, merging observations with the new framework provided by the GR theory. The cosmological solutions of general relativity were first found, independently, by A. Friedmann, G. Lemaitre, H. P. Robertson, and A. G. Walker and constitute what is known as the Friedmann-Robertson-Walker (FRW) model (see Refs.~\cite{kolb1990early, WeinbergBook} for exhaustive reviews on the subject). 

\section{General relativity}\label{general_relativity}

The insight behind GR consists of equating the effects of the gravitational force on a test particle with the acceleration that such a test particle would experience on a nontrivial manifold. This idea is encoded in the weak equivalence principle, which states that local effects of motion in a curved space-time are indistinguishable from those of an accelerated observer in a flat space-time, and it is depicted by the famous elevator thought-experiment. With this key concept in mind, Einstein derived the famous formula
\begin{equation}\label{einstein_equation}
\mathcal{R}_{\mu\nu} - g_{\mu\nu}\,\mathcal{R} = 8\pi\,G\,T_{\mu\nu} + g_{\mu\nu}\,\Lambda.
\end{equation}
The term $\Lambda$ appearing in Eq.~(\ref{einstein_equation}) is a constant that might play an important role in the description of the universe at present times. We will discuss the effects of this term in depth in Secs.~\ref{term_Lambda} and~\ref{Particular solutions of the Friedmann equation}.

We first discuss the left-hand side of Eq.~(\ref{einstein_equation}). Here, $g_{\mu\nu}$ is the metric of the manifold considered, from which we can derive the Ricci tensor $\mathcal{R}_{\mu\nu}$ via the Riemann tensor $\mathcal{R}^\sigma_{\mu\nu\rho}$ as
\begin{equation}
\mathcal{R}_{\mu\nu} = \mathcal{R}^\sigma_{\mu\nu\sigma}.
\end{equation}
The Riemann tensor $\mathcal{R}^\sigma_{\mu\nu\rho}$ is defined through the Christoffel symbol (Levi-Civita connection) $\Gamma^\sigma_{\mu\nu}$ as
\begin{equation}
\mathcal{R}^\sigma_{\mu\nu\rho} = \partial_\mu\,\Gamma^\sigma_{\nu\rho} - \partial_\nu\,\Gamma^\sigma_{\mu\rho} + \Gamma^\gamma_{\mu\rho}\,\Gamma^\sigma_{\nu\gamma} - \Gamma^\gamma_{\nu\rho}\,\Gamma^\sigma_{\mu\gamma}.
\end{equation}
Here, the Levi-Civita connection is given by
\begin{equation}
\Gamma^\sigma_{\mu\nu} = \frac{1}{2}\,g^{\sigma\rho}\,\left(\partial_\nu\,g_{\rho\mu} + \partial_\mu\,g_{\nu\rho} - \partial_\rho\,g_{\mu\nu}\right),
\end{equation}
and $\partial_\mu$ means differentiation with respect to the manifold coordinate $x^\mu$.

In Eq.~(\ref{einstein_equation}), the curvature is $\mathcal{R} = g^{\mu\nu}\,\mathcal{R}_{\mu\nu}$, with the inverse of the metric $g^{\mu\nu}$ satisfying
\begin{equation}
g_{\mu\sigma}\,g^{\sigma\nu} = \delta_\mu^\nu.
\end{equation}
Sometimes the combination
\begin{equation}
G_{\mu\nu} = \mathcal{R}_{\mu\nu} - g_{\mu\nu}\,\mathcal{R},
\end{equation}
is referred to as the Einstein tensor.

On the right-hand-side of Eq.~(\ref{einstein_equation}), the term $T_{\mu\nu}$ describes the energy-momentum content at each point of space-time and is known as the stress-energy tensor. Since $T_{\mu\nu}$ depends implicitly on the metric tensor, in GR a feedback mechanism appears in which the metric structure of space-time and the matter content of the universe mutually influence each other, a statement that can be summed up with the paraphrase by Wheeler: {\it space-time tells matter how to move; matter tells space-time how to curve}. In the following, we will always use the case for an isotropic perfect fluid of density field $\rho$, pressure $p$ and four-velocity $u_\mu$, whose stress tensor reads
\begin{equation} \label{perfect_fluid}
T_{\mu\nu} = \rho\,u_\mu\,u_\nu + p\,(u_\mu\,u_\nu - g_{\mu\nu}).
\end{equation}
We remark that the density $\rho$ and the pressure $p$ are time-dependent quantities, although in the text, we do not indicate such dependence explicitly. In Sec.~\ref{The Friedmann equations}, we will discuss how the density and the pressure for a specific fluid are related by a specific equation of state.

\section{General relativity and cosmology} \label{General relativity and cosmology}

\subsection{The Friedmann-Robertson-Walker metric} \label{The Friedmann-Robertson-Walker metric}

In the FRW metric, the Copernican principle that demotes Earth's position in the Solar system is extended in a cosmological framework, by stating that no point in the universe is special with respect to the other ones. This concept, called the cosmological principle, can be restated by posing that the universe is isotropic with respect to any point, or equivalently, that it is homogeneous and isotropic with respect to one point. Isotropy and homogeneity can be understood in terms of cosmic inflation, see Sec.~\ref{inflation}, and lead to the FRW universe, with line element
\begin{equation} \label{FRW_metric}
ds^2 = dt^2 - a^2(t)\,\left[\frac{dr^2}{1-k\,r^2} + r^2\,d\theta^2 + r^2\,\sin^2\theta\,d\phi^2\right].
\end{equation}
Here $a(t)$ is the scale factor of the universe depending on the cosmic time $t$, and the term in square brackets indicates the spatial metric, which is specialized to represent a three-manifold of constant curvature $k$. The only possibilities for a spatial manifold of constant curvature are zero curvature with $k=0$ (flat universe), positive curvature with $k = +1$ (closed universe), and negative curvature with $k = -1$ (open universe). Sometimes the conformal time is defined,
\begin{equation}
d\tau = \frac{dt}{a(t)},
\end{equation}
so that the FRW metric is written in the conformal form,
\begin{equation}
ds^2 = a^2(t(\tau))\,\left(d\tau^2 - \frac{dr^2}{1-k\,r^2} - r^2\,d\theta^2 - r^2\,\sin^2\theta\,d\phi^2\right).
\end{equation}

\subsection{The Friedmann equations}\label{The Friedmann equations}

The Friedmann equations are a set of two equations that relate the expansion of the universe to its energy density and pressure content. In this section, we expose their derivation from the Einstein equation and some of their uses.

In the formulas used for computations, we trade Newton's gravitational constant $G$, appearing in Eq.~(\ref{einstein_equation}), with the Planck mass $M_{\rm Pl}$, using the relation
\begin{equation}
G = \frac{1}{M_{\rm Pl}^2}.
\end{equation}
For numerical computations, we will use $M_{\rm Pl} = 1.221\times 10^{19} {\rm ~GeV}$.

\subsubsection{Derivation of the Friedmann equations} \label{Derivation of the Friedmann equations}

It can be shown that the 00 and 11 components of the Einstein Eq.~(\ref{einstein_equation}), expressed on the FRW metric in Eq.~(\ref{FRW_metric}) and with the source in Eq.~(\ref{perfect_fluid}), give the relations
\begin{equation} \label{friedmann1}
\left(\frac{\dot{a}(t)}{a(t)}\right)^2 = \frac{8\pi}{3\,M_{\rm Pl}^2}\,\rho - \frac{k}{a^2} + \frac{\Lambda}{3},
\end{equation}
and
\begin{equation} \label{friedmann2}
\frac{\ddot{a}(t)}{a(t)} = -\frac{4\pi}{M_{\rm Pl}^2}\,\left(\rho+3p\right) + \frac{\Lambda}{3}.
\end{equation}
A dot over a quantity will always indicate a total derivative with respect to the cosmic time $t$, so $\dot{a}(t) = da(t)/dt$. Because of isotropy, the 22 and 33 components of Eq.~(\ref{einstein_equation}) give the same expression as Eq.~(\ref{friedmann2}), while the off-diagonal terms are all zero due to homogeneity. The role of the term $\Lambda$, also known as the vacuum energy term, will be discussed in detail in Sec.~\ref{term_Lambda}.

In a realistic cosmological model, more types of fluids are present. Examples include the nonrelativistic matter fluid, of density $\rho_m$ and pressure $p_m = 0$, and the radiation fluid, of density $\rho_r$ and pressure $p_r = \rho_r/3$. When more than one fluid are present in the theory, we define the total energy densities $\rho_{\rm tot}$ and the total pressure $p_{\rm tot}$ of the system as
\begin{equation}
\rho_{\rm tot} = \sum_i\,\rho_i,\quad p_{\rm tot} = \sum_i\,p_i,
\end{equation}
where $\rho_i$ and $p_i$ are the density and the pressure of the $i$-th fluid. To completely specify the system, we also need an equation of state linking the density and the pressure of each fluid,
\begin{equation} \label{equation_of_state}
p_i = w_i\,\rho_i,
\end{equation}
with the parameter $w_i$ specifying the type of fluid. For example, $w_m = 0$ is for a nonrelativistic fluid and $w_r = 1/3$ is for a relativistic fluid.

In the multiple fields case, we substitute $\rho \to\rho_{\rm tot}$ and $p \to p_{\rm tot}$ in the Friedmann Eqs.~(\ref{friedmann1}) and~(\ref{friedmann2}). This is the case we will discuss from now on. Defining the Hubble rate
\begin{equation} \label{hubble_rate}
H(t) = \frac{\dot{a}(t)}{a(t)},
\end{equation}
we obtain the Friedmann Eqs.~(\ref{friedmann1}) and~(\ref{friedmann2}) in the form
\begin{equation} \label{friedmann_equation}
H^2(t) = \frac{8\pi}{3\,M_{\rm Pl}^2}\,\rho_{\rm tot} - \frac{k}{a^2(t)} +\frac{\Lambda}{3},
\end{equation}
and
\begin{equation} \label{energy_conservation}
\dot{\rho}_{\rm tot} = -3H(t)\,(p_{\rm tot}+\rho_{\rm tot}).
\end{equation}

Unless otherwise specified, we will always derive our results in the next chapters in the case of flat geometry $k=0$ and negligible cosmological constant $\Lambda = 0$, which are suitable approximations in the cases we consider. For these reasons, we will refer to the Friedmann equation in the form
\begin{equation} \label{friedmann}
H^2(t) = \frac{8\pi}{3\,M_{\rm Pl}^2}\,\rho_{\rm tot}.
\end{equation}
We will keep the parameters $k$ and $\Lambda$ for the rest of the discussion in the current chapter.

\subsubsection{The term $\Lambda$} \label{term_Lambda}

In cosmology, the constant term $\Lambda$ is usually identified with the vacuum energy, responsible for the accelerated expansion we observe at the present time. Using Eq.~(\ref{friedmann1}), we can associate to the cosmological constant $\Lambda$ the energy density and the pressure 
\begin{equation}
\rho_\Lambda = \frac{\Lambda\,M_{\rm Pl}^2}{8\pi}, \quad p_\Lambda = -\rho_\Lambda.
\end{equation}
From now on, we include, in the definition of the terms $\rho_{\rm tot}$ and $p_{\rm tot}$, the contributions from $\rho_\Lambda$ and $p_\Lambda$, respectively. With these definitions, the Friedmann Eq.~(\ref{friedmann_equation}) is rewritten as
\begin{equation} \label{friedmann_equation_Lambda}
H^2(t) = \frac{8\pi}{3\,M_{\rm Pl}^2}\,\rho_{\rm tot}(t) - \frac{k}{a^2(t)}.
\end{equation}
From Eq.~(\ref{equation_of_state}), we see that the equation of state for the cosmological constant term has $w_\Lambda = -1$. Models in which only a cosmological constant is present have been studied since the early days of relativistic cosmology, and fall under the name of Lemaitre models.

To sum up, in order to include the contribution from the vacuum energy in the Friedmann equations, one has to add the energy density $\rho_\Lambda$ to the sum $\rho_{\rm tot}$, and similarly the pressure $p_\Lambda$ to the sum $p_{\rm tot}$, keeping in mind that the equation of state for the vacuum energy is $\rho_\Lambda = -p_\Lambda$.

\subsubsection{Equivalent forms of the Friedmann equations} \label{Equivalent forms of the Friedmann equations}

The Friedmann Eq.~(\ref{friedmann_equation_Lambda}) is particularly suitable for solving specific problems in which the content of the universe is specified and for explaining the thermal history of the universe. Defining the critical density,
\begin{equation} \label{critical_density}
\rho_{\rm crit} = \frac{3\,H^2(t)\,M_{\rm Pl}^2}{8\pi},
\end{equation}
and the cosmological density ratios,
\begin{equation}
\Omega_i(t) = \frac{\rho_i}{\rho_{\rm crit}},\quad \Omega_{\rm tot}(t) = \frac{\rho_{\rm tot}}{\rho_{\rm crit}} = \sum_i\,\Omega_i(t),
\end{equation}
we cast the Friedmann Eq.~(\ref{friedmann_equation}) in the alternative form
\begin{equation} \label{friedmann_equation1}
\frac{k}{a^2(t)} = H^2(t)\,(\Omega_{\rm tot}(t)-1).
\end{equation}
Eq.~(\ref{friedmann_equation1}) shows that the universe is
\begin{center}
\begin{enumerate}[{$\bullet$}]
\item open ($k = -1$) when $\Omega_{\rm tot}(t) < 1$,
\item flat ($k = 0$) when $\Omega_{\rm tot}(t) = 1$,
\item closed ($k = +1$) when $\Omega_{\rm tot}(t) > 1$.
\end{enumerate}
\end{center}

The second, alternative formulation of the Friedmann equation we show now is particularly suitable for computing the scale factor $a(t)$ given the energy content of the universe at a specific time, like for example at the present time $t_0$. We indicate the values of the density ratios at the present time with the index $0$, and the present value of the critical density is
\begin{equation}
\rho_c \equiv \rho_{\rm crit\,0} = \frac{3\,H_0^2\,M_{\rm Pl}^2}{8\pi} = 1.878\times 10^{-29}\,h^2\,{\rm g/cm^3},
\end{equation}
where the Hubble expansion rate at present time and the value of $h$ are \cite{Komatsu:2010fb}
\begin{equation}
H_0 = 100\,h\,{\rm km\,s^{-1}\,Mpc^{-1}} = (70.2 \pm 1.4)\,{\rm km\,s^{-1}\,Mpc^{-1}}.
\end{equation}
Introducing the density of curvature
\begin{equation}
\Omega_k(t) = -\frac{k}{a^2(t)\,H^2(t)},
\end{equation}
Eq.~(\ref{friedmann_equation}) is cast in the form
\begin{equation}\label{friedmann_omega}
\frac{H^2(t)}{H_0^2} = \Omega_r\,\left(\frac{a_0}{a(t)}\right)^4 + \Omega_m\,\left(\frac{a_0}{a(t)}\right)^3 + \Omega_k\,\left(\frac{a_0}{a(t)}\right)^2 + \Omega_\Lambda + \Omega_w\,\left(\frac{a_0}{a(t)}\right)^{3(1+w)}.
\end{equation}
Here and in the following, we have used the notation $\Omega_i = \Omega_i(t_0)$. We have included the possibility to include in the theory a fluid of density ratio $\Omega_w$ with generic equation of state $p = w\,\rho$ and $w$ unspecified. Notice that setting $a(t) = a(t_0) = a_0$ in Eq.~(\ref{friedmann_omega}) gives the constraint
\begin{equation}
\Omega_r + \Omega_m + \Omega_k + \Omega_\Lambda + \Omega_w = 1.
\end{equation}

\subsection{Redshifts} \label{Redshifts}

The redshift $z$ describes the lowering in the frequency of cosmological photons that have been emitted in the past with some frequency $\nu$ and are received at the present time with frequency $\nu_0$, and is defined as
\begin{equation}
1+z = \frac{\nu_i}{\nu_0}.
\end{equation}
Since it can be shown that the momentum of free particles moving on the FRW metric decreases with $1/a(t)$, as well as lengths and wavelength stretch proportionally to $a(t)$, the redshift at some time $t$ is related to the scale factor at that time by
\begin{equation}\label{definition_redshift}
1+z = \frac{a(t_0)}{a(t)}.
\end{equation}
A precise relation between redshift $z$ and time $t$ is given below in Eq.~(\ref{time_redshift}). Since the scale factor $a(t) \to 0$ for $t \to 0$ (Big Bang) and $a(t) \to a_0$ at present time, we also have $z \to +\infty$ when $t \to 0$ and $z = 0$ at the present time.

Using Eq.~(\ref{definition_redshift}), one can trade the time or scale factor dependence appearing in some equation with a $z$-dependence. In this view, a useful form of Eq.~(\ref{friedmann_omega}) that provides the Hubble expansion rate in terms of the redshift $z$ is
\begin{equation}\label{friedmann_omega_redshift}
H(z) = H_0\,\left[\Omega_r\,(1+z)^4 + \Omega_m\,(1+z)^3 + \Omega_k\,(1+z)^2 + \Omega_\Lambda + \Omega_w\,(1+z)^{3(1+w)}\right]^{1/2}.
\end{equation}

\subsection{Particular solutions of the Friedmann equation} \label{Particular solutions of the Friedmann equation}

We show the explicit solution of Eq.~(\ref{friedmann}) in various cosmologically interesting scenarios. In general, we expect the energy density $\rho_i$ that dominates the universe at some particular time to depend on the scale factor as
\begin{equation}
\rho_i = \rho_{i0}\,\left(\frac{a_0}{a}\right)^{3(1+w_i)}.
\end{equation}
The Friedmann Eq.~(\ref{friedmann}) with $\rho = \rho_i$ is written in the form
\begin{equation}
\left(\frac{\dot{a}}{a}\right)^2 = \frac{8\pi}{3\,M_{\rm Pl}^2}\,\rho_{i0}\,\left(\frac{a_0}{a}\right)^{3(1+w_i)} =  H_0^2\,\Omega_i\,\left(\frac{a_0}{a}\right)^{3(1+w_i)},
\end{equation}
with solution
\begin{equation}
a(t) = a_0\,\begin{cases}
\frac{3(1+w_i)}{2}\,\sqrt{\Omega_i}\,\left(H_0\,t\right)^{\frac{2}{3(1+w_i)}}, \quad\hbox{with $w_i \neq -1$},\\
{\rm Exp}\left[H_0\,t\,\sqrt{\Omega_i}\,\right], \quad\hbox{with $w_i = -1$}.
\end{cases}
\end{equation}
In particular, the solutions when the universe is dominated by matter ($w_m = 0$), radiation ($w_r = 1/3$), and vacuum ($w_\Lambda = -1$) are, respectively,
\begin{equation}
a(t) \propto t^{2/3}\quad\hbox{for matter domination},
\end{equation}
\begin{equation} \label{radiation_domination}
a(t) \propto t^{1/2}\quad\hbox{for radiation domination},
\end{equation}
\begin{equation} \label{friedmann_inflation}
a(t) \propto e^{\,t\,\sqrt{\Lambda/3}}\quad\hbox{for vacuum domination}.
\end{equation}
Notice that when $w_i = -1$, the scale factor expands exponentially. A family of cosmological models that use such accelerated expansion in the very early stages of the universe has been used in order to solve serious problems linked with the standard Big Bang model and fall under the name of inflation. We will discuss these problems and the solution posed by the inflationary scenarios in Sec.~\ref{inflation}. 

When more than one fluid is present, a general expression for the cosmological time $t$ as a function of the redshift $z$ can be found by integrating Eq.~(\ref{friedmann_omega_redshift}),
\begin{equation} \label{time_redshift}
t = \int_z^{+\infty}\,\frac{dz}{(1+z)\,H(z)};
\end{equation}
we derived Eq.~(\ref{time_redshift}) setting our initial time $t = 0$ at redshift $z = +\infty$. The age of the universe $t_0$ is given by Eq.~(\ref{time_redshift}) by setting $z = 0$, so
\begin{equation}
t_0 = \int_0^{+\infty}\,\frac{dz}{(1+z)\,H(z)}.
\end{equation}
For a universe where only matter is present we have
\begin{equation}\label{age_universe}
H_0\,t_0 = \begin{cases}
\frac{1}{1-\Omega_m} - \frac{\Omega_m}{2(1-\Omega_m)^{3/2}}\,{\rm ArcCosh}\left(\frac{2}{\Omega_m}-1\right), \quad\hbox{for $\Omega_m < 1$},\\
\frac{2}{3},\quad\quad\quad\quad\quad\quad\quad\quad\quad\quad\quad\quad\quad\quad\quad\quad\quad\,\hbox{for $\Omega_m = 1$},\\
\frac{-1}{\Omega_m-1}+\frac{\Omega_m}{2(\Omega_m-1)^{3/2}}\,{\rm ArcCos}\left(\frac{2}{\Omega_m}-1\right),\quad\,\,\,\,\hbox{for $\Omega_m > 1$}.
\end{cases}
\end{equation}
This equation, for $\Omega_m = 1$ and with $H_0 = 70 {\rm ~km/s/Mpc}$ yields $t_0 = 9.3$ Gyr, which underestimates the effective age of the universe by almost 35\%. To obtain the correct result for $t_0$, we need to include the dark energy contribution in Eq.~(\ref{time_redshift}); in fact, considering a flat ($k = 0$) universe where matter $\Omega_m$ and vacuum energy $1-\Omega_m$ are present (with negligible radiation), we obtain
\begin{equation} \label{age_universe_1}
H_0\,t_0 = \frac{1}{3\sqrt{1-\Omega_m}}\,\ln\left[\frac{2(1+\sqrt{1-\Omega_m})}{\Omega_m}-1\right].
\end{equation}
Eq.(\ref{age_universe_1}) with $H_0 = 70$ km/s/Mpc and $\Omega_m = 0.25$ gives the correct result for the age of the universe, $t_0 = 13.7$ Gyr. In Sec.~\ref{dark energy}, we will discuss how current observations motivate a nonzero value of $\Omega_\Lambda$.

\section{Cosmic inflation} \label{inflation}

The standard Big Bang theory yields striking successes in explaining a large number of cosmological observations. However, in order for this model to actually be consistent, the universe has to emerge from the Big Bang with very specific initial conditions, in order to match measurements of the quantities we observe at the present time. The most challenging problems that the standard Big Bang theory faces are the horizon, flatness, and unwanted relics problems. The inflationary scenario was introduced in order to solve these specific problems \cite{Kazanas:1980tx, Starobinsky:1980te, Guth:1980zm, Sato:1981ds, Albrecht:1982wi, Linde:1981mu}, and it also proved to be a valid model in which the seed of local inhomogeneities form \cite{Mukhanov:1981xt, Guth:1982ec, Hawking:1982cz, Starobinsky:1982ee, Bardeen:1983qw}. For reviews of the inflationary mechanism, see Refs.~\cite{linde1990particle, kolb1990early}.

\subsection{The flatness problem} \label{The flatness problem}

If the initial value of the total energy density of the universe slightly differs from the critical density, or $|\Omega_{\rm tot}(t) - 1| \neq 0$ in Eq.~(\ref{friedmann_equation1}), any deviation from unity will eventually be amplified by the present time; alternatively, it is required that the universe be extremely flat at early times, in order to explain the close-to-flat geometry we observe today.

To illustrate the flatness problem in cosmology, we assume that the universe had an initial deviation from unity $\Omega_{\rm tot, i} -1$. 
Indicating with $\Omega_{\rm tot} = \Omega_{\rm tot}(t_0)$ the value of $\Omega_{\rm tot}(t)$ at the present time $t_0$, we compute the deviation of $\Omega_{\rm tot}$ from unity by considering Eqs.~(\ref{friedmann_equation1}) in the form $k = \dot{a}^2\,(\Omega_{\rm tot, i}-1)$: assuming radiation domination at all times $a(t) \sim t^{1/2}$, we obtain
\begin{equation} \label{radiation_domination_density_perturbations}
|\Omega_{\rm tot, i}-1| = |\Omega_{\rm tot}-1|\,\left(\frac{t_{\rm Pl}}{t_0}\right),
\end{equation}
where $t_{\rm Pl} = 1/M_{\rm Pl} = 5.3\times 10^{-44}$ s is the Planck time and $t_0 = 13.7$ Gyr is the age of the universe. Numerically it is
\begin{equation}
|\Omega_{\rm tot, i}-1| \approx 10^{-60}\,|\Omega_{\rm tot}-1|.
\end{equation}
Measurements of the anisotropies in the Cosmic Microwave Background Radiation (CMBR) by the WMAP7+BAO+SN data constraint $|\Omega_{\rm tot}-1|$ at 95\% C.L. as \cite{Komatsu:2010fb}
\begin{equation}
-0.0178 < \Omega_{\rm tot}-1 < 0.0063;
\end{equation}
we state that $|\Omega_{\rm tot}-1| \lesssim 0.01$, implying that $\Omega_{\rm tot, i}$ initially differed from one by one part over $10^{62}$, any initial deviation of $\Omega_{\rm tot, i}$ from unity being amplified at the present time according to Eq.~(\ref{radiation_domination_density_perturbations}) above. Summing up, we have that the primordial universe has to exit from the Big Bang being extremely flat in order to explain the tiny anisotropy in the density observed. Inflation solves this problem because, during the inflationary period, the value on the right hand side of Eq.~(\ref{friedmann_equation1}) drops exponentially due to the exponential increase of the scale factor. Inflation has to last sufficiently long in order to solve the flatness problem: a parameter used to describe such a requirement is the number of e-folds $N_e$, defined as the logarithm of the ratio of the scale factors at the time when inflation ends $t_{\rm end}$ and at the beginning of inflation $t_i$,
\begin{equation}
N_e \equiv \ln \,\frac{a(t_{\rm end})}{a(t_{\rm Pl})} = \int_{t_{\rm Pl}}^{t_{\rm end}}\,H(t)\,dt.
\end{equation}
Sufficient inflation requires $N_e \gtrsim 60$.

\subsection{The horizon problem}

The horizon problem deals with the fact that a large degree of homogeneity is observed in the sky, although most of the patches have never been in causal contact before. For example, inhomogeneities in the CMBR are observed only at a $\delta T/T \sim 10^{-5}$ scale, even though the sky contains about a hundred patches that never interact causally in the standard picture. A very simplified view of this problem can be sketched by computing the age of the universe using Eq.(\ref{age_universe_1}) with $H_0 = 70$ km/s/Mpc and $\Omega_m = 0.25$, which yields $t_0 = 20$ Gyr. This computation overestimates the effective age of the universe $t_0  =14.3$ Gyr and implies that in the past, there was not enough time for the universe to be in causal contact. An early inflationary period would solve such a problem because in this scenario, an initial region that was in causal contact before inflation is then expanded and appears at the present time as distinct patches not causally connected.

\subsection{Unwanted relics}

Generic Grand Unification Theories (GUT) predict heavy particles and topological defects to be copiously produced in the early universe (see Ref.~\cite{Vilenkin:1984ib} for a review), to the extent that these relics and the particles emitted by them would eventually dominate the expansion of the universe \cite{Battye:1994au}. Since such relics are not observed in the present universe, it is believed that some other mechanisms have diluted their number to the present day. One viable explanation for this fact is that defects were produced before or during inflation, with monopoles being separated from each other (or ``washed out'') as the accelerated expansion progressed and their number density being consistently reduced to a safe cosmological value.

\subsection{Small-scale structures} \label{small-scale structures}

Since the standard FRW model describes a homogeneous and isotropic universe, it does not account for the structures we observe today at various scales, like stars, galaxies, and clusters of galaxies. For this reason, perturbations in the FRW metric have been long studied, see Refs.~\cite{Bardeen:1980kt, Kodama:1985bj}, and following the evolution of these perturbations, it is possible to explain the spectrum of inhomogeneities we observe in the CMBR. The initial power spectrum of perturbations that describes such variations in the densities will be reviewed in Sec.~\ref{Fluctuations during inflation}.

\subsection{Inflation building} \label{inflation_building}

A period of inflation is defined as a period of accelerated expansion: using Eq.~(\ref{friedmann2}) with $p = w\,\rho$, it follows that it must be
\begin{equation} \label{condition_inflation}
1 + 3w < 0.
\end{equation}
If the condition in Eq.~(\ref{condition_inflation}) is satisfied, a solution to the Friedmann Eq.~(\ref{friedmann1}) when $k = 0$ and $\rho$ is constant is (see Eq.~(\ref{friedmann_inflation}) above)
\begin{equation}\label{scale_factor_inflation}
a(t) = a(t_i)\,{\rm Exp}\left(t\,\sqrt{\frac{8\pi}{3\,M_{\rm Pl}^2}\,\rho}\,\right).
\end{equation}
An exponential growth like in Eq.~(\ref{scale_factor_inflation}) can be obtained from the Friedmann equation whenever the density, and thus the Hubble, are constant. This situation can be achieved in a $\Lambda$-dominated universe, the so-called Lemaitre models, or when we have some form of energy that is dominating the expansion rate of the universe whose energy density $\rho$ is constant. It is generally believed that a constant term $\Lambda$ might account for the accelerated expansion at the present time, see Sec.~\ref{dark energy}, whereas in order to give a microscopic explanation of primordial inflation, the early domination of an exotic form of energy has been invoked. It is usually postulated that the expansion rate of the universe during the inflation epoch is dominated by a hypothetical scalar field called the inflaton $\phi$. In most inflation models, the inflaton is initially stuck in a high energy state and it is the slow release of energy that governs the inflationary period. The dynamics of the inflation is governed by some flat potential $U(\phi)$, so that the energy density and pressure associated with this field are
\begin{equation}
\rho_\phi = \frac{1}{2}\,\dot{\phi}^2 + U(\phi),
\end{equation}
\begin{equation}
p_\phi = \frac{1}{2}\,\dot{\phi}^2 - U(\phi).
\end{equation}
If the field does not develop enough kinetic energy, $p_\phi = -\rho_\phi$ and $w_\phi = -1$, meeting the condition in Eq.~(\ref{condition_inflation}) for inflation to occur. If the value of the potential $U(\phi)$ remains approximately unchanged during the field evolution, then the density and the pressure do not change as well during the inflationary period and Eq.~(\ref{scale_factor_inflation}) applies with $\rho_\phi = U(\phi)$. A precise mechanism for inflation is yet to be found, the major problems facing inflaton models being new fine-tuning troubles concerning the self-interaction of the inflaton field itself. Moreover, the inflaton field is, in the simplest models, a scalar field, and motivations from string theories justify the use of such fields, although fundamental scalar fields are yet to be discover experimentally.  

\section{Fluctuations during inflation} \label{Fluctuations during inflation}

As mentioned in Sec.~\ref{small-scale structures}, one attractive feature of inflation is that scalar and tensor perturbations emerge during this epoch: these features later evolve into fluctuations in the primordial density and gravitational waves that might lead an imprint in the CMBR anisotropy and on the large scale structures \cite{Mukhanov:1981xt, Guth:1982ec, Hawking:1982cz, Starobinsky:1982ee, Bardeen:1983qw}. Each fluctuation is characterized by a power spectrum and a spectral index, respectively $\Delta^2_{\mathcal{R}}(k)$, $n_s$ for density perturbations and $\Delta^2_{\mathcal{T}}(k)$, $n_T$ for tensor perturbations.

\subsection{The scalar power spectrum} \label{The scalar power spectrum}

The spectrum of the adiabatic density perturbations generated by inflation is specified by the power spectrum $\Delta^2_{\mathcal{R}}(k)$ that depends mildly on the comoving wavenumber $k$ accordingly to a spectral index $n_s$  and its tilt $dn_s/d\ln k$ as \cite{Kosowsky:1995aa}
\begin{equation} \label{curvature_perturbations}
\Delta^2_{\mathcal{R}}(k) \equiv \frac{k^3\,P_{\mathcal{R}}(k)}{2\pi^2} = \Delta_{\mathcal{R}}^2(k_0)\,\left(\frac{k}{k_0}\right)^{n_s(k)-1}.
\end{equation}
The function $\Delta_{\mathcal{R}}^2(k)$ describes the contribution to the total variance of the primordial density perturbation due to perturbations at a given scale per logarithmic interval in $k$ \cite{Komatsu:2008hk}. The WMAP collaboration reports the combined measurement from WMAP7+BAO+SN of $\Delta_{\mathcal{R}}^2(k_0)$ at the reference wavenumber $k=k_0 = 0.002 {\rm ~Mpc^{-1}}$ \cite{Komatsu:2010fb}, 
\begin{equation} \label{constraint_power_spectrum}
\Delta^2_{\mathcal{R}}(k_0) = (2.430 \pm 0.091) \times 10^{-9},
\end{equation}
where the uncertainty refers to a 68$\%$ likelihood interval. The RHS of Eq.~(\ref{curvature_perturbations}) is evaluated when a given comoving wavelength crosses outside the Hubble radius during inflation, and the LHS when the same wavelength re-enters the horizon. In Eq.~(\ref{curvature_perturbations}), we have used the notation in Refs.~\cite{Komatsu:2008hk, Komatsu:2010fb} for the density perturbations. Other authors use the symbol $P_\mathcal{R}(k)$ for our $\Delta^2_\mathcal{R}(k)$ and $\mathcal{R}_k^2$ for our $P_\mathcal{R}(k)$, and might differ by factors of $2\pi^2$.

From a theoretical computation that describes the fluctuations in the inflaton field $\phi$, the scalar power spectrum has the form
\begin{equation}\label{scalar_spectrum}
\Delta^2_{\mathcal{R}}(k) = \left(\frac{H}{\dot{\phi}}\right)^2\,\langle|\delta \phi|^2\rangle,
\end{equation}
where $\dot{\phi}$ is the time derivative of the inflaton field and $\langle|\delta \phi|^2\rangle$ describes the variance of the fluctuations in the inflaton scalar field, related to the spectrum of fluctuations. The LHS of Eq.~(\ref{scalar_spectrum}) is computed at the time at which the largest density perturbations on observable scales are produced, corresponding $N_e$ e-foldings before the end of inflation. For any massless and nearly-massless fields, the theory of quantum fluctuations predicts \cite{Guth:1982ec}
\begin{equation} \label{quantum_fluct}
\langle|\delta \phi|^2\rangle_{\rm quantum} = \left(\frac{H}{2\pi}\right)^2.
\end{equation}
Since the Hubble rate $H$ is approximately constant during inflation, see Sec.~\ref{inflation_building}, we can substitute $H$ in Eq.~(\ref{quantum_fluct}) with its value when inflation ends $H_I$, with the index $I$ standing for ``Inflation'':
\begin{equation}\label{quantum_fluctuations}
\langle|\delta \phi|^2 \rangle_{\rm quantum} = \left(\frac{H_I}{2\pi}\right)^2.
\end{equation}
The Hubble expansion rate at the end of inflation $H_I$ is bound by the WMAP measurements, as we will discuss in Sec.~\ref{Bounds on the Hubble rate at the end of inflation}, and parametrizes the effectiveness of inflation. 
Using Eqs.~(\ref{scalar_spectrum}) and~(\ref{quantum_fluctuations}), the scalar power spectrum when quantum fluctuations in the inflaton field dominate is
\begin{equation} \label{power_spectrum_cool}
\Delta^2_{\mathcal{R}}(k_0) = \left(\frac{H_I^2}{2\pi\dot{\phi}}\right)^2.
\end{equation}

\subsection{The scalar spectral index} \label{The scalar spectral index}

The scalar spectral index $n_s$ describes the mild dependence of the scalar power spectrum on the wavenumber $k$, as in Eq.~(\ref{curvature_perturbations}). We expand the spectral index around the reference scale $k_0$ as
\begin{equation}
n_s(k) = n_s + \frac{1}{2}\,\tau\,\ln \frac{k}{k_0},
\end{equation}
where $n_s \equiv n_s(k_0)$, and the spectral tilt $\tau$ is
\begin{equation} \label{spectral_tilt}
\tau = \frac{dn_s(k)}{d\ln k/k_0}\bigg|_{k=k_0}.
\end{equation}
Using Eq.~(\ref{curvature_perturbations}), the scalar spectral index is \cite{Kosowsky:1995aa}
\begin{equation}\label{derivative_power_spectrum}
n_s-1 = \frac{\partial}{\partial \ln k/k_0}\,\ln\frac{\Delta^2_{\mathcal{R}}(k)}{\Delta^2_{\mathcal{R}}(k_0)}.
\end{equation}

\subsection{The tensor power spectrum} \label{The tensor power spectrum}

Fluctuations in the gravitational wave field are statistically described by a power spectrum for tensor perturbations $\Delta^2_h(k)$. Writing this spectrum in a similar fashion to $\Delta^2_{\mathcal{R}}(k_0)$, we have
\begin{equation} \label{tensor_perturbations}
\Delta^2_{\mathcal{T}}(k) \equiv \frac{k^3\,P_{\mathcal{T}}(k)}{2\pi^2} = \Delta_{\mathcal{T}}^2(k_0)\,\left(\frac{k}{k_0}\right)^{n_T},
\end{equation}
where the tensor spectral index $n_T$ is assumed to be independent of $k$, because current measurement cannot constraint its scale dependence.

WMAP does not constraint $\Delta_{\mathcal{T}}^2(k_0)$ directly, but rather the tensor-to-scalar ratio
\begin{equation} \label{def_r}
r \equiv \frac{\Delta_{\mathcal{T}}^2(k_0)}{\Delta_{\mathcal{R}}^2(k_0)}.
\end{equation}
which qualitatively measures the amplitude of gravitational waves per density fluctuations. The WMAP5+BAO+SN measurement constrains the tensor-to-scalar ratio as \cite{Komatsu:2010fb}
\begin{equation} \label{measure_r}
r < 0.20 \quad \hbox{at 95$\%$ C.L.}
\end{equation}

\subsection{Bounds on the Hubble rate at the end of inflation} \label{Bounds on the Hubble rate at the end of inflation}

We combine the results from the WMAP-7 plus BAO and SN in Eqs.~(\ref{constraint_power_spectrum}) and~(\ref{measure_r}) to obtain an upper bound on the spectrum of primordial gravitational waves,
\begin{equation}
\Delta^2_h(k_0) \lesssim 4.86\times 10^{-10}.
\end{equation}
Expressing $\Delta^2_h(k_0)$ in terms of $H_I$,
\begin{equation}
\Delta^2_h(k_0) = \frac{2 H^2_I}{\pi^2 M^2_{Pl}},
\end{equation}
leads to an upper bound on $H_I$,
\begin{equation} \label{HI_bound}
H_I < 6.0 \times 10^{14}{\rm ~GeV}.
\end{equation}
A lower limit on $H_I$ comes from requiring the Universe to be radiation-dominated at $T \simeq 4\,$MeV, so that primordial nucleosynthesis can take place \cite{Kawasaki:1999na, Kawasaki:2000en, Hannestad:2004px}. Equating the highest temperature of the radiation
\begin{equation} \label{maximum_T}
T_{MAX} \sim (T^2_{RH} H_I M_{Pl})^{1/4},
\end{equation}
to the smallest allowed reheating temperature $T_{RH} = 4\,$MeV gives
\begin{equation}
H_I > H(T_{RH}) = 7.2 \times 10^{-24}{\rm ~GeV}.
\end{equation}

\subsection{Exiting inflation}

After a certain period in which the universe experienced an exponential growth, a transition towards the standard cosmology occurred. In the literature, this transition period goes under the general name of ``reheating'', the designation referring to the idea that it was at this time that most of the particles and radiation that formed the primordial soup were created. Since the detailed mechanism behind inflation is still obscure, the theory of the reheating process is yet to be specified in its entirety. Generally speaking, inflation ends when the potential of the inflaton field is no longer flat enough for the exponential solution to occur: the inflaton field starts rolling down and oscillates around a minimum of its potential $U(\phi)$, with damped oscillations. The energy stored in the inflaton field is transferred through these damped oscillations into Standard Model particles and possibly other exotic particles, which make up the primordial soup \cite{Dolgov:1982th, Abbott:1982hn}. An alternative mechanism is the decay of the inflaton through a broad parametric resonance into intermediary particles, which then decay into Standard Model particles \cite{Dolgov:1989us, Traschen:1990sw, Kofman:1994rk}. In the standard cosmological scenario, the universe quickly becomes radiation-dominated, with Eq.~(\ref{radiation_domination}) describing the growth of the scale factor with time, $a(t) \sim t^{1/2}$. 

\section{Thermal history of the universe}

Right after the end of inflation and the subsequent re-ionization, the early universe was filled with a hot plasma of Standard Model particles and possibly dark matter and other exotic particles and forms of energy. At such high temperature, most interaction rates were capable of keeping these constituents in thermal equilibrium; a specific particle $i$ would go out of the thermal equilibrium when its annihilation rate into other particles $\Gamma_i$ falls below the Hubble expansion rate $H(T)$ at some temperature $T_{{\rm dec},i}$ defined via
\begin{equation}\label{decoupling_temperature}
H(T_{{\rm dec},i}) \approx \Gamma_i.
\end{equation}
From that moment on, the number density of such a relic is fixed to its value at temperature $T_{{\rm dec},i}$.

\subsection{Statistical mechanics at thermal equilibrium}

Because of the thermal equilibrium existing among particles participating in this primordial soup, we can use statistical tools for describing the properties of each species. Here we review these methods, following the treatments in Refs.~\cite{kolb1990early, Bergstrom:1999kd}. In statistical mechanics, the properties of a species $i$ of mass $m_i$ in a thermal bath at temperature $T$ are described by a distribution function over the momentum ${\bf p}$,
\begin{equation}
f_i({\bf p}) = \left[{\rm Exp}\left(\frac{E_i-\mu_i}{T}\right)\pm1\right]^{-1},
\end{equation}
where $E_i = \sqrt{{\bf p}^2 + m_i^2}$i is the total energy, $\mu_i$ is the chemical potential, and the minus sign (plus sign) specifies bosons (fermions). An additional number that characterizes the species is the number $g_i$ of degrees of freedom, describing the possible number of polarization of the species. In terms of these quantities, the number density $n_i$ and energy density $\rho_i$ are, respectively,
\begin{equation}
n_i = \frac{g_i}{(2\pi)^3}\,\int\,d^3p\,f_i({\bf p}),
\end{equation}
and
\begin{equation}
\rho_i = \frac{g_i}{(2\pi)^3}\,\int\,d^3p\,E_i\,f_i({\bf p}).
\end{equation}
The pressure of the $i$ fluid is obtained from the equation of state
\begin{equation}
p_i = w_i\,\rho_i,
\end{equation}
where $w_i$ is specified by the type of fluid itself, see Sec.~\ref{Derivation of the Friedmann equations}.

These quantities can be readily computed in the nonrelativistic $T \ll m_i$ and ultrarelativistic $T \gg m_i$ cases. In the former case we obtain, for both boson and fermion particles
\begin{equation}
n_i = g_i\,\left(\frac{m_i\,T}{2\pi}\right)^3\,e^{-m_i/T},\quad \hbox{for $T \ll m_i$},
\end{equation}
and
\begin{equation}
\rho_i = m_i\,n_i = g_i\,m_i\,\left(\frac{m_i\,T}{2\pi}\right)^3\,e^{-m_i/T},\quad \hbox{for $T \ll m_i$}.
\end{equation}
In the ultra-relativistic case, different results are obtained for the two statistics,
\begin{equation} \label{number_density_statmech}
n_i = \begin{cases}
\frac{\zeta(3)}{\pi^2}\,g_i\,T^3,\quad \hbox{for $T \gg m_i$ and Bose-Einstein statistics},\\
\frac{3}{4}\,\frac{\zeta(3)}{\pi^2}\,g_i\,T^3,\quad \hbox{for $T \gg m_i$ and Fermi-Dirac statistics},\\
\end{cases}
\end{equation}
where $\zeta(z)$ is the Riemann zeta function of argument $z$, and
\begin{equation} \label{energy_density_statmech}
\rho_i = \begin{cases}
\frac{\pi^2}{30}\,g_i\,T^4,\quad \hbox{for $T \gg m_i$ and Bose-Einstein statistics},\\
\frac{7}{8}\,\frac{\pi^2}{30}\,g_i\,T^3,\quad \hbox{for $T \gg m_i$ and Fermi-Dirac statistics}.\\
\end{cases}
\end{equation}
Notice that the number density in the Fermi-Dirac statistics differs from the one in the Bose-Einstein statistics for a factor $3/4$, and similarly the result for the energy density in the two cases differs by a factor $7/8$.

\subsection{Conservation of the number and entropy densities}

We now quote some important results concerning the evolution of a system of particles in thermal equilibrium, remanding to Ref.~\cite{Bergstrom:1999kd} for further details. The second Friedmann Eq.~(\ref{energy_conservation}) can be cast in the form
\begin{equation}
\frac{d}{dt}\,\left[a^3\,\frac{p+\rho}{T}\right] = 0,
\end{equation}
where the expression between square brackets is identified with the total entropy of the universe within a volume $a^3$ and $\rho = \sum_i\rho_i$. Introducing the entropy density
\begin{equation}
s(T) = \frac{p+\rho}{T},
\end{equation}
we have that the entropy density scales with the Hubble volume $a^3$. Using the fact that for a relativistic bath of particles $p = \rho/3$ and using the expressions for the various $\rho_i$ in Eq.~(\ref{energy_density_statmech}), we obtain
\begin{equation}\label{entropy_density}
s(T) = \frac{2\pi^2}{45}\,g_{*S}(T)\,T^3,
\end{equation}
with the entropy degrees of freedom $g_{*S}(T)$ at temperature $T$ being defined in a similar way as $g_*(T)$ in Eq.~(\ref{relativistic_dof}),
\begin{equation}\label{entropy_dof}
g_{*S}(T) \equiv \sum_{i\in {\rm bosons}}\,g_i\,\left(\frac{T_{{\rm dec},i}}{T}\right)^4 + \frac{3}{4}\sum_{j\in {\rm fermions}}\,g_j\,\left(\frac{T_{{\rm dec},j}}{T}\right)^4.
\end{equation}
For $T \gtrsim 1{\rm~MeV}$, we essentially have $g_{*S}(T) \approx g_*(T)$. From the conservation of the entropy in a comoving volume $a^3(T)$ and using Eq.~(\ref{entropy_density}), we derive the relation between the scale factor and temperature as
\begin{equation}\label{scale_factor}
g_*(T)\,T^3\,a^3(T) = {\rm~constant},
\end{equation}
valid only when no release of entropy occurs.

{}If there is no release of entropy and $g_{*S}(T)$ is constant, the total number density is also conserved,
\begin{equation}
\frac{dn}{dt} = 0.
\end{equation}
The two conditions for the conservation of the number and the entropy densities can be cast in a single expression \cite{kolb1990early},
\begin{equation}\label{conservation_comoving_number}
\delta\left(\frac{n}{s}\right) = 0.
\end{equation}

\subsection{Application to the radiation-dominated universe}\label{Application to the radiation-dominated universe}

We now turn our attention to a thermal bath of particles in cosmology. Considering an ensemble of relativistic particle species in thermal equilibrium, we evaluate the Hubble rate $H_{\rm rad}(T)$ from Eq.~(\ref{friedmann}) using the expression for the energy density in Eq.~(\ref{energy_density_statmech}) as
\begin{equation}\label{hubble_radiation.1}
H_{\rm rad}^2 = \frac{8\pi}{3\,M_{\rm Pl}}\,\sum_i\,\rho_i = g_*(T)\,\frac{8\pi^3}{90\,M_{\rm Pl}}\,T^4,
\end{equation}
where we defined the number of relativistic degrees of freedom at temperature $T$ for an ensemble of species $i$ that decouple at temperature $T_{{\rm dec},i}$ in Eq.~(\ref{decoupling_temperature}) as
\begin{equation}\label{relativistic_dof}
g_*(T) \equiv \sum_{i\in {\rm bosons}}\,g_i\,\left(\frac{T_{{\rm dec},i}}{T}\right)^4 + \frac{7}{8}\sum_{j\in {\rm fermions}}\,g_j\,\left(\frac{T_{{\rm dec},j}}{T}\right)^4.
\end{equation}
Numerically, the function that gives the number relativistic degrees of freedom as a function of temperature for values of $T$ around the QCD scale $\Lambda_{\rm QCD}$ can be approximated by a step function,
\begin{equation} \label{relativistic_dof_numerical}
g_*(T) =
\begin{cases}
61.75, & {\rm for}\,\, T \gtrsim \Lambda_{\rm QCD},\\
10.75, & {\rm for}\,\, \Lambda_{\rm QCD} \gtrsim T \gtrsim 4{\rm MeV},\\
 3.36, & {\rm for}\,\, T \lesssim 4{\rm MeV}.
\end{cases}
\end{equation}
When all standard model particles can be treated as relativistic, we have $g_*(T) = 106.75$, while the number of relativistic degrees of freedom in the Minimal Supersymmetric Standard Model (MSSM) is $g_*(T) = 228.75$.

We simplify Eq.~(\ref{hubble_radiation.1}) by writing
\begin{equation}\label{hubble_radiation}
H_{\rm rad} = \sqrt{g_*(T)\,\frac{8\pi^3}{90}}\,\frac{T^2}{M_{\rm Pl}} = 1.66\,\sqrt{g_*(T)}\,\frac{T^2}{M_{\rm Pl}}.
\end{equation}
This expression describes the Hubble rate when the universe is dominated by radiation.

\subsection{Matter-dominated universe}

As the universe cools down, relativistic particles lose momentum due to the redshift effect and eventually become nonrelativistic. For a stable nonrelativistic particle, the number density scales with $n \sim a^{-3} \sim T^3$, with the latter relation coming from Eq.~(\ref{scale_factor}). Knowing the matter energy density at the present time $\rho_M$, the energy density at a time $t$ is
\begin{equation}
\rho_{\rm matter} = \rho_M\,\left(\frac{a_0}{a}\right)^3,
\end{equation}
from which the Hubble rate for a matter-dominated universe scales as $H(T)\sim T^{3/2}$.

\section{Content of the universe at the present time}

Even though we are still lacking a fundamental theory for inflation, the inflationary scenario has been embedded in the history of the cosmos due to its capability in solving all of the problems posed in Sec.~\ref{inflation} and make predictions on the amplitude of the seeds for inhomogeneities. This paradigm on the history of the universe explains the exceptionally flat universe we live in, $\Omega_{\rm tot,0} \approx 1$, as evidenced from distinct measurements on the CMBR, the baryon acoustic oscillations and the redshift of supernovae. These same measurements also point out that the expansion of the universe is accelerating at the present time, a fact that is in sharp contrast with the naive expectation that the universe be matter-dominated at the present time. Clearly, the specific content of the universe is yet to be determined, although in the last decade we have been able to determine some general features of the different fluids that make it up. Here, we review the most important components of the present total energy density.

\subsection{Baryons}

In cosmology, the term ``baryon'' indicates the totality of the Standard Model species, and not only the color-neutral bound system made of three quarks in the acceptation of particle physics. The WMAP-7 data constrain the density of baryons today as
\begin{equation}
\Omega_b = 0.0458 \pm 0.0016.
\end{equation}

\subsection{Dark matter}

It has long been known that baryons only account for a very small fraction of the present energy density. This conclusion was first obtained from considering the rotation curves of outer objects in galaxies, which reveal that the average galactic mass not only consists of dust and gases, but also and for its most part of a nonluminous halo of unknown composition, hence dubbed dark matter.

The majority of the dark matter observed has to be in the form of CDM, which means that this exotic component has to be nonrelativistic at the time of galaxy formation. The fact that dark matter is a nonrelativistic field has been established with the first results on the CMBR anisotropy from COBE. In the CDM theory, small structures clump and grow hierarchically ``from the bottom up'', forming larger structures. This scenario is opposite to the Hot Dark Matter (HDM) paradigm, in which larger structure form earlier and subsequently fragment, following a ``top - down'' evolution. The predictions of the CDM model are in general agreement with the observations, whereas the HDM paradigm disagrees with large-scale structure observations. The WMAP 7-years data, once combined with the BAO and SNe data, yield the value
\begin{equation}\label{CDM}
\Omega_{\rm CDM} = 0.229 \pm 0.015.
\end{equation}
for the totality of CDM observed in the present universe.

Although the CDM paradigm explains current data and the evolution of large-scale structures, there are some major discrepancies with observation of other features within galaxies and clusters of galaxies. In particular, CDM models predict that the density distribution of dark matter halos be much more peaked than what is inferred by the rotation curves of galaxies, the so-called cuspy halo problem. Moreover, CDM models predict a large amount of low angular momentum dust, in contrast with observations.

\subsection{Dark energy} \label{dark energy}

The dark matter and baryon components are not sufficient to explain the flatness observed, since the total abundance of nonrelativistic matter only accounts for about 26\% of the total content of the Universe. Instead, what comes out of the measurements on the content of the universe reveals that a large fraction of the present energy density is due to the so-called dark energy, responsible for the current period of accelerated expansion. In some models, dark energy is identified with the constant $\Lambda$ appearing in the Friedmann Eq.~(\ref{friedmann_equation}) and which would lead to an accelerated expansion as discussed in Sec.~\ref{Particular solutions of the Friedmann equation}. For anthropic reasons for this choice, see Ref.~\cite{Weinberg:1987dv}. Another popular explanation for the dark energy introduces a new light scalar field whose equation of state, similarly to the inflaton field $\phi$, resembles that of a cosmological constant. Mechanisms of this latter type include quintessence \cite{Ratra:1987rm, Wetterich:1987fm}, or the landscape of string theory (see Ref.~\cite{Susskind:2003kw}).

Evidence that the universe is experiencing a period of accelerated expansion comes from measuring the distances of type Ia supernovae (SNe). In fact, the lifespan of a SN is directly correlated with its luminosity, so that SNe can be used as standard candles to measure distances of neighboring stars to us via the ``cosmic distance ladder'' technique. Experiments reporting distances of SNe have shown that a nonzero cosmological term $\Omega_\Lambda$ better fits data than a vanishing $\Omega_\Lambda$. For a flat universe, the favored region of the parameter space has
\begin{equation}
\Omega_\Lambda = 0.725 \pm 0.016.
\end{equation}

%% file: chap3.tex
\chapter{Elements of quantum field theory and QCD}\label{Elements of field theory and QCD}

\fixchapterheading

\section{Elements of group theory} \label{Elements of group theory}

Group theory is a branch of mathematics devoted to the study of groups. A group $G$ is a mathematical structure consisting of a set $V$ together with an operation $\cdot$ that acts on a pair of elements in $V$. For this reason, the particular group structure is often explicitly indicated by writing $G = G(V,\,\cdot)$. In order for $G$ to qualify as a group, the operation $\cdot$ must satisfy the following conditions:
\begin{enumerate}
\item {\it Closure}: $\forall \,x, y \in V, x\cdot y \in V$;
\item {\it Associativity}: $\forall\, x, y, z \in V, (x\cdot y)\cdot z = x\cdot (y\cdot z)$;
\item {\it Identity}: $\exists !\, e \equiv \,x \cdot e = e\cdot x = e\, \forall x \in V$;
\item {\it Invertibility}: $\forall x \in\, V\, \exists !\, x^{-1} \in V \equiv x\cdot x^{-1} = x^{-1}\cdot x = e$.
\end{enumerate}
If further the {\it commutativity} relation holds, the group $G(V,\,\cdot)$ is called Abelian.

Of central importance in the context of Quantum Field Theory (QFT) is the concept of the compact Lie group. The adjective {\it compact} refers to the compactness of the topology of the group, namely the topological space associated to the group $G$ is compact; the notion of the {\it Lie} group refers to the smoothness of the differentiable manifold associated to $G$. Due to the importance of the subject, the theory of compact Lie groups is particularly well-developed.

The properties of a compact Lie group assure that the group $G$ contains elements $x(\alpha)$ that are arbitrarily close to the identity $e$ for small values of the group parameter $\alpha$. For continuous groups such as Lie groups, we can write
\begin{equation}\label{close_unity}
x(\alpha) = e + i\,\alpha^a\,T^a + \mathcal{O}(\alpha^2),
\end{equation}
where the index $a$ runs over the dimension of the group and the $T^a$'s are called the generators of the group. These generators satisfy the commutation relation
\begin{equation}\label{structure_constant}
[T^a, T^b] = i\,f^{abc}\,T^c,
\end{equation}
and the coefficients $f^{abc}$ are called the structure constants of the group. If the set of generators cannot be further divided into two subsets of mutually commuting generators, the generated group is called simple. For compact Lie groups that are simple, a complete classification of these groups is known \cite{AdamsBook}, and it is indicated as the Cartan classification system. Given two $N$-dimensional complex vectors $u$ and $v$, a general transformation is, respectively, a
\begin{enumerate}

\item {\it unitary transformations} if it preserves the inner product $u^*_a\, v_a$. A unitary transformation belongs to the so-called special unitary group in $N$ dimensions, $SU(N)$. The generators $t^a$ of this group are $N\,\times\, N$ traceless Hermitian matrices, and there are $N^2-1$ of these generators.

\item {\it orthogonal transformations} if it preserves the inner product $u_a \,\delta_{ab}\,v_b$.  An orthogonal transformation belongs to the so-called orthogonal group in $N$ dimensions, $SO(N)$. The elements of this group are $N\,\times\, N$ orthogonal matrices of determinant one, and there are $N(N-1)/2$ generators.

\item {\it symplectic transformations} if it preserves the inner product $u_a \,E_{ab}\,v_b$, with $E_{ab}$ being the symplectic matrix. A symplectic transformation belongs to the so-called symplectic group in $N$ dimensions ($N$ even), $Sp(N)$. There exist $N(N+1)/2$ generators.
\end{enumerate}
There also exist five {\it exceptional} Lie groups that are simple but do not belong to the above classification. These exceptional groups are of interest in unified theories, but they will not be treated in this thesis.

Summing up, the dimension of the simple Lie groups, not including the exceptional groups, can be written as
\begin{equation}
d(G) = 
\begin{cases}
N^2-1 \quad\hbox{for $SU(N)$},\\
N(N-1)/2 \quad\hbox{for $SO(N)$},\\
N(N+1)/2 \quad\hbox{for $Sp(N)$}.
\end{cases}
\end{equation}

In the following, we will specify the relations in Eq.~(\ref{structure_constant}) for some groups of interest in particle physics, namely SU(2), SO(3), SU(3), and the homogeneous Lorentz group SO(1,3).
\newpage

\section{Lagrangian functions for scalar and spinor fields} \label{Lagrangian functions for scalar and spinor fields}

\subsection{The Lorentz group}

As discussed in Sec.~\ref{Elements of group theory}, a group can be defined through the algebra of its generators. For the case of the Lorentz group SO(1,3), we have that the six generators for the angular momentum $J_i$ and for the boost $K_i$, with $i \in \{1,2,3\}$, satisfy the commutation relations:
\begin{equation} \label{def_SO3}
[J_i,J_j] = i\epsilon_{ijk}\,J_k,
\end{equation}
\begin{equation} \label{commutator1.1}
[K_i,K_j] = -i\epsilon_{ijk}\,J_k,
\end{equation}
\begin{equation} \label{commutator1.2}
[J_i,K_j] = i\epsilon_{ijk}\,K_k.
\end{equation}
We see from Eq.~(\ref{def_SO3}) that the three $J_i$ form a closed subalgebra of the Lorentz group, more precisely the algebra that defines the SU(2) or SO(3) Lie groups. Representations for both these groups are three hermitian matrices of dimension $2n + 1$, with $n$ assuming integer values when referring to a representation of the SO(3) group, or $n$ taking half-integer values when representing the SU(2) group. Nevertheless, each specific value of $n$ in physics refers to a different particle that is associated with this group. For each different choice of $n$, the wave function for such a particle will transform differently when a Lorentz transformation is applied. We will focus only on the cases $n = 0$ (boson) and $n=1/2$ (Weyl spinor), with other values of $n$ not being treated here.

\subsection{The Lorentz transformation}

In the following, we denote with ${\Lambda^\mu}_\nu$ the matrix describing the homogeneous linear transformation of a four-vector $x^\mu$,
\begin{equation}\label{lorentz_transform}
x^\mu \to x'^\mu = {\Lambda^\mu}_\nu\,x^\nu.
\end{equation}
Imposing the invariance of the pseudo-length of the four-vector,
\begin{equation}
x^2 = x^\mu\,x_\mu = \eta_{\mu\nu}\,x^\mu\,x^\nu = x_0^2-x_1^2-x_2^2-x_3^2,
\end{equation}
we find the relation
\begin{equation} \label{Lambda_relation}
{\Lambda_\sigma}^\mu\,{\Lambda_\rho}^\nu\,\eta_{\mu\nu} = \eta_{\sigma\rho}.
\end{equation}
A Lorentz transformation, represented by the matrix $\Lambda$, belongs to the Lorentz group SO(1,3), and thus the product of two Lorentz matrices is a Lorentz matrix; the identity $\delta^\mu_\nu$ belongs to the group and a specific transformation $\Lambda$ has an inverse $\Lambda^{-1}$.

Taking the determinant of Eq.~(\ref{Lambda_relation}), we see that 
\begin{equation}
{\rm Det}\,\Lambda = \pm 1,
\end{equation}
so that Lorentz transformations are called {\it proper} if ${\rm Det}\,\Lambda = +1$ and {\it improper} if ${\rm Det}\,\Lambda = -1$. A further, independent subdivision can be made by noticing that Eq.~(\ref{Lambda_relation}) implies that
\begin{equation}
\left({\Lambda^0}_0\right)^2 = 1 + \left({\Lambda^i}_0\right)^2,
\end{equation}
so that a transformation is {\it orthochronous} when ${\Lambda^0}_0 \geq 1$, and nonorthochronous when ${\Lambda^0}_0 \leq -1$.

{}Transformations that are both proper and orthochronous are connected to the identity; in particular, we can write an infinitesimal transformation as
\begin{equation}\label{infinitesimal_Lorentz}
{\Lambda^\mu}_\nu = \delta^\mu_\nu + {\epsilon^\mu}_\nu,
\end{equation}
where the matrix of infinitesimal quantities ${\epsilon^\mu}_\nu$ is forced by Eq.~(\ref{Lambda_relation}) to be antisymmetric:
\begin{equation}
\epsilon_{\mu\nu} = -\epsilon_{\nu\mu}.
\end{equation}
These are six parameters, corresponding to the three spatial rotations and the three boosts. Explicitly, a Lorentz transformation that rotates the system by an infinitesimal angle $\delta \theta$ about the direction ${\bf \hat{n}}$ and gives an infinitesimal boost along ${\bf \hat{n}}$ with rapidity $\delta \eta$ to the system has parameters
\begin{equation}
\epsilon_{ij} = -\epsilon_{ijk}\,\hat{n}_j\,\delta\theta,\quad \epsilon_{i0} = \hat{n}_i\,\delta \eta.
\end{equation}

\subsection{Equation of motion for a scalar field} \label{Equation of motion for a scalar field}

We consider a generic complex scalar field $\Phi = \Phi(x)$. The term ``scalar'' indicates that, when a Lorentz transformation $\Lambda$ is applied to the coordinate four-vector $x$ as in Eq.~(\ref{lorentz_transform}), the scalar field transforms as
\begin{equation}
\Phi(x) \to \Phi'(x) = \Phi(\Lambda^{-1}\,x').
\end{equation}
The Lagrangian density for a classical scalar field $\Phi$ moving in a potential $U(\Phi)$ is
\begin{equation}
\mathcal{L} = \partial^\mu\Phi^*\,\partial_\mu\Phi - U(\Phi),
\end{equation}
and the equation of motion is
\begin{equation} \label{scalar_eq_motion}
\partial^\mu\partial_\mu\,\Phi - \frac{\partial U(\Phi)}{\partial \Phi^*} = 0.
\end{equation}
In particular, when the potential is quadratic in the field,
\begin{equation}
U(\Phi) = m^2\,|\Phi|^2,
\end{equation}
Eq.~(\ref{scalar_eq_motion}) is cast as the Klein-Gordon equation
\begin{equation} \label{scalar_eq_motion1}
\left(\partial^\mu\partial_\mu - m^2\right)\,\Phi(x) = 0.
\end{equation}
A similar equation holds for the complex conjugate field $\Phi^*(x)$.

\subsection{Equation of motion for a spinor field} \label{Equation of motion for a spinor field}

\subsubsection{Representation of dimension two for the Lorentz transformation} \label{Left- and right-hand Lorentz transformations}

We focus again on Eqs.~(\ref{def_SO3})-(\ref{commutator1.2}) that give the algebra of the Lorentz group in terms of the generators. We define two sets of linear combinations of these generators,
\begin{equation}
N_i = \frac{1}{2}\,\left(J_i - i\,K_i\right),
\end{equation}
and
\begin{equation}
N_i^\dag = \frac{1}{2}\,\left(J_i + i\,K_i\right),
\end{equation}
so that the Lorentz algebra is rewritten as
\begin{equation}
[N_i,N_j] = i\epsilon_{ijk}\,N_k,
\end{equation}
\begin{equation}
[N_i^\dag,N_j^\dag] = -i\epsilon_{ijk}\,N_k^\dag,
\end{equation}
\begin{equation}
[N_i,N_j^\dag] = 0.
\end{equation}
This means that the algebra of each set $N_i$ and $N_i^\dag$ is separately closed, and the Lorentz algebra consists of two sets of SU(2) algebras, whose generators are hermitian conjugate. For this reason, a Lorentz transformation is specified by two indices $n_1$, $n_2$, and the representation has dimension $(2n_1 + 1)\times(2n_2+1)$. The convention for Weyl spinors is that the choice $(n_1 = 1/2, n_2 = 0)$ leads to a left-handed transformation, while $(n_1 = 1/2, n_2 = 0)$ yields a right-handed transformation. These are two distinct representations of the Lorentz group with dimension two.

\subsubsection{Weyl spinors}

Because of the two different representations of the Lorentz group with the same dimension of two, two distinct type of spinors are possible, called the left-handed and right-handed Weyl spinors. These two representations are distinguished by the action of the boost transformations $K_i$, which in this representation are
\begin{equation}
K_i^{\pm} = \pm\frac{i}{2}\,\sigma_i,
\end{equation}
with the plus (minus) sign acting on the left (right) handed Weyl spinor. For both spinors, the generators of the angular momentum are given by
\begin{equation}
J_i = \frac{1}{2}\,\sigma_i.
\end{equation}
The nonzero entries of the left-handed generator $T_L^{\mu\nu}$ are defined as
\begin{equation}
T_L^{ij} = \epsilon^{ijk}\,J_k = \frac{1}{2}\,\epsilon^{ijk}\,\sigma_k,\quad\hbox{and}\quad T_L^{k0} = K^{k+} = \frac{i}{2}\,\sigma_i.
\end{equation}

We now discuss the action of a generic Lorentz transformation $\Lambda$ on these spinors, considering first a left-handed Weyl spinor. In the following, we write the left-handed spinor as $\chi_\alpha$, with the spinor index $\alpha$ raised and lowered by the spinor metric tensor $\epsilon_{\alpha\beta}$. Under the transformation in Eq.~(\ref{lorentz_transform}), the left-handed Weyl spinor transforms as
\begin{equation}
\chi_\alpha(x) \to \chi'_\alpha(x) = {L_\alpha}^\beta(\Lambda)\,\chi_\beta(\Lambda^{-1}x).
\end{equation}
Here, $L(\Lambda)$ is a matrix in the (1/2, 0) representation of the Lorentz group: for the infinitesimal Lorentz transformations in Eq.~(\ref{infinitesimal_Lorentz}), it is
\begin{equation}
{L_\alpha}^\beta(1+\epsilon) = {\delta_\alpha}^\beta + \frac{i}{2}\,{\left(T_L^{\mu\nu}\right)_\alpha}^\beta\,\epsilon_{\mu\nu},
\end{equation}
where we have indicated explicitly the spinor structure of the generator $T_L$, that is, the entries of the specific matrix with given $\mu$ and $\nu$.

The right-handed Weyl spinor is defined as
\begin{equation}
\chi_{\dot{\alpha}}^\dag = [\chi_\alpha(x)]^\dag.
\end{equation}
A dotted index $\dot{\alpha}$ is used for the right-handed spinor, in order to distinguish it from the undotted index for the left-handed spinor. The metric tensor used for raising and lowering dotted indices is $\epsilon^{\dot{\alpha}\dot{\beta}}$. Under a Lorentz transformation, the right-handed Weyl spinor transforms as
\begin{equation}
\chi_{\dot{\alpha}}(x) \to \chi'_{\dot{\alpha}}(x) = {R_{\dot{\alpha}}}^{\dot{\beta}}(\Lambda)\,\chi_{\dot{\beta}}(\Lambda^{-1}x),
\end{equation}
where $R(\Lambda)$ is a matrix in the (0,1/2) representation of the Lorentz group. For an infinitesimal Lorentz transformation, we have
\begin{equation}
{R_{\dot{\alpha}}}^{\dot{\beta}}(1+\epsilon) = {\delta_{\dot{\alpha}}}^{\dot{\beta}} + \frac{i}{2}\,{\left(T_R^{\mu\nu}\right)_{\dot{\alpha}}}^{\dot{\beta}}\,\epsilon_{\mu\nu},
\end{equation}
where $T_R^{\mu\nu}$ is a set of matrices, linked to the set of $T_L^{\mu\nu}$ matrices by
\begin{equation}
{\left(T_R^{\mu\nu}\right)_{\dot{\alpha}}}^{\dot{\beta}} = -\left[{\left(T_L^{\mu\nu}\right)_\alpha}^\beta\right]^*.
\end{equation}

\subsubsection{Equation of motion for the Weyl spinor} \label{Equation of motion for the Weyl spinor}

Under specific conditions, it is possible to study the equation of motion for just one spinor, say the left-handed spinor $\chi_\alpha(x)$. The Lagrangian is \cite{SrednickiBook}
\begin{equation}\label{Lagrangian_LHS_Weyl}
\mathcal{L}_L = i\chi_{\dot{\alpha}}^\dag\,\left(\sigma^\mu\right)^{\dot{\alpha}\beta}\,\partial_\mu\chi_\beta + \frac{m}{2}\,\left(\chi^\alpha \chi_\alpha + \chi^{\dag\dot{\alpha}}\chi^\dag_{\dot{\alpha}}\right),
\end{equation}
with the set of matrices
\begin{equation}
\sigma^\mu = (I_2, {\bf \sigma}), \quad \bar{\sigma}^\mu = (I_2, -{\bf \sigma}).
\end{equation}
From Eq.~(\ref{Lagrangian_LHS_Weyl}), we derive two equations of motion, one for $\chi_\alpha$ and one for $\chi_{\dot{\alpha}}^\dag$ as
\begin{equation}
i\left(\sigma^\mu\right)_{\alpha\dot{\beta}}\,\partial_\mu\,\chi^{\dag\dot{\beta}} - m\,\chi_\alpha = 0,
\end{equation}
and
\begin{equation}
i\left(\bar{\sigma}^\mu\right)^{\dot{\alpha}\beta} \,\partial_\mu\,\chi_\beta -m\,\chi^{\dag\dot{\alpha}} = 0.
\end{equation}
These last two equations can be combined in the matrix form
\begin{equation}
\left(\begin{array}{cc} -m\,{\delta_\alpha}^\beta & i\left(\sigma^\mu\right)_{\alpha\dot{\beta}}\,\partial_\mu\\ i\left(\bar{\sigma}^\mu\right)^{\dot{\alpha}\beta} \,\partial_\mu & -m\,{\delta^{\dot{\alpha}}}_{\dot{\beta}} \end{array} \right) \, \bigg(\begin{array}{c} \chi_\beta \\ \chi^{\dag\dot{\beta}} \end{array} \bigg) = 0.
\end{equation}

\subsection{Equation of motion for the Majorana spinor} \label{Equation of motion for the Majorana spinor}
We define the four-spinor
\begin{equation}
\Psi_M = \left(\begin{array}{c} \chi_\alpha \\ \chi^{\dag\dot{\alpha}} \end{array} \right),
\end{equation}
which is known in the literature as the Majorana spinor. The key element characterizing the Majorana spinor is that it is invariant under the charge conjugation operation, that is, when we invert the electric charge of the particle, we obtain the same spinor. In order for this to be consistent, the electric charge of a Majorana field must be zero. To see this in more detail, we introduce two operations that act on generic four-spinors $\Psi$.

{}A bar over a spinor $\Psi$ indicates the operation
\begin{equation}
\bar{\Psi} = \Psi^\dag\,\beta,
\end{equation}
with the matrix
\begin{equation} \label{beta_matrix}
\beta =  \left(\begin{array}{cc} 0 & {\delta^{\dot{\alpha}}}_{\dot{\beta}}\\  {\delta_\alpha}^\beta & 0 \end{array} \right).
\end{equation}
For the Majorana field,
\begin{equation}\label{bar_majorana}
\bar{\Psi}_M = \left(\chi^\alpha, \chi^\dag_{\dot{\alpha}}\right).
\end{equation}
We also introduce the charge conjugation operator $\hat{C}$, whose matrix representation in four dimensions reads
\begin{equation}
\mathcal{C} =  \left(\begin{array}{cc} \epsilon_{\alpha\beta} & 0\\  0 & \epsilon^{\dot{\alpha}\dot{\beta}} \end{array} \right).
\end{equation}
Given a generic four-spinor $\Psi$, the charge-conjugated spinor is
\begin{equation}
\Psi^C = \mathcal{C}\,\bar{\Psi}^T.
\end{equation}
For a Majorana field, we obtain indeed $\Psi^C_M = \Psi_M$. Examples of Majorana particles are the chargeless supersymmetric partners of the Higgs, $Z^0$ and photon. A question mark is still posed on the nature of the neutrino, which may also be described by a Majorana field.

{}Defining the Dirac matrices in the chiral representation
\begin{equation}
\gamma^\mu = \left(\begin{array}{cc} 0 & \sigma^\mu\\ \bar{\sigma}^\mu &0 \end{array} \right),
\end{equation}
we have the Dirac equation for a Majorana field
\begin{equation}\label{Eq_Majorana}
\left(i\gamma^\mu\partial_\mu - m\right)\,\Psi_M = 0.
\end{equation}
This equation of motion can be derived from the Lagrangian
\begin{equation} \label{Majorana_Lagrangian}
\mathcal{L}_M = \frac{i}{2}\,\Psi_M^T\,\mathcal{C}\,\gamma^\mu\partial_\mu\Psi_M - \frac{1}{2}\,m\,\Psi_M^T\,\mathcal{C}\,\Psi_M,
\end{equation}

\subsection{Equation of motion for the Dirac spinor} \label{Equation of motion for the Dirac spinor}

In order to describe spin one-half particles that possess charge, such as electrons, we consider a four-spinor in which two different and unrelated Weyl spinors $\chi_\alpha$, $\xi^{\dag\dot{\alpha}}$ appear. We construct the Dirac four-spinor as
\begin{equation}
\Psi_D =  \left(\begin{array}{c} \chi_\alpha \\ \xi^{\dag\dot{\alpha}} \end{array} \right).
\end{equation}
Under charge conjugation, the Dirac field does not transform into itself; in fact we have
\begin{equation}
\Psi^C_D =  \left(\begin{array}{c} \xi_\alpha \\ \chi^{\dag\dot{\alpha}} \end{array} \right) \neq \Psi_D,
\end{equation}
so that nonzero values of the electric charge are allowed. The Dirac field satisfies the equation
\begin{equation}
\left(\begin{array}{cc} -m\,{\delta_\alpha}^\beta & i\left(\sigma^\mu\right)_{\alpha\dot{\beta}}\,\partial_\mu\\ i\left(\bar{\sigma}^\mu\right)^{\dot{\alpha}\beta} \,\partial_\mu & -m\,{\delta^{\dot{\alpha}}}_{\dot{\beta}} \end{array} \right) \, \bigg(\begin{array}{c} \chi_\beta \\ \xi^{\dag\dot{\beta}} \end{array} \bigg) = 0,
\end{equation}
which is put in the compact form
\begin{equation}\label{Eq_Dirac}
\left(i\gamma^\mu\partial_\mu - m\right)\,\Psi_D = 0.
\end{equation}
The fact that the Dirac field allows us to describe particles with a nonzero charge can be seen from the Dirac Lagrangian,
\begin{equation} \label{Dirac_Lagrangian}
\mathcal{L}_D = i\bar{\Psi}_D\,\gamma^\mu\partial_\mu\Psi_D - m\,\bar{\Psi}_D\Psi_D,
\end{equation}
which is invariant under a U(1) transformation of parameter $\omega$,
\begin{equation}\label{U1_transformation}
\Psi_D \to e^{i\omega}\,\Psi_D,\quad \bar{\Psi}_D \to e^{-i\omega}\,\bar{\Psi}_D.
\end{equation}

\section{Application of group theory to QFT}

We now turn our attention to gauge theories in Quantum Field Theory (QFT), focusing on Dirac fields. For this reason, we drop the index $D$ and simply denote with $\Psi$ a collection of one or more Dirac fields. The Dirac Lagrangian in Eq.~(\ref{Dirac_Lagrangian}), describing a free field theory for a massless complex-valued collection of Dirac fields, reads
\begin{equation} \label{Lagrangian}
\mathcal{L} = i\,\bar{\Psi}(x)\,\gamma^\mu\,\partial_\mu\,\Psi(x).
\end{equation}
We assume that the collection of fields $\Psi(x)$ is invariant under a global symmetry that generalizes Eq.~(\ref{U1_transformation}),
\begin{equation}\label{global_symmetry}
\Psi(x) \to U(\alpha^a)\,\Psi(x),
\end{equation}
where the transformation of the field can be written in terms of a set of constant parameter $\alpha^a$ as
\begin{equation}
U({\bf \alpha}) = {\rm exp}\left(i \alpha^a\,t^a\right).
\end{equation}
For example, if $\Psi(x)$ represents a single Dirac field, then the Lagrangian in Eq.~(\ref{Lagrangian}) is invariant under the $U(1)$ transformation in Eq.~(\ref{U1_transformation}), while if $\Psi(x)$ represents a doublet of Dirac fields then the Lagrangian in Eq.~(\ref{Lagrangian}) is invariant under a $SU(2)$ transformation,
\begin{equation}
\Psi(x) \to {\rm exp}\left(i{\bf \alpha}\,\frac{\bf \sigma}{2}\right)\,\Psi(x),
\end{equation}
where ${\bf \sigma} = \{\sigma^1, \sigma^2, \sigma^3\}$ are the usual $2\times2$ $\sigma$-matrices.

In QFT, a general recipe is adopted in order to include the interaction of the Dirac field $\Psi(x)$ with a gauge field $A^a_\mu$. We first promote the symmetry from global to local by making the parameters ${\bf \alpha}$ depending on the space-time coordinate $x$,
\begin{equation}
{\bf \alpha} \to {\bf \alpha}(x).
\end{equation}
The transformation of the Dirac field $\Psi(x)$ under this local symmetry reads
\begin{equation} \label{local_symmetry}
\Psi(x) \to U(\alpha^a(x))\,\Psi(x) = {\rm exp}\left(i \alpha^a(x)\,t^a\right)\,\Psi(x).
\end{equation}
This local transformation resembles that in Eq.~(\ref{global_symmetry}), the main difference being that the parameters $\alpha^a(x)$ now depend on the specific space-time point $x$ at which the transformation takes place. This justifies the use of the adjective ``local'' to describe such kinds of transformation. In order for this transformation to be imposed on the Lagrangian in Eq.~(\ref{Lagrangian}), we define a covariant derivative
\begin{equation}
D_\mu = \partial_\mu - ig\,A^a_\mu\,t^a,
\end{equation}
where $g$ is the coupling of the Dirac field to the gauge field. The Lagrangian in Eq.~(\ref{Lagrangian}) then reads
\begin{equation} \label{Lagrangian_gauge}
\mathcal{L} = \bar{\Psi}(x)\,i\,\gamma^\mu\,D_\mu\,\Psi(x) = \bar{\Psi}(x)\,i\,\gamma^\mu\,( \partial_\mu - ig\,A^a_\mu\,t^a)\,\Psi(x).
\end{equation}
We see that the introduction of the covariant derivative introduces in the Lagrangian above the interaction term
\begin{equation}
\mathcal{L}_{\rm int} =  g\,\bar{\Psi}(x)\,\gamma^\mu\,A^a_\mu\,t^a\,\Psi(x).
\end{equation}
The requirement that the Lagrangian in Eq.~(\ref{Lagrangian_gauge}) be invariant with respect to the local transformation in Eq.~(\ref{local_symmetry}) imposes the gauge field $A^a_\mu$ to transform as
\begin{equation}
A^a_\mu \to A^a_\mu +\frac{1}{g}\,\partial_\mu\alpha^a\,t^a + f^{abc}\,A^b_a\,\alpha^c,
\end{equation}
and $f^{abc}$ are the structure constants defined in Eq.~(\ref{structure_constant}) for the group $G$ with generators $t^a$.

Finally, we can write the complete Lagrangian for a Dirac multiplet $\Psi(x)$ of mass matrix $M$ and belonging to an irreducible representation of a gauge group $G$, known as the Yang-Mills Lagrangian, as
\begin{equation} \label{Lagrangian_YM}
\mathcal{L}_{YM} = \bar{\Psi}(x)\,(i\,\gamma^\mu\,D_\mu-M)\,\Psi(x) - \frac{1}{4}\,F^{\mu\nu}_a\,F^{a}_{\mu\nu}.
\end{equation}
In the Yang-Mills Lagrangian, the index $a$ is intended summed over the generators of the group $G$, while we have introduced the field strength
\begin{equation}
F^a_{\mu\nu} = \partial_\mu\,A^a_\nu - \partial_\nu\,A^a_\mu + g\,f^{abc}\,A^b_\mu\,A^c_\nu.
\end{equation}
In Eq.~(\ref{Lagrangian_YM}), we have included the mass term $M\,\bar{\Psi}\Psi$ which is manifestly invariant under the symmetry in Eq.~(\ref{local_symmetry}), and the term
\begin{equation}
\mathcal{L}_{\rm Gauge} = -\frac{1}{4}\,F^{\mu\nu}_a\,F^{a}_{\mu\nu},
\end{equation}
that describes the self-interaction and the dynamics of the gauge field.

\section{Quantum Chromodynamics}

We apply the tools developed in the last sections to describe the theory of Quantum Chromodynamics (QCD), and provide a theory to describes the mechanism of the strong interactions. The QCD theory is nonabelian since gluons carry color charge causing them to interact with each other in a more complicated structure: in fact, the underlying gauge group for QCD is the nonabelian group $G = SU(3)$. As we will see, a yet-to-be-solved problem arises in QCD, the so-called strong Charge-Parity (CP) problem. In fact, as in the theory of weak interactions, the CP symmetry is expected to be violated in strong interactions, whereas experiments show that the CP symmetry is preserved by strong interactions to a high degree of precision. The lack of a CP-violating term then requires a fine tuning of QCD parameters in its mathematical description.

\subsection{The QCD Lagrangian}

We denote the QCD gauge field (the gluon tensor field) living in the adjoint representation of the group as $G^a_{\mu\nu}$, where $a \in\{1, ..., 8\}$. The Lagrangian obtained using the $SU(3)$ gauge group with $N_f$ quark (the QCD Lagrangian) is \cite{PeskinSchroeder}
\begin{equation} \label{Lagrangian_QCD}
\mathcal{L}_{QCD} = \bar{q}\,(i\,\gamma^\mu\,D_\mu - M)\,q - \frac{1}{4}\,G^{\mu\nu\,a}\,G^{a}_{\mu\nu},
\end{equation}
where $q$ is the collection of quark spinor fields, $M$ is the quark mass matrix \cite{Nakamura:2010zzi} and $G^a_{\mu\nu}$ is the gluon tensor field, which can be written in terms of the QCD field strength $A_\mu^a$ and of the structure constant $f^{abc}_{SU(3)}$ of the $SU(3)$ group as
\begin{equation}
G^a_{\mu\nu} = \partial_\mu\,A^a_\nu - \partial_\nu\,A^a_\mu + g\,f^{abc}_{SU(3)}\,A^b_\mu\,A^c_\nu.
\end{equation}
The covariant derivative $D_\mu$ appearing in Eq.~(\ref{Lagrangian_QCD}) is defined in terms of the Gell-Mann matrices $t^a_{SU(3)}$ by
\begin{equation}
D_\mu = \partial_\mu - i\,g\,A^a\,t^a_{SU(3)}.
\end{equation}
In the limit of vanishing quark masses $M \to 0$, the QCD Lagrangian is invariant under the global transformation $U_L(N_f) \otimes U_R(N_f)$. However, such symmetry implies that hadrons come in doublets \cite{Poroski}, which is not seen experimentally. In fact, it turns out that the axial part of the chiral symmetry is spontaneously broken into the subgroup $SU(N_f)_A \otimes U(1)_A$. Because of the symmetry breaking, nearly massless Goldstone bosons arise: these bosons are experimentally identified with the pseudo-scalar octet formed by the $u$, $d$, and $s$ quarks and explain the $SU_A(N_f)$ breaking. However, there exists no candidate in the particle spectrum that would play the role of the Goldstone boson associated with the spontaneous breakdown of $U(1)_A$. In fact, such a particle should be a light pseudo-scalar boson of mass $m_a < \sqrt{3}\,m_\pi$: the latter requirement excludes the $\eta'$ meson which would otherwise be a suitable candidate having the proper quantum numbers. The absence of this light pseudo-scalar particle is known as the $U(1)_A$ problem \cite{Weinberg:1975ui}.

\subsection{Solving the $U(1)_A$ problem} \label{solving_U1A}

As proposed by t'Hooft \cite{PhysRevD.14.3432, PhysRevLett.37.8}, the $U(1)_A$ problem might be solved by adding a term 
 to the QCD Lagrangian in Eq.~(\ref{Lagrangian_QCD}) that explicitly breaks the $U(1)_A$ symmetry,
\begin{equation}
\mathcal{L}_{QCD} \to \mathcal{L}_{QCD}  + \mathcal{L}_{\tilde{\theta}},
\end{equation}
with
\begin{equation} \label{L_theta}
\mathcal{L}_{\tilde{\theta}} = \frac{g^2}{32\pi^2}\,\tilde{\theta}\,G^{\mu\nu\,a}\,\tilde{G}^{a}_{\mu\nu}.
\end{equation}
Here, we have defined the dual of the field strength tensor
\begin{equation}
\tilde{G}^a_{\mu\nu} = \frac{1}{2}\,\epsilon_{\mu\nu\sigma\rho}\,G^{\sigma\rho\,a},
\end{equation}
with $\epsilon_{\mu\nu\sigma\rho}$ the Levi-Civita completely antisymmetric tensor having $\epsilon_{0000}=1$, $g$ is a coupling constant, and $\tilde{\theta}$ is an observable \cite{Jackiw:1975fn, Callan:1976je} that might take values form zero to $2\pi$. The origin of the variable $\tilde{\theta}$ in the QCD Lagrangian can be explained by the following argument. The classical gluon field equations admit an anti-instanton solution, which satisfies the antiduality condition
\begin{equation}
G_{\mu\nu} = \tilde{G}_{\mu\nu},
\end{equation}
and which has an integer index
\begin{equation} \label{U_axial_wn}
n = \frac{1}{32\pi^2}\,\int\,d^4x\,G^a_{\mu\nu} \,\tilde{G}^{\mu\nu\,a}.
\end{equation}
The angular variable $\theta$ parametrizes the linear combination of different $|n\rangle$-vacua in the theory corresponding to different values of the integer $n$, with the particular linear combination known as the $\theta$-vacuum,
\begin{equation}
|\tilde{\theta}\rangle = \sum_{n=-\infty}^{+\infty}\,e^{-in\tilde{\theta}}\,|n\rangle.
\end{equation}
Here, every value of $n$ characterizes the winding number of the $U(1)_A$ symmetry in Eq.~(\ref{U_axial_wn}). Although the term in Eq.~(\ref{L_theta}) is a surface term, it is possible to show that if the $U(1)_A$ problem is solved and none of the quark is massless, a nonzero value of $\tilde{\theta}$ implies that the CP symmetry is broken \cite{Shifman:1979if}. Thus, QCD physics depends on the value of $\tilde{\theta}$, or better, when electroweak interactions are included, on the combination
\begin{equation} \label{def_tilde_theta}
\bar{\theta} = \tilde{\theta} + \theta_{\rm weak},
\end{equation}
with the extra term arising in the QCD Lagrangian of the form
\begin{equation}
\mathcal{L}_\theta = \frac{g^2}{32\pi^2}\,\theta_{\rm weak}\,G^{\mu\nu\,a}\,\tilde{G}^{a}_{\mu\nu}.
\end{equation}
The parameter $\theta_{\rm weak}$ is expressed in terms of the quark mass matrix $M$ as
\begin{equation}
\theta_{\rm weak} = {\rm arg}\left({\rm Det}\, M\right).
\end{equation}
Eventually, the QCD Lagrangian including the 'tHooft and electroweak interaction terms reads
\begin{equation} \label{QCD_complete}
\mathcal{L}_{\rm QCD+\bar{\theta}} = \mathcal{L}_{\rm QCD}  + \mathcal{L}_{\bar{\theta}} =  \mathcal{L}_{\rm QCD} +  \frac{g^2}{32\pi^2}\,\bar{\theta}\,G^{\mu\nu\,a}\,\tilde{G}^{a}_{\mu\nu}.
\end{equation}
\newpage

\subsection{The strong CP problem}\label{The strong CP problem}

As we discussed in Sec.~\ref{solving_U1A}, it is expected that the CP symmetry is violated by strong interactions due to the CP-breaking term $\mathcal{L}_{\bar{\theta}}$. However, experimentally no violation of the CP symmetry is observed, and the strong interactions preserve this symmetry to a high degree of accuracy. This problem in merging theoretical motivations and experimental observations is known in the literature as the strong CP problem, which can be restated by asking why strong interactions do not violate the CP symmetry when CP violation is not forbidden in the theory.

The most stringent constrain on the violation of the CP symmetry by the strong interaction comes from the measurement of the electric dipole moment of the neutron, defined as
\begin{equation}
d_N \sim \bar{\theta}\,\frac{e\,m_{ud}}{m_N^2},
\end{equation}
where $e$ is the absolute value of the electron charge, $m_N$ is the neutron mass and $m_{ud}$ is given in terms of the masses of the up and down quarks $m_u$, $m_d$ as
\begin{equation}
m_{ud} = \frac{m_u\,m_d}{m_u+m_d}.
\end{equation}
The latest experimental bound on the neutron electric dipole moment is \cite{PhysRevLett.97.131801}
\begin{equation} \label{bound_NEDM}
|d_N| < 2.9\times 10^{-26}\,e\,{\rm cm}\quad\hbox{at 90\% C.L.},
\end{equation}
while we take the value of $d_N$ from the review by Kim and Carosi \cite{Kim:2008hd},
\begin{equation} \label{theory_NEDM}
d_N = 4.5\times 10^{-15}\,\bar{\theta}\,e\,{\rm cm}.
\end{equation}
Eqs.~(\ref{bound_NEDM}) and~(\ref{theory_NEDM}) imply the constraint
\begin{equation}\label{bound_tilde_theta}
|\bar{\theta}| < 0.7 \times 10^{-11}.
\end{equation}
From the definition of $\bar{\theta}$ in Eq.~(\ref{def_tilde_theta}), we see that this quantity is the sum of the two terms $\tilde{\theta}$ and $\theta_{\rm weak}$, whose physical origin is completely unrelated; it is thus a mystery why two quantities which are naturally of order one have to cancel out in a sum with such a high degree of accuracy as in Eq.~(\ref{bound_tilde_theta}).
\newpage
\subsection{Possible solutions to the strong CP problem} \label{Possible solutions to the strong CP problem}

As of today, three main solutions to the strong CP problem have been proposed. These are, respectively: calculable $\bar{\theta}$, massless up quark, the axion particle. As the axion is the most favorable  solution and the subject of this thesis, we will briefly review here only the first two cases and concentrate on the axion throughout the subsequent chapters.

\subsubsection{Calculable $\bar{\theta}$}

The underlying idea of the proposed calculable $\bar{\theta}$ is to impose CP invariance in the QCD Lagrangian, setting $\tilde{\theta} = 0$. Taking into account that the only source of CP violation comes from weak interactions, as explicit in phenomena like the neutral K-meson oscillations and the B-meson decay, the term $\theta_{\rm weak}$ is still nonzero and can be calculable in an underlying theory, with a small value constrained by observations as in Eq.~(\ref{bound_tilde_theta}). An example of such a model is the Barr-Nelson CP violating model, in which heavy singlet quarks are introduced and whose vacuum expectation value lies much above the electroweak scale. These quarks mix with the ordinary light quarks after symmetry breaking, so that at low energy they can be integrated out and the light quarks acquire the required amount of CP symmetry through the Cabibbo-Kobayashi-Maskawa mixing. In this theory, the condition ${\rm arg}({\rm Det}\, M) = 0$ is imposed at tree level. A nonzero value of $\theta_{\rm weak}$ arises through higher-loop corrections, $\theta_{\rm weak} \approx \alpha_{\rm loop}$, with the fine structure constant $\alpha_{\rm loop}$ bound by the stringent bound in Eq.~(\ref{bound_tilde_theta}).

\subsubsection{Massless up quark}

If the lightest quark (the up quark) is massless, it is possible to eliminate $\bar{\theta}$ by a rotation of the quark fields $q$,
\begin{equation}
q \to {\rm Exp}\,\left(\frac{i}{2}\,\bar{\theta}\gamma^5\right)\,q,
\end{equation}
so that $\bar{\theta}$ is no longer an observable. In fact, the QCD Lagrangian in Eq.~(\ref{QCD_complete}) transforms under the rotation above as
\begin{equation}
\mathcal{L}_{\rm QCD+\bar{\theta}} \to \mathcal{L}_{\rm QCD},
\end{equation}
making the term $\mathcal{L}_{\bar{\theta}}$ disappear. The massless up quark possibility has been ruled out by measurements, as quoted in Eq.~(\ref{quark_ratios}) below that states $m_u/m_d \equiv z_d = 0.568 \pm 0.042$.

%% file: chap4.tex
\chapter{The axion} \label{The axion}

\fixchapterheading

\section{The axion as a solution to the strong\protect\\ CP problem}

The third and the most compelling solution to the strong CP problem discussed in Secs.~\ref{The strong CP problem} and~\ref{Possible solutions to the strong CP problem} introduce an additional chiral symmetry that promotes the $\bar{\theta}$ parameter to a dynamical field: the dynamics of this new field naturally relaxes its expectation values towards arbitrarily small values, thus eliminating the CP problem. This mechanism is achieved by introducing a new global, chiral symmetry $U(1)_{\rm PQ}$ which is spontaneously broken at an unknown energy scale $f_a$, known as the PQ energy scale. The axion is the Goldstone boson resulted from such symmetry breaking.

The Lagrangian term resulting after the spontaneous symmetry breaking of the $U(1)_{\rm PQ}$ symmetry reads
\begin{equation} \label{Lagrangian_axion}
\mathcal{L}_a = -\frac{1}{2}\,\left(\partial^\mu a\right)\,\left(\partial_\mu a\right) + \frac{g^2}{32\pi^2}\,\frac{a}{f_a/N}\,G^{\mu\nu\,a}\,\tilde{G}^{a}_{\mu\nu},
\end{equation}
where $a\equiv a(x)$ is the axion field and $N = \sum_f\,X_f$ is a new parameter called the PQ color anomaly, defined as the sum of the PQ charges $X_f$ over the fermions in the theory. In Eq.~(\ref{Lagrangian_axion}), the first term represents the kinetic term, while the second term describes the interaction of the axion with the gluon field.

When considering the Lagrangian
\begin{equation}
\mathcal{L}_{{\rm QCD}+\bar{\theta}+a} = \mathcal{L}_{\rm QCD} + \mathcal{L}_{\bar{\theta}}+ \mathcal{L}_a,
\end{equation}
an effective potential $V_{\rm eff}(a)$ for the axion appears, whose minimum is reached when
\begin{equation} \label{minimum_axion}
\langle a(x) + \frac{f_a}{N}\,\bar{\theta} \rangle = 0.
\end{equation}
The expectation value of the axion field in $\mathcal{L}_a$ at its minimum cancels out the $\bar{\theta}$ term in $\mathcal{L}_{\bar{\theta}}$, thus eliminating the CP-violating term. Since we can shift the axion field by a constant amount without changing the physics, we define the (axion) misalignment angle
\begin{equation}
\theta = \bar{\theta} + \frac{a}{f_a/N}.
\end{equation}
Summing up, with the PQ solution to the strong CP problem, the free parameter $\bar{\theta}$ has been traded for a dynamical field $a(x)$ that evolves to its CP-conserving minimum $\theta = 0$ through the spontaneous breaking of a $U(1)$ symmetry. Essentially, $\theta$ can be seen as the phase of a new complex scalar field
\begin{equation}
\Phi_a = |\Phi_a|\,e^{i\theta},
\end{equation}
that takes a nonzero expectation value at the energy scale $f_a/N$. The axion potential thus has the usual Mexican hat form 
\begin{equation}\label{axion_potential_PQ}
V_{\rm PQ}(\theta) = \frac{\lambda}{4}\,\left(|\Phi_a|^2-\left(\frac{f_a}{N}\right)^2\right)^2 + m_a^2(T)\,f_a^2\,(1-\cos\theta),
\end{equation}
where $\lambda$ is a coupling constant. Here, the second term in Eq.~(\ref{axion_potential_PQ}) comes from nonperturbative QCD effects associated with instantons \cite{Gross:1980br}, that break the $U(1)_{\rm PQ}$ symmetry down to a $Z(N)$ discrete subgroup \cite{Sikivie:1982qv}. As we will see later in Sec.~\ref{Finite temperature effects}, the presence of this second term justifies the fact that the axion mass $m_a(T)$ depends on the temperature of the plasma considered \cite{Vafa:1984xg}. After the spontaneous breaking of the PQ symmetry, we are thus left with the residual potential (but see Refs.~\cite{Fugleberg:1998kk, Gabadadze:2000vw, Gabadadze:2002ff} for different potentials)
\begin{equation} \label{axion_potential1}
V(\theta) = m_a^2(T)\,f_a^2\,(1-\cos\theta).
\end{equation}
This potential is approximately quadratic for small values of $\theta$, but quickly shows deviations from a pure quadratic potential for $\theta \sim 1$.

\section{On the mass of the axion}

\subsection{Axion mass at zero temperature}

We can expand the term $\mathcal{L}_{\bar{\theta}} + \mathcal{L}_a$ around the minimum in Eq.~(\ref{minimum_axion}) to obtain the value of the axion mass as
\begin{equation}
m_a^2 = -\frac{g^2}{32\pi^2}\,\frac{N}{f_a}\,\frac{\partial}{\partial a}\left\langle G^{\mu\nu\,a}\,\tilde{G}^{a}_{\mu\nu}\right\rangle\bigg|_{\langle a(x)\rangle = -f_a\,\bar{\theta}/N}\,\,.
\end{equation}
The axion mass has been computed in Ref.~\cite{Weinberg:1977ma} including the axion mixing with the other neutral pions through the axion-gluon-gluon coupling, and is written in terms of the quark ratios $z_d = m_u/m_d$, $z_s = m_u/m_s$ and of the $\pi^0$ mass and decay constant $m_{\pi^0}$ and $f_{\pi^0}$ as
\begin{equation} \label{axion_mass_mixing}
m_a = \frac{f_{\pi^0}\,m_{\pi^0}}{f_a/N}\,\sqrt{\frac{z_d}{(1+z_d)(1+z_d+z_s)}},
\end{equation}
Using the experimental values \cite{Leutwyler:1996qg}
\begin{equation} \label{quark_ratios}
z_d = 0.568 \pm 0.042,\quad z_s = 0.029 \pm 0.003,
\end{equation}
and the values \cite{Amsler:2008zzb}
\begin{equation}
m_{\pi^0} = (134.9766 \pm 0.0006) {\rm ~MeV},\quad f_{\pi^0} = \frac{1}{\sqrt{2}}\,(130 \pm 5) {\rm ~MeV},
\end{equation}
we obtain the numerical value
\begin{equation} \label{eq:axionmass}
m_a =  (5.91 \pm 0.53) {\rm \mu eV}\,\frac{1}{f_{a,12}/N}.
\end{equation}
Here and in the following, a number $y$ indexing some quantity with units of energy indicates that such quantity has been divided by $10^y$ GeV: for example, in the equation above, it is $f_{a,12} = f_a/10^{12}$ GeV.

{}The axion mass in Eq.~(\ref{eq:axionmass}) is exact in the context of the simplest hadronic axion model \cite{Kim:1979if, Shifman:1979if} (see Sec.~\ref{The KSVZ model}), where Standard Model fermions do not possess a PQ charge, and axions only interact with an exotic heavy quark. Above the QCD phase transition nonperturbative effect arise, giving the axion a temperature-dependent mass as we now discuss.

\subsection{Finite temperature effects} \label{Finite temperature effects}

In a quark-gluon plasma of finite temperature $T$, nonperturbative effects modify the properties of the components of the plasma itself. Nonperturbative effects in QCD have been studied via lattice simulations, see ex. Ref.~\cite{Chu:1994vi}, or phenomenology, see ex. Refs.~\cite{Diakonov:1985eg, Shuryak:1987jb}. For axions, the mass acquires a temperature dependence at sufficiently high $T$ due to the instanton effects: In the high temperature limit, $T \gg \Lambda_{\rm QCD}$, the axion mass is given by an integral over noninteracting instantons of all sizes \cite{Gross:1980br} as (see also Refs.~\cite{Turner:1985si, Fox:2004kb})
\begin{equation} \label{gross_axion_mass}
m_a^2(T) \approx \hat{c}\,\left(\frac{33-2N_f}{6}\right)^{6}\,\hat{\chi}^{N_f}\,\frac{\Lambda_{\rm QCD}^{4-N_f}}{(f_a/N)^2}\,\left(\frac{\Lambda_{\rm QCD}}{T}\right)^{7+\frac{N_f}{3}}\,{\rm Det}\left(M\right)\,\mathcal{I}(T),
\end{equation}
with
\begin{equation}\label{I_T}
\mathcal{I}(T) = \int_0^{+\infty}\,dx\,x^{6+\frac{N_f}{3}}\,\left(\ln\frac{T}{\Lambda_{\rm QCD}\,x}\right)^{6-\hat{a}}\,e^{-f(x)},
\end{equation}
\begin{equation}
f(x) = \frac{\pi^2}{3}\,\left(6+N_f\right)\,x^2 +\frac{9-N_f}{6}\,\left[\frac{12\,\hat{\alpha}}{\left(1+\hat{\gamma}\,(\pi\,x)^{-3/2}\right)^{8}}- \ln\left(1+\frac{\pi^2}{3}\,x^2\right)\right],
\end{equation}
and the numerical values $\hat{c} = 0.130078$, $\hat{\chi} = 1.33876$, $\hat{\alpha} = 0.01289764$, $\hat{\gamma} = 0.15858$ and $\hat{a} = (153-19 N_f)/(33-2N_f)$. The parameter $N_f$ gives the number of quarks that are relativistic at temperature $T$, that is, whose mass is $\lesssim T$. It is worth stressing here that, using different computation techniques, various authors have obtained different results for $m_a(T)$. For later convenience, we parametrize the general expression for the temperature-dependent mass with a general broken-law function,
\begin{equation} \label{axion_mass_temperature0}
m_{a}(T) = m_a\,
\begin{cases}
a \,\big(\frac{\Lambda_{\rm QCD}}{T}\big)^c\,(1-\ln\frac{\Lambda_{\rm QCD}}{T})^d ,& T\gtrsim \Lambda_{\rm QCD},\\
\quad 1, & T\lesssim \Lambda_{\rm QCD},
\end{cases}
\end{equation}
with the parameter $c = 7/2 + N_f/6$ from Eq.~(\ref{gross_axion_mass}). Before showing our computation for the parameters $a$ and $d$, we will briefly review the literature on the subject for the nontrivial case $T \gtrsim \Lambda_{\rm QCD}$.

Earlier work on the temperature-dependent axion mass are in Refs.~\cite{Preskill:1982cy, Abbott:1982af, Dine:1982ah}, where the authors use the method by Gross, Pisarski, and Yaffe in Ref.~\cite{Gross:1980br} in the dilute gas approximation. Preskill {\it et al.} \cite{Preskill:1982cy} obtain, for the case $N_f = 3$, the expression
\begin{equation}
m_a(T) = 2\times 10^{-2}\,\frac{\Lambda_{\rm QCD}^{1/2}}{f_a}\,\sqrt{{\rm Det}\left(M\right)}\,\left(\frac{\Lambda_{\rm QCD}}{\pi\,T}\right)^4\,\left[9\ln\left(\frac{\pi T}{\Lambda_{\rm QCD}}\right)\right]^3,
\end{equation}
which we put in the form of the first line in Eq.~(\ref{axion_mass_temperature0}) with $\Lambda_{\rm QCD} = 200$ MeV and ${\rm Det}\left(M\right) = 1000{\rm ~MeV^3}$ around $T \sim 4\Lambda_{\rm QCD}$ \cite{Preskill:1982cy} as
\begin{equation}\label{preskill_mass}
m_a(T) = a_{\rm preskill} \,m_a\,\left(\frac{\rm \Lambda_{\rm QCD}}{T}\right)^4,
\end{equation}
with $a_{\rm preskill} = 0.035$.

Sikivie \cite{Sikivie:2006ni} quotes the general result from Refs.~\cite{Preskill:1982cy, Abbott:1982af, Dine:1982ah} as
\begin{equation}
m_a(T) = 4\times 10^{-9}{\rm ~eV}\,\left(\frac{10^{12}{\rm ~GeV}}{f_a}\right)\,\left(\frac{\rm GeV}{T}\right)^4.
\end{equation}
We notice that this expression can be written as
\begin{equation}
m_a(T) = a_{\rm sikivie} \,m_a\,\left(\frac{\Lambda_{\rm QCD}}{T}\right)^4,
\end{equation}
with $a_{\rm sikivie} = 0.417/\Lambda_{0.2}^4$ and $\Lambda_{0.2} = \Lambda_{\rm QCD}/200$MeV.

Turner \cite{Turner:1985si} uses the tools developed in Ref.~\cite{Gross:1980br} to compute $m_a(T)$ in the case $T \gg \Lambda_{\rm QCD}$ and gives the general expression
\begin{equation}\label{turner_mass}
m_a(T) = m_a\,a\,\Lambda_{0.2}^b\,\left(\frac{\Lambda_{\rm QCD}}{T}\right)^c\,\left[1-\ln\frac{\Lambda_{\rm QCD}}{T}\right]^d,
\end{equation}
where $m_a \equiv m_a(T=0)$ is given in Eq.~(\ref{eq:axionmass}). Table.~\ref{turner_table} shows the dependence of the parameters $a$, $b$, $c$, and $d$ on the number $N_f$ of relativistic quarks at the temperature $T_1$ at which the axion field starts oscillating, see Eq.~(\ref{definition_T1})

\begin{table}[h!]
\caption{Values of the parameters $a$, $b$, $c$, and $d$ appearing in Eq.~(\ref{turner_mass}), as a function of the number of relativistic degrees of freedom $N_f$.}
\begin{center}
\begin{tabular}{|c|c|c|c|c|}
\hline
$N_f$ & $a$ & $b$ & $c$ & $d$\\
\hline
1 & 0.277   & 3/2  & 3.67 & 0.84\\
2 & 0.0349 & 1     & 3.83 & 1.02\\
3 & 0.0256 & 1/2  & 4.0   & 1.22\\
4 & 0.0421 & 0     & 4.17 & 1.46\\
5 & 0.118   & -1/2 & 4.33 & 1.74\\
6 & 0.974   & -1    & 4.5   & 2.07\\
\hline
\end{tabular}
\label{turner_table}
\end{center}
\end{table}
The exact value of $N_f$ is not well determined, because the value of $T_1$ is of the order of the current mass of the charm quark, $m_c \sim 1{\rm ~GeV}$, so it is possible for it to be $N_f = 3$ or $N_f = 4$. Because of this uncertainty, Ref.~\cite{Turner:1985si} considers the function
\begin{equation}
m_a(T) = 7.7\times 10^{-2\pm 0.5}\,m_a\,\left(\frac{\Lambda_{\rm QCD}}{T}\right)^{3.7\pm0.1},
\end{equation}
where the uncertainties interpolate between the various values of the parameters in Table~\ref{turner_table} for different plausible values of $N_f$.

DeGrand {\it et al.} \cite{DeGrand:1985uq}, using the formula in Ref.~\cite{Gross:1980br}, quote an expression for the axion mass valid for $T \gg \Lambda_{\rm QCD}$, as
\begin{equation}
m_a(T) = 15\,\frac{\Lambda_{\rm QCD}^2}{f_a}\,\sqrt{\frac{{\rm Det}\left(M\right)}{\Lambda_{\rm QCD}}}\,\left(\frac{\Lambda_{\rm QCD}}{\pi\,T}\right)^4\,\left[\ln\frac{\pi\,T}{\Lambda_{\rm QCD}}\right]^3.
\end{equation}
This is basically the same expression found by Preskill in Eq.~(\ref{preskill_mass}).

Bae, Huh, and Kim \cite{Bae:2008ue} improved the results by Turner by considering updated values of current quark masses and the effects of the QCD phase transition. They parametrized the axion mass with
\begin{equation} \label{bae_mass}
m_a^2(T) = \alpha_{\rm inst}\,\frac{{\rm GeV}^4}{f_a^2\,(T/{\rm GeV})^\nu}.
\end{equation}
For the parameters $\alpha_{\rm inst}$ and $\nu$, the authors find a $\Lambda_{\rm QCD}$ dependence like in Table~\ref{bae_table}

\begin{table}[h!]
\caption{Values of the parameters $\alpha_{\rm inst}$ and $\nu$ in Eq.~(\ref{bae_mass}) for different values of $\Lambda_{\rm QCD}$. From Ref.~\cite{Bae:2008ue}}
\begin{center}
\begin{tabular}{|c|c|c|}
\hline
$\Lambda_{\rm QCD}$ & $\alpha_{\rm inst}/10^{-12}$ & $\nu$\\
\hline
320 & 0.9967& 6.967\\
380 & 3.964& 6.878\\
440 & 12.74& 6.789\\
\hline
\end{tabular}
\label{bae_table}
\end{center}
\end{table}
Wantz and Shellard \cite{Wantz:2009it, Wantz:2009mi}, within the interacting instanton liquid model \cite{Schafer:1996wv}, obtained a numerical expression for the temperature dependence of $m_a(T)$ valid at all temperatures; in particular, they were able to follow the $m_a(T)$ function around the QCD scale $\Lambda_{\rm QCD}$, where analytic methods break. The analytic fit in Refs.~\cite{Wantz:2009it, Wantz:2009mi} reports, for $T \gtrsim \Lambda_{\rm QCD}$,
\begin{equation}
m_a^2(T) = \frac{\alpha_a\,\Lambda_{\rm QCD}^4}{f_a^2\,(T/\Lambda_{\rm QCD})^\nu},
\end{equation}
with $\alpha_a = 1.68\times 10^-7$ and $\nu = 6.68$.

\begin{figure}[h!]
\includegraphics[width=13cm]{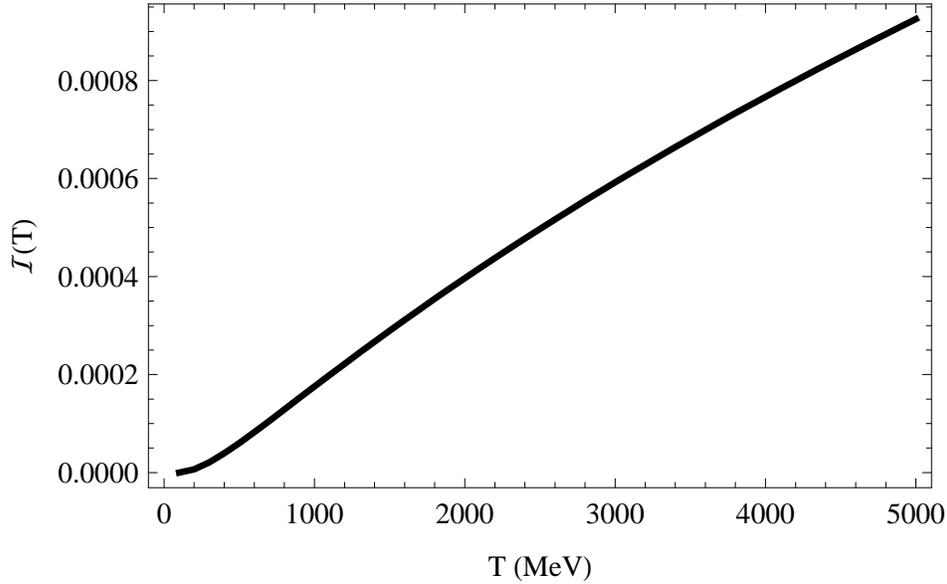}
\caption[The function $\mathcal{I}(T)$ in Eq.(\ref{I_T}).]{The function $\mathcal{I}(T)$ in Eq.(\ref{I_T}) as a function of temperature $T$ for $\Lambda_{\rm QCD}= 200$ MeV.}
\label{plot_I}
\end{figure}

Here, we consider the case $N_f = 3$, because, as we will show later, the axion field starts oscillating at a temperature $T_1$ that lies below the GeV; with only three quarks being relativistic at $T_1$ we have $N_f = 3$ and thus, from Eq.~(\ref{gross_axion_mass}) and from Table~\ref{turner_table}, the axion mass decreases with $T$ as $T^{-4}$ to the leading order. For this reason, we set $c = 4$ in Eq.~(\ref{axion_mass_temperature0}). To compute the parameters $a$ and $d$, we proceed as follows. Based on the latest available data for the masses of the quarks $u$, $d$, $s$, which are the relativistic quarks at temperature $T_1$, we write the determinant of the quark mass matrix as
\begin{equation}
{\rm Det}\left(M\right) = m_u\,m_d\,m_s = \frac{m_u^3}{z_d\,z_s} = (1.64 \pm 0.29)\times 10^3 {\rm ~MeV}^3\,\left(\frac{m_u}{\rm 3\,MeV}\right)^3,
\end{equation}
where $z_d$ and $z_s$ are given in Eq.~(\ref{quark_ratios}) and $m_u$ lies in the range (1.7 - 3.3) MeV \cite{Nakamura:2010zzi}: here we take the value $m_u = 3$ MeV. To evaluate $m_a(T)$ we use Eq.~(\ref{gross_axion_mass}) with $N_f = 3$. We are particularly interested in evaluating the function $\mathcal{I}(T)$, whose plot is shown in Fig.~\ref{plot_I}.

Since the integral in the function $\mathcal{I}(T)$ in Eq.~(\ref{gross_axion_mass}) depends on the temperature, we fit the numerical solution of such function and obtain
\begin{equation}
\mathcal{I}(T) \approx 5.84\times 10^{-6}\,\left(1+\ln\frac{T}{\Lambda_{\rm QCD}}\right)^{2.41}.
\end{equation}
Using this fit for $\mathcal{I}(T)$ with the values of the parameters as above and $N_f = 3$, we find for the temperature-dependent axion mass in Eq.~(\ref{gross_axion_mass}) the expression
\begin{equation}\label{axion_mass_new}
m_a(T) = (0.017\pm 0.003)\,m_a\,\Lambda_{0.2}^{1/2}\,\left(\frac{\Lambda_{\rm QCD}}{T}\right)^{4}\,\left(1-\ln\frac{\Lambda_{\rm QCD}}{T}\right)^{1.2},\quad\hbox{for $T \gtrsim \Lambda_{\rm QCD}$},
\end{equation}
while $m_a(T) = m_a$ in Eq.~(\ref{eq:axionmass}) for $T \lesssim \Lambda_{\rm QCD}$. The uncertainty in the prefactor $a = 0.017\pm 0.003$ is due to the uncertainties in $z_d$ and $z_s$ only, since we neglected the uncertainty in the mass of the up quark. Comparing our Eq.~(\ref{axion_mass_new}) above with the result by Turner in Table~\ref{turner_table}, we see that we confirm the exponent $d = 1.2$, but we differ from Turner's result because of the prefactor $a = 0.017$, which is about 50\% lower than $a = 0.0256$ in Table~\ref{turner_table}. We attribute this discrepancy to the different value of the up quark mass used in Ref.~\cite{Turner:1985si}. Notice that the value of our prefactor $a$ is compatible with the value $a = 0.018$ used in Refs.~\cite{Beltran:2006sq, Fox:2004kb, Hertzberg:2008wr, Visinelli:2009zm}.

For $T$ of the order of $\Lambda_{\rm QCD}$, we neglect the logarithmic term; in the following, we will always assume this is the case, adopting the following expression for the axion mass
\begin{equation} \label{axion_mass_temperature}
m_{a}(T) = m_a\,
\begin{cases}
a\,\big(\frac{\Lambda_{\rm QCD}}{T}\big)^4 ,& T\gtrsim \Lambda_{\rm QCD},\\
\quad 1, & T\lesssim \Lambda_{\rm QCD},
\end{cases}
\end{equation}
with $a = 0.017\pm 0.003$.

\section{Coupling of standard model particles \protect\\with the axion}

Thanks to the Lagrangian term $\mathcal{L}_a$, axions couple at tree level to gluons and photons. However, depending on the specific particle model, axions may couple differently to the Standard Model fermions and other gauge bosons, and the coupling to photons might be suppressed.

\subsection{Coupling of axions to gluons} \label{Coupling of axions to gluons}

In all axions models, axions couple to gluons through the second term in the Lagrangian in Eq.~(\ref{Lagrangian_axion}),
\begin{equation}
\mathcal{L}_{ag} = \frac{\alpha_s}{8\pi}\,\frac{a}{f_a/N}\,G^{\mu\nu\,a}\,\tilde{G}^{a}_{\mu\nu},
\end{equation}
where we included the expression for the fine structure constant of strong interactions $\alpha_s = g^2/4\pi$. Thanks to this coupling to gluons, the axion mixes with pions and acquires the zero-temperature axion mass in Eq.~(\ref{axion_mass_mixing}).

\subsection{Coupling of axions to photons} \label{Coupling of axions to photons}

\begin{figure}[tb!]
\begin{center}
\includegraphics[width=5cm]{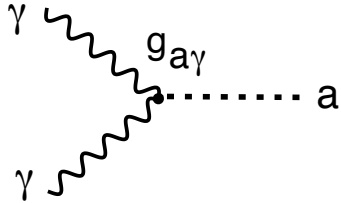}
\caption{The axion-photon vertex in an axion theory in which $g_{a\gamma} \neq 0$.}
\label{axion_photon_img}
\end{center}
\end{figure}

The axion-photon coupling is described by the term
\begin{equation}
\mathcal{L}_{a\gamma\gamma} = -\frac{g_{a\gamma}}{4}\,a\,F_{\mu\nu}\tilde{F}^{\mu\nu} = g_{a\gamma}\,a\,{\bf E}\cdot {\bf B}.
\end{equation}
where $g_{a\gamma}$ is the axion-photon coupling, given by
\begin{equation} \label{axion_photon_coupling}
g_{a\gamma} = \frac{\alpha}{2\pi\,f_a}\,\left|\frac{E}{N}-\frac{2(4+z+z_s)}{3(1+z-z_s)}\right| = \frac{\alpha}{2\pi\,f_a}\,\left|\frac{E}{N}-1.92 \pm 0.08\right|,
\end{equation}
where $\alpha \approx1/137$ is the fine structure constant. The ratio $E/N$ can be written in terms of the PQ charges $X_f$, the electric charges $Q_f$ and the number of colors in a multiplet $D_f$ as
\begin{equation} \label{def_E/N}
\frac{E}{N} = \frac{\sum_f \,X_f\,D_f\,Q_f^2}{\sum_f\,X_f}.
\end{equation}
Depending on the model, $g_{a\gamma}$ can be enhanced when the number of quarks $D_f$ is large, or suppressed when $E/N \sim 2$. In the following, we indicate the Feynman diagram associated to the Lagrangian term in Eq.~(\ref{axion_photon_coupling}) with the graph shown in Fig.~\ref{axion_photon_img}.

Models in which there exists an axion-photon coupling at tree level predict that the Primakoff effect for axions might have an important role in a number of astrophysical processes like the cooling of stars, and enable us to attempt the detection of axions through a ``shining through a wall'' experiment.

\subsection{Coupling of axions to fermions} \label{Coupling of axions to fermion}

\begin{figure}[tb!]
\begin{center}
\includegraphics[width=5cm]{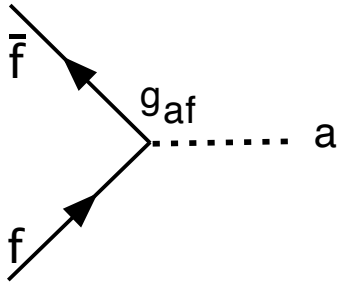}
\caption[Coupling of an axion with a fermionic field $f$.]{The Feynman diagram showing the axion-fermion vertex for a fermion field $f$, in an axion theory in which $g_{af} \neq 0$.}
\label{axion_fermion_img}
\end{center}
\end{figure}

In the Lagrangian term describing the interaction of axions with the fermionic field $\Psi_f$, the axion field always appears in the derivative term $\partial_\mu a(x)$, so that the invariance $a \to a + {\rm const.}$  proper to Nambu-Goldstone bosons is imposed:
\begin{equation} \label{axion_fermion_interaction}
\mathcal{L}_{aff} = \frac{C_f}{2f_a}\,\bar{\Psi}_f\,\gamma^\mu\,\gamma_5\,\Psi_f\,\partial_\mu\,a.
\end{equation}
In the expression above, $\Psi_f$ is the fermionic field of mass $m_f$ and $C_f$ is a model-dependent parameter that gives the PQ charge of the fermion $f$. Writing the interaction Lagrangian as
\begin{equation}
\mathcal{L}_{aff} = -i\,\frac{C_f\,m_f}{2f_a}\,\bar{\Psi}_f\,\gamma_5\,\Psi_f\,a,
\end{equation}
it appears that the combination
\begin{equation}\label{coupling_fermion}
g_{af} = \frac{C_f\,m_f}{f_a},
\end{equation}
plays the role of a Yukawa interaction, with $\alpha_{af} \equiv g_{af}^2/4\pi$ being analogous to the fine structure constant in this model. Fig.~\ref{axion_fermion_img} shows the Feynman diagram associated with $\mathcal{L}_{aff}$ for the fermion $f$.

\section{Axion models}

\subsection{The ``visible'' axion model}

The first viable axion model is known as the PQWW model from the initials of the authors Peccei and Quinn \cite{Peccei:1977hh1}, Weinberg \cite{Weinberg:1977ma}, and Wilczek \cite{Wilczek:1977pj}, or the ``visible'' axion because this theory would lead to sizable effects in various physical phenomena due to the strong axion interaction with light and light mesons. In the PQWW model, the PQ energy scale $f_a$ is related to the weak interaction scale $\Lambda_{\rm weak} \sim 250$GeV. A pair of Higgs fields $\Phi_1$ and $\Phi_2$ is introduced, the first giving mass to the up quark and the second to the down quark, with each Higgs field having a vacuum expectation value $\langle\Psi_{1,2}\rangle = \Lambda_{1,2}/\sqrt{2}$, with $\sqrt{\Lambda_1^2+\Lambda_2^2} = \Lambda_{\rm weak}$. From these premise it is possible to bound the PQ scale as $f_a \lesssim 42$ GeV and so the axion mass $m_a \gtrsim 200$ keV.

The PQWW axion model has long been abandoned because its very premise $f_a \sim \Lambda_{\rm weak}$ has been ruled out by various astrophysical considerations (see Sec.~\ref{Astrophysical bounds on axions}) and because this model predicts a lifetime of various mesons like $K^+$ or J/$\Psi$ that would be too short compared to observations.

\subsection{Models for the ``invisible'' axion}

After experiments showed that the PQ energy scale cannot be related to the weak interaction scale, authors have either considered $f_a$ as a free parameter or they have considered more complicated models in which $f_a$ relates to the Grand Unification Theory (GUT) scale $\Lambda_{\rm GUT} \sim 10^{16}$ GeV or to the Planck scale  $\Lambda_{\rm Planck} \sim 10^{19}$ GeV. In the first case, the axion is studied in a model-free theory and its generic properties are constrained by astrophysical considerations, direct and indirect searches, and accelerator experiments. In the second case, the axion is embedded in a more fundamental theory such as a GUT or a string theory.

All these models share a PQ scale much higher than the $\Lambda_{\rm weak}$ energy scale, so that the coupling of axions to Standard Model particles is considerably weaker, thus making the experimental detection of the axion even more challenging.

\subsubsection{The KSVZ model} \label{The KSVZ model}

The first invisible axion model has been proposed by Kim \cite{Kim:1979if} and by Shifman, Vainshtein, and Zakharov \cite{Shifman:1979if}, and it is thus known as the KSVZ model. Since, in this model, the axion does not couple with SM fermions at tree level, being the $C_f = 0$ in Eq.~(\ref{axion_fermion_interaction}) for SM fermions, in the literature this model is also referred to as the hadronic axion model. The KSVZ axion couples at tree level to gluons and to an exotic heavy quark $Q$; interaction with the other SM particles occur at one loop through the intermediation of gluons and of the new $Q$ quark. Depending on the charge of $Q$, the ratio $E/N$ in Eq.~(\ref{def_E/N}) may take a value in the range $[0, 6]$. In particular, the KSVZ model allows the value $E/N = 2$ at which the axion-photon coupling $g_{a\gamma}$ in Eq.~(\ref{axion_photon_coupling}) is suppressed.

\subsubsection{The DFSZ model}

In the DFSZ model, introduced by Dine, Fischler, Srednicki \cite{Dine:1981rt}, and by Zhitnitskii \cite{Zhitnitsky:1980tq}, the axion couples at tree level to SM photons and charged leptons, besides nucleons. One of the main purposes for the introduction of this model, and an advantage for it, is that it can be easily embedded in GUT models. Any GUT model predicts $E/N = 8/3$, and so a precise value of the axion-photon coupling,
\begin{equation}
g_{a\gamma} = (8.67\pm 0.93)\,\times\,\frac{10^{-4}}{f_a}.
\end{equation}
The model also predicts the value of the axion-electron coupling $C_e$ appearing in Eq.~(\ref{axion_fermion_interaction}) with $f = e$, as
\begin{equation}\label{coupling_electron}
C_e = \frac{\cos^2\beta_H}{3},
\end{equation}
with $\cot\beta_H$ the ratio of two Higgs vacuum expectation values of this model. The coupling of axions to nucleons are also calculable in the model, and they are related by generalized Goldberger-Treiman relations to nucleon axial-vector current matrix elements. Finally, the axion coupling to the up and down quarks are
\begin{equation}
C_u = \frac{\sin^2\beta_H}{3},\quad\hbox{and}\quad C_d = \frac{\cos^2\beta_H}{3}.
\end{equation}

\subsection{Lifetime of the axion} \label{Lifetime of the axion}

Similarly to the neutral pion, the coupling of the axion to two photons arises through the electromagnetic anomaly of the PQ symmetry, and the decay of an axion into two photons is allowed with lifetime
\begin{equation} 
\tau_{a\to 2\gamma} = \frac{64\pi}{g_{a\gamma}^2\,m_a^3} = (6\pm 1) \times 10^{24}\,{\rm s}\,\frac{({\rm eV}/m_a)^{5}}{[(E/N-1.92\pm 0.08)/0.75]^2}.
\end{equation}
In the last expression, we used the values of $f_a$ and $g_{a\gamma}$ in Eqs.~(\ref{eq:axionmass}) and~(\ref{axion_photon_coupling}). In analogy with the theory of pion decay, the lifetime of the axion depends on the fifth power of the axion mass. When considering light axions of mass $m_a \lesssim 1$ eV, the lifetime $\tau_{a\to 2\gamma} \approx 10^{25}$ s implies that these particles are cosmologically stable.

\section{Astrophysical bounds on axions} \label{Astrophysical bounds on axions}

If axions existed, they would be produced in the hot plasma that constitutes stars and other astrophysical objects. The presence of axions in this dense environment would open up additional channels for the occurrence of well-studied astrophysics processes, and thus it would alter star evolution.

\subsection{Constraints from the cooling time of white dwarfs}

When helium-burning stars reach their latest stages of  helium consumption, they ascend in the Hertzsprung-Russell (HR) diagram through the red giant branch and evolve to the asymptotic giant branch (AGB). It is then possible that a AGB star, consisting of a degenerate carbon-oxygen core and an outer helium-burning shell, to evolve into a white dwarf star by first cooling down due to neutrino emission and later by surface photon emission.

If axions existed, an additional channel for the cooling of AGB into a white dwarf star exists given by the axion bremsstrahlung process
\begin{equation}
e + Ze \to e + Ze+ a.
\end{equation}
Computing the theoretical luminosity function by taking into account the processes described above and comparing it to the observed cooling rate derived from the measured decrease of the rotational period $\dot{P}/P$, it is possible to constrain any new physics contribution to the cooling process and in particular the contribution from axion bremsstrahlung.

The constraint on the axion-electron coupling thus obtained is
\begin{equation} \label{constraint_gae}
g_{ae} < 1.3\times 10^{-13},
\end{equation}
which set the most stringent bound on the axion-electron coupling constant. We derive a constraint on the PQ scale $f_a$ by inserting the expression for $g_{ae}$ in Eq.~(\ref{coupling_fermion}) and~(\ref{coupling_electron}) for $C_e$ in the bound in Eq.~(\ref{constraint_gae}), to obtain
\begin{equation} \label{astrophysical_bound}
f_a > 1.3\times 10^{8}{\rm~GeV}\,\cos^2\beta_H,
\end{equation}
where $\cot\beta_H$ is the ratio of the two Higgs vacuum expectation values.

{}Recently, a method for constraining the axion-photon coupling $g_{a\gamma}$ using the amount of linear polarization in the radiation emerging from magnetic white dwarfs has been proposed \cite{Gill:2011yp}. This method sets an upper limit on $g_{a\gamma}$ that depends on the axion mass $m_a$. For $m_a \lesssim 10^{-4}$ eV, the authors in Ref.~\cite{Gill:2011yp} derive the bound $g_{a\gamma} < 10^{-10}{\rm ~GeV^{-1}}$: this bound is much more stringent than the one obtained using measurements on the lifetime of horizontal branch stars in globular clusters, see Sec.~\ref{Axions and globular clusters}.

\subsection{Constraints from SN1987A} \label{Constraints from SN1987A}

A supernova (SN) event of type II consists of a core collapse of a massive star which subsequently leads to a proto-neutron star. Axions may be produced in SN events via axion-nucleon bremsstrahlung
\begin{equation}
N + N \to N + N+ a,
\end{equation}
and this additional process can affect the cooling time of the SN and shorten the duration of the burst. The pattern of the cooling time as a function of the axion-neutron coupling $g_{aN}$ is described in Ref.~\cite{Raffelt:2006cw}. For very low values of the coupling, the axion emission does not affect the burst duration. As $g_{aN}$ increases, the burst duration shortens because the emission of bremsstrahlung axions increases, reaching its minimum value when the mean free path of axions in the medium is of the order of the size of the SN. For even larger values the axions are trapped in the medium and the axion emission decreases, until the burst duration becomes unaffected by the presence of these new particles.

In 1987 it was possible to test models of supernovae explosions, due to the observation of the emission from a distant supernova named 1987A. The flux of antineutrinos $\bar{\nu}_e$ coming from SN1987A was detected at both Kamiokande II and Irvine-Michigan-Brookhaven experiments, allowing us to compare the data with theoretical expectations. Due to this analysis, it is possible to exclude axion models with an axion-nucleon coupling in the range
\begin{equation}
3\times 10^{-10} \lesssim g_{aN} \lesssim 3\times 10^{-7}.
\end{equation}
This range corresponds to the high yield emission of axions excluded by the statistical analysis on the SN1987A data.

\subsection{Axions from the Sun} \label{Axions from the Sun}

In the Sun, axions might be produced through the Primakoff process: an incoming photon interacts with the electromagnetic field of a nearby electron or nucleus, converting into an axion, see the Feynman diagram in Fig.~\ref{axion_photon_img}. Since this new channel can in principle shorten the lifetime of the Sun, it is possible to constrain the coupling of axions to photons, electrons, and nucleus by imposing that the presence of axions does not spoil the standard solar model (SSM).

To see this in mode details, we consider solar limits on the axion-photon coupling $g_{a\gamma}$. in the approximation in which the electron or nucleus is nonrelativistic, the rate of conversion of photons of energy $E$ into axions of the same energy in a nondegenerate plasma of temperature $T$ is
\begin{equation}
\Gamma_{\gamma\to a} = \frac{g_{a\gamma}^2\,T\,k_s^2}{32\pi}\,\left[\left(1+\frac{k_s^2}{4E^2}\right)\log\left(1+\frac{4E^2}{k_s^2}\right)-1\right].
\end{equation}
Here, the screening factor $k_s$ is given by the Debye-Huckel formula,
\begin{equation}
k_s^2 = 4\pi\,\alpha\,\frac{n_B}{T}\,\left(Y_e+\sum_j\,Z_j^2\,Y_j\right),
\end{equation}
where $n_B$ is the nondegenerate baryon density and $Y_e$, $Y_j$ are the fractions of electrons and nuclear species $j$ per baryon. The energy loss rate in axions per unit volume is then
\begin{equation}
Q_{\gamma\to a} = g_\gamma\int\frac{d^3\,k}{(2\pi)^3}\,\frac{E}{e^{-E/T}-1}\,\Gamma_{\gamma\to a},
\end{equation}
where $g_\gamma = 2$ and ${\bf k}$ is the three-momentum of the incoming photon of energy $E$. Eventually, using the solar temperature distribution obtained from the SSM, the flux of solar axions on Earth is evaluated as
\begin{equation}\label{axion_flux_Sun}
L_a = 1.7\times10^{-3}\,\left(\frac{g_{a\gamma}}{10^{-10}{\rm~GeV^{-1}}}\right)^2\,L_\Sun,
\end{equation}
with $L_\Sun$ indicating the Sun luminosity. Here, we use the conservative limit on the solar axion luminosity $L_a \lesssim 0.2\,L_\Sun$ \cite{Schlattl:1998fz} which yields
\begin{equation}
g_{a\gamma} \lesssim 1.1\times 10^{-9}\,{\rm ~GeV^{-1}}.
\end{equation}

Using the fact that axions affect the sound speed-profile of the Sun, the authors in Ref.~\cite{Schlattl:1998fz} obtained a similar bound,
\begin{equation}
g_{a\gamma} \lesssim 1.0\times 10^{-9}\,{\rm ~GeV^{-1}}.
\end{equation}
Raffelt and Gondolo \cite{Gondolo:2008dd} derived a restrictive bound on $g_{a\gamma}$ using the sensitive dependence of the production rate of neutrinos from $^8B$. Using the measurements of the solar neutrino flux by the Sudbury Neutrino Observatory, the authors in Ref.~\cite{Gondolo:2008dd} obtain
\begin{equation}
g_{a\gamma} \lesssim 7\times 10^{-10}\,{\rm ~GeV^{-1}}.
\end{equation}
To conclude, we remark here that the best constraint on $g_{a\gamma}$ from astrophysical measurements comes from considering the production of axions in globular clusters.

\subsection{Axions and globular clusters} \label{Axions and globular clusters}

A globular cluster (GC) is a gravitationally-bound ensemble of stars which formed at about the same epoch. Since these stars share the same age, but not other physical parameters like the mass or the surface temperature, a GC is particularly suitable for testing models of stellar evolution. A typical method used for comparing stars within the same GC is to plot the color luminosity vs. the total brightness of each star in a color-magnitude diagram. The typical quantities used are the color $B-V$ and the brightness $V$.

When axions are included in the model, helium-burning stars may consume helium faster than their expected rate because of the extra axion-production channel.  The lifetime of stars which are in the horizontal branch is reduced by a factor
$$\left[1+\frac{3}{8}\left(\frac{g_{a\gamma}}{10^{-10}{\rm ~GeV^{-1}}}\right)^2\right]^{-1},$$
due to the production of axions. When a statistically significant ensemble of helium-burning stars is considered, it is found that the lifetime of these stars agrees with theoretical expectations within 10\%, which in turns leads to the bound 
\begin{equation}
g_{a\gamma} \lesssim 1\times 10^{-10}\,{\rm ~GeV^{-1}}.
\end{equation}
This bound on the axion-photon coupling is a much more stringent bound than those obtained from considerations on the solar activity.

\section{Direct axion searches}

As first proposed by Sikivie \cite{Sikivie:1983ip}, it is possible to directly search for axions in laboratories by using the conversion of an axion into a photon in the presence of an external electromagnetic field. This process, which is the inverse process of the Primakoff effect, has been extensively used in a number of experiments. Direct searches of axions fall into three primary categories: search of axions of galactic origin (axion haloscope), produced in the Sun (axion helioscope), and pure laboratory experiments, in which virtual axions are produced by shining lasers in strong magnetic fields.

\subsection{Axion haloscope}

Axion haloscope searches rely on the possibility that relic axions from the Big Bang would be gravitationally bound to the Milky way, having a nonrelativistic velocity $v$ with dispersion $\Delta v \approx 10^{-3}\,c$, corresponding to the rotational velocity of a virialized object in the vicinity of the solar system. This would correspond to an axion mean energy and energy dispersion
\begin{equation} 
E \approx m_a\,c^2\,\left(1+\frac{v^2}{2c^2}\right),\quad\hbox{and}\quad\Delta E = m_a\,v\,\Delta v.
\end{equation}
In order to detect these particles, a sensitive technique known as the microwave cavity has been developed. In such a cavity, a strong electromagnetic field is produced, with a frequency related to the size of the cavity. For a given frequency, there exists a narrow range of the axion mass $m_a$ for which the axion would interact with the electromagnetic field and convert into a light pulse which would be eventually detected by a receiver. In order to search for different values of the axion mass, the size of the microwave cavity is adjustable. The search of axions through a microwave cavity assumes that the totality of the CDM is in the form of axions. As we will explore in depth in this thesis, this is possible if the mass of the axion is of the order of the $\mu{\rm ~eV}$, corresponding to the energy of a microwave with a wavelength of approximately $10{\rm~cm}$.

The probability that an axion of energy $E$ converts into a detectable photon while traveling in the homogeneous and transverse magnetic field $B$ of a microwave cavity is
\begin{equation} \label{probability_conversion}
P_{a\to \gamma} = \frac{(g_{a\gamma}\,B\,L/2)^2}{L^2\,(q^2+\Gamma^2/4)}\,\left[1+e^{-\Gamma L} - 2\,E^{-2\Gamma L}\,\cos(q\,L)\right),
\end{equation}
where $L$ is the length of the path, $\Gamma$ is the inverse absorption length for photons in the medium and $q$ is the momentum transfer between the axion and photon. In the vacuum, the momentum transfer $q$ reads
\begin{equation} \label{transfer_momentum}
q = \left|\frac{m_a^2-m_\gamma^2}{2E}\right|,
\end{equation}
where the result is expressed in terms of the effective photon mass in a plasma with electrons density $n_e$,
\begin{equation}
m_\gamma = \sqrt{\frac{4\pi\,\alpha\,n_e}{m_e}}.
\end{equation}
Notice that the result in Eq.~(\ref{transfer_momentum}) is similar to that obtained in the neutrino oscillations theory. When the absorption $\Gamma$ is neglected, Eq.~(\ref{probability_conversion}) reduces to
\begin{equation} \label{probability_conversion1}
P_{a\to \gamma} = \left(\frac{g_{a\gamma}\,B}{q}\right)^2\,\sin^2\left(\frac{q\,L}{2}\right),
\end{equation}

The first experiments of this type were performed at Brookhaven Laboratories \cite{Wuensch:1989sa} and at the University of Florida \cite{Hagmann:1990tj}: the axion mass is excluded from the former and the latter experiments in the ranges $5.4 \div 5.9{\rm ~\mu eV}$ and $4.5 \div 16.3{\rm ~\mu eV}$, respectively. As of today, the most sensitive axion haloscope is the Axion Dark Matter eXperiment (ADMX) at Laurence Livermore National Laboratory (LLNL). In its first direct scan of the parameter space, ADMX directly excluded axions with mass in the range \cite{Asztalos:2003px}
\begin{equation}\label{axion_mass_bound_ADMX}
1.9 {\rm ~\mu eV} < m_a < 3.3 {\rm ~\mu eV}.
\end{equation}
A later upgrade of the experiment \cite{Asztalos:2009yp} exploited a Superconducting QUantum Interference Device (SQUID), which replaced the microwave receiver and allowed us to extend the scan to higher values of the axion mass. The improved ADMX excluded the region \cite{Asztalos:2009yp}
\begin{equation}
3.3 {\rm ~\mu eV} < m_a < 3.53 {\rm ~\mu eV},
\end{equation}
so that the overall region excluded by the ADMX experiment so far is
\begin{equation} \label{exclusion_haloscope}
1.9 {\rm ~\mu eV} < m_a < 3.53 {\rm ~\mu eV}.
\end{equation}

\subsection{Axion helioscope}

The Sun produces axions through nuclear interactions within its core; the expected axion flux at Earth due to the solar activity is expressed in Eq.~(\ref{axion_flux_Sun}). The search for solar axions is mainly conducted via two types of experiments, which respectively exploit the intense Coulomb field in a crystal to convert axions into photons (Bragg scattering experiments), or use intense magnetic fields that point at the Sun in which solar axions can convert into photons (magnet helioscope experiments).

\subsubsection{Search of solar axions through Bragg scattering}

This type of experiment takes advantage of the fact that the mean energy of solar axions is around 4keV, so that the axion wavelength is of the same order as the lattice spacing in a typical crystal. The expected Bragg scattering of the resulting photon would increase the signal of the axion interaction with the lattice to around $10^4$ with respect to the interaction of an impurity in the same crystal. Moreover, the signal would be distinctive because of the relative movement of the Sun within 24 hours (daily modulation of the signal). Various experiments undergoing such a search use a different crystal such as Sodium Iodide at DAMA \cite{Bernabei:2001ny}, Germanium at SOLAX \cite{Gattone:1997hf} and COSME \cite{Morales:2001we}, or Germanium and Silicon at CDMS \cite{PhysRevLett.103.141802}. All of these experiments constrain the axion-to-photon coupling to about the same level, $g_{a\gamma} \lesssim 2\times 10^{-9}{\rm ~GeV^{-1}}$, which however is more than one order of magnitude above the bound from haloscope searches.

\subsubsection{Solar axions through axion telescopes}

The main component of an axion telescope consists of a magnet in which the strong magnetic field can be pointed at the Sun. An axion in the strong magnetic field may convert to a low-energy X-ray, through a mechanism similar to the conversion of CDM axions in a microwave cavity. In fact, the probability of conversion is given by Eq.~(\ref{probability_conversion}) also in this situation, provided that the same conditions in which the latter equation has been derived are met.

Searches of solar axions via this method were conducted by Lazarus and collaborators \cite{PhysRevLett.69.2333} and by the Tokyo Axion Helioscope collaboration \cite{Moriyama:1998kd}: the latter group constrained $g_{a\gamma} < 6\times 10^{-10}{\rm ~GeV^{-1}}$ for axion masses $m_a < 0.03$ eV.

A parallel search has been conducted by CERN Axion Solar Telescope (CAST) experiment, which to date is the most sensitive axion helioscope in use. In fact, CAST in its two phases has been able to constrain $g_{a\gamma}$ more stringently than astrophysical bounds, setting $g_{a\gamma} < 8.8\times 10^{-11}{\rm ~GeV^{-1}}$ at 95\% C.L. and for $m_a < 0.02$ eV in Phase I \cite{Zioutas:2004hi}, and $g_{a\gamma} \lesssim 2.2\times 10^{-10}{\rm ~GeV^{-1}}$ at 95\% C.L. and for the mass range $0.02 {\rm ~eV} < m_a < 0.4$ eV during Phase II \cite{Arik:2008mq}.

\subsection{Production of axions by laser}

Axion searches conducted in laboratories make use of an intense laser beam that might partially convert into axions or other pseudo-scalar particles in the presence of a strong magnetic field, via the Primakoff effect. We briefly discuss two different types of experiments that exploit this technique.

\subsubsection{Polarization experiments}

When axions are produced by the interaction of a strong laser beam with an external magnetic field, the polarization of the laser beam is affected because of such a nonzero coupling with the pseudo-scalar particles. Moreover, when a photon-axion-photons reconversion takes place, the emerging photon has a component of the ${\bf E}$ and ${\bf B}$ fields retarded with respected to a freely-propagating photon. This signal had been claimed by the Polarizzazione del Vuoto LASer (PVLAS) collaboration \cite{Cantatore:2008zz}, but the event was later labeled as artificial.

\subsubsection{Regeneration experiments}

When a polarized laser beam propagating in a constant transverse magnetic field is blocked by a thick absorber (a wall), it is possible to detect photons of the same wavelength as the laser beam on the other side of the blocking medium itself. This process is called photon regeneration, and the experiment is colloquially referred to as the {\it shining light through a wall} experiment. This unusual phenomenon is possible if some photons from the laser beam convert into axions in the magnetic field, before being absorbed by the wall. In this case, the produced axions are able to travel through the wall with little absorption. Applying a second magnetic field on the other side of the wall makes it possible for some axions to reconvert into photons through the inverse Primakoff process.

This type of experiments has been conducted by the Brookhaven-Fermilab-Rutherford-Trieste collaboration \cite{Cameron:1993mr} and at CERN with the Optical Search for QED vacuum birefringence, Axions and photon Regeneration (OSQAR) experiment. In addition, the shining light through a wall experiment is able to put limits on more general models in which photons couple to axion-like particles (ALPs), theoretical particles that share similar properties with the axion but for which the mass-energy scale relation in Eq.~(\ref{eq:axionmass}) does not apply. We will talk in detail about ALPs in Sec.~\ref{Natural warm inflation}. The bound on the ALP-photon coupling constant $G_{a\gamma}$ from this experiment is
\begin{equation}
G_{a\gamma} \lesssim 1\times 10^{-10}{\rm ~GeV^{-1}}.
\end{equation}

%% file: chap5.tex
\chapter{Revising the axion as the cold dark matter} \label{Revising the axion as the cold dark matter}

\fixchapterheading
\section{Introduction} \label{introduction_V}
The recent measurements by the WMAP mission \cite{Komatsu:2008hk} have established the relative abundance of dark and baryonic matter in our universe with great precision. About 84$\%$ of the nonrelativistic content in the universe is in the form of CDM, whose composition is yet unknown. One of the most promising hypothetical particles proposed for solving the enigma of the dark matter nature is the axion \cite{Weinberg:1977ma, Wilczek:1977pj}. This particle was first considered in 1977 by R. Peccei and H. Quinn \cite{Peccei:1977hh} in their proposal to solve the strong-CP problem of the QCD theory. Although the original PQ axion is by now excluded, other axion models are still viable \cite{Kim:1979if, Shifman:1979if, Dine:1982ah, Zhitnitsky:1980tq}.

The hypothesis that the axion can be the dark matter particle has been studied in various papers (see e.g. \cite{Preskill:1982cy, Abbott:1982af, Dine:1982ah, Stecker:1982ws, Turner:1990uz, Lyth:1992tw, Beltran:2006sq, Hertzberg:2008wr} and the reviews in \cite{Fox:2004kb, Sikivie:2006ni}). Here we examine the possibility that the invisible axion may account for the totality of the observed CDM, in light of the WMAP5 mission, baryon acoustic oscillations (BAO), and supernovae (SN) data. We also upgrade the treatment of anharmonicities in the axion potential, which we find important in certain cases. We consider invisible axion models, in which the breaking scale of the PQ symmetry $f_a$ is well above the electroweak scale. The axion parameter space is described by three parameters, the PQ energy scale $f_a$, the Hubble parameter at the end of inflation $H_I$, and the axion initial misalignment angle $\theta_i$.

\section{Production of axions in the early universe}

\subsection{Thermal production of axions} \label{Thermal production of axions}

In the early universe, a population of thermal axions originates together with standard model particles and possibly other exotic components. Scattering and annihilation processes keep this population of axions in thermal equilibrium with the remaining of the hot plasma at temperature $T$, with the axion number density $n^{\rm th}_a$ following the Boltzmann equation
\begin{equation}\label{boltzmann_eq}
\frac{d\,n_a^{\rm th}}{dt} + 3H(t)\,n_a^{\rm th} = \Gamma_{\rm ann}\,\left(n_a^{\rm eq,th} - n_a^{\rm th}\right).
\end{equation}
Here, the number density of axions at thermal equilibrium is, see Eq.~(\ref{number_density_statmech}),
\begin{equation}
n_a^{\rm eq,th} = \frac{\zeta(3)}{\pi^2}\,T^3.
\end{equation}
In Eq.~(\ref{boltzmann_eq}), the rate at which axions annihilate and are created in the plasma is
\begin{equation} \label{rate_ann}
\Gamma_{\rm ann} = \sum_i\,n_i\,\langle\sigma_i\,v\rangle,
\end{equation}
where $n_i$ is the number density of the particle species $i$. The index $i$ runs over processes of the type $a + i \rightleftharpoons$ other particles, with cross section $\sigma_i$, and with $v$ the relative velocity between $i$ and the axion. Finally, the angular brackets in Eq.~(\ref{rate_ann}) indicate an average over the momentum distributions of the particles involved. Using the conservation of the number density at equilibrium,
\begin{equation}\label{boltzmann_eq_equilibrium}
\frac{d\,n_a^{\rm eq,th}}{dt} + 3H(t)\,n_a^{\rm eq,th} = 0,
\end{equation}
we can rewrite Eq.~(\ref{boltzmann_eq}) as
\begin{equation}\label{boltzmann_eq1}
\frac{d}{dt}\left[a^3(t)\,\left(n_a^{\rm eq,th}-n_a^{\rm th}\right)\right] = -\Gamma_{\rm ann}\,a^3(t)\,\left(n_a^{\rm eq,th} - n_a^{\rm th}\right).
\end{equation}
The solution of Eq.~(\ref{boltzmann_eq1}) implies that a thermal population of axions reaches its equilibrium value $n_a^{\rm eq,th}$ exponentially fast whenever
\begin{equation}
\Gamma_{\rm ann} > H(t).
\end{equation}
The time $t_{\rm chem}$ at which axions chemically decouple from the plasma is then defined with $\Gamma_{\rm ann} = H(t_{\rm chem})$.

Depending on the axion model, axions couple differently to standard model fermions, photons, and exotic particles, whereas the coupling to gluons is model-independent and is given in Eq.~(\ref{Lagrangian_axion}). Thermal axions were first studied by Turner \cite{Turner:1986tb}, who considered the Primakoff and photoproduction processes. Given a heavy quark $Q$ of mass $m_Q$, whenever the temperature of the plasma $T \lesssim m_Q$ the Primakoff process dominates and axions are created via $q + \gamma \rightleftharpoons q + a$ ($q$ is a light quark), with cross section
\begin{equation}
\sigma_{\rm Primakoff} = \frac{\alpha^3}{f_a^2}.
\end{equation}
At higher temperatures $T \gtrsim m_Q$, photoproduction of axions via $Q + \gamma \rightleftharpoons Q + a$ dominates with cross section
\begin{equation}
\sigma_{\rm \gamma-prod} = \alpha \,\left(\frac{m_Q}{f_a \,T}\right)^2.
\end{equation}
Ref.~\cite{Turner:1986tb} concluded that a thermal population of axions exists today whenever $f_a \lesssim 10^9{\rm ~GeV}$; moreover, for $f_a \lesssim 4\times 10^8$ GeV, the thermal population is greater than the nonthermal.

These considerations were revised in Ref.~\cite{Masso:2002np}, where the authors extended the work in Ref.~\cite{Turner:1986tb} by including other processes than Primakoff and photoproduction, using the model-independent axion-gluon vertex derived from Eq.~(\ref{Lagrangian_axion}). In detail, the authors in Ref.~\cite{Masso:2002np} include the following reactions:
$$a + g \rightleftharpoons q + \bar{q}$$
$$a + q \rightleftharpoons g+ q$$
$$a + g \rightleftharpoons g +g$$
These processes have a cross section of the order of $\sigma_{\rm agg} = \alpha_s^3/f_a^2$. With these additional reactions included, axions thermalize if $f_a \lesssim 10^{12}{\rm ~GeV}$ \cite{Masso:2002np}, which is a broader condition than the one found in Ref.~\cite{Turner:1986tb} because axions are tighter bound to the primordial plasma.

Assuming that axions thermalize at temperature $T_{\rm th}$, their number density today is obtained by assuming the conservation of the comoving number density,
\begin{equation}
n_a^{\rm th}(T_0) = n_a^{\rm th}(T_{\rm th})\,\left(\frac{a(T_{\rm th})}{a(T_0)}\right)^3 = 7.5\,{\rm ~cm^{-3}}\,\frac{107.75}{g_*(T_{\rm th})},
\end{equation}
where $g_{*S}(T)$ denotes the relativistic number of degrees of freedom at temperature $T$. This thermal population of axions is very diluted today, once compared for example to the CMB photon number density $n_\gamma(T_0) \approx 410 {\rm ~cm^{-3}}$.

\subsection{Axions from the misalignment mechanism} \label{Axions from the misalignment mechanism}

\subsubsection{Equation of motion on the FRW metric}

The generic Friedmann-Robertson-Walker (FRW) metric is written as
\begin{equation}
ds^2 = dt^2 - a^2(t)\,d{\bf x}\cdot d{\bf x},
\end{equation}
where $a(t)$ is the scale factor, which relates to the Hubble expansion rates by $H(t) = \dot{a}(t)/a(t)$, and ${\bf x}$ are comoving spatial coordinates. On this metric, the angular variable $\theta = \theta(x)$ has equation of motion
\begin{equation}
D_\mu\,\partial^\mu\,\theta(x) + \frac{1}{f_a^2}V'(\theta) = \ddot{\theta} +3H(t)\,\dot{\theta} - \frac{1}{a^2(t)}\nabla_x^2\,\theta + \frac{1}{f_a^2}V_\theta = 0,
\end{equation}
where a dot indicates a derivative with respect to $t$ and $V_\theta = \partial V/\partial \theta$. Using the axion potential in Eq.~(\ref{axion_potential1})
\begin{equation} \label{axion_potential}
V(\theta) = m_a^2(T)\,f_a^2\,(1-\cos\theta),
\end{equation}
that originates from nonperturbative effects after the spontaneous breaking of the $U(1)_{\rm PQ}$ symmetry, we obtain
\begin{equation}
\ddot{\theta} +3H(t)\,\dot{\theta} - \frac{1}{a^2(t)}\nabla_x^2\,\theta + m_a^2(T)\,\sin\theta = 0.
\end{equation}
From now on, we only focus on the mode with zero momentum, for which $\nabla_x^2\,\theta = 0$ and which constitutes the CDM component. A general treatment on axionic modes with nonzero momenta can be found in Ref.~\cite{Sikivie:2006ni}. For zero modes, it is
\begin{equation} \label{axion_eq_motion}
\ddot{\theta} + 3H(t)\,\dot{\theta} + m_a^2(T)\,\sin\theta = 0.
\end{equation}
In the early universe, when the plasma temperature is $T \gtrsim \Lambda_{\rm QCD}$, the axion mass and hence the potential is negligible, see Eq.~(\ref{axion_mass_temperature}), and a solution to Eq.~(\ref{axion_eq_motion}) is $\theta = \theta_i = $constant, with $\theta_i$ being the initial value of the misalignment field. As the temperature drops below $\Lambda_{\rm QCD}$, or, more precisely, when the axion field starts to oscillate at a temperature $T_1 = O({\rm GeV})$ given by \cite{Turner:1985si}
\begin{equation} \label{definition_T1}
3H(T_1) \approx m_a(T_1),
\end{equation}
the mass term in Eq.~(\ref{axion_eq_motion}) becomes important and the solution is no longer trivial.

The axion field starts oscillating in the nonperturbative potential $V(\theta)$ with initial condition $\theta = \theta_i$. We can solve the equation above analytically by assuming that the axion mass $m_a$ is constant and that $\theta_i$ be much smaller than one, so that $\sin\theta \approx \theta$ at all times. For a generic cosmology in which the scale factor depends on time as
\begin{equation}
a(t) \sim t^\beta,
\end{equation}
and thus the Hubble parameter is
\begin{equation}
H(t) = \frac{\dot{a}(t)}{a(t)} = \frac{\beta}{t},
\end{equation}
we obtain
\begin{equation} \label{approx_solution}
\theta(x) = 2^{1/4}\,\Gamma\left(\frac{5}{4}\right)\,\theta_i\,\tau_a^{\frac{1- 3 \beta}{2}} J_{\frac{3 \beta-1}{2}}(\tau_a), 
\end{equation}
with $\tau_a = m_a\,t$ and where $J_\nu(x)$ is the Bessel function of the first kind of order $\nu$ in the variable $x$ and $\Gamma(x)$ is the Euler Gamma function.

\begin{figure}[h!]
\begin{center}
\includegraphics[width=13cm]{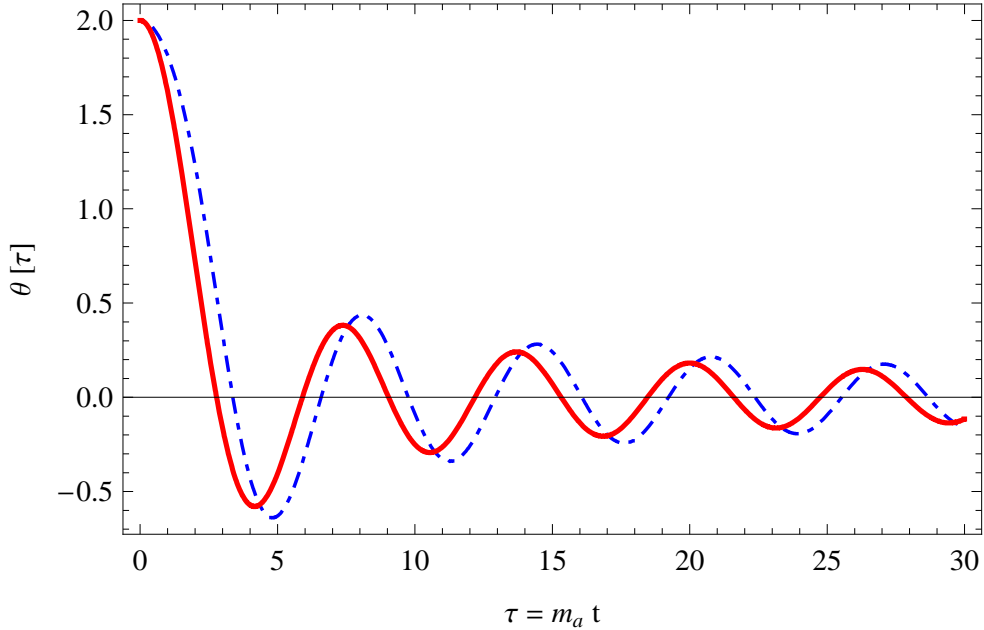}
\caption[Numerical (solid line) and approximated (dashed line) solutions for $\theta(\tau)$.]{Comparison between the numerical solution of Eq.~(\ref{axion_eq_motion}) (red solid line) and the approximate solution given in Eq.~(\ref{approx_solution}) (blue dashed line), with initial condition $\theta_i = 2$.}
\label{motionplot}
\end{center}
\end{figure}

The general solution to Eq.~(\ref{axion_eq_motion}) that does not assume $\theta_i \ll 1$ and takes into account a generic temperature-dependent mass $m_a(T)$ can be found only numerically. In Fig.~\ref{motionplot}, we compare the numerical solution of Eq.~(\ref{axion_eq_motion}) (solid red line) with the approximate solution expressed in Eq.~(\ref{approx_solution}) (blue dashed line) for $\theta_i = 2$. We see that anharmonicities sensibly modify the behavior of the field dynamics.

\subsubsection{Computing the oscillation temperature} \label{Computing the oscillation temperature}

When the potential term in Eq.~(\ref{axion_eq_motion}) can no longer be neglected, coherent oscillations in the axion field begin. This happens when the temperature of the universe drops to the value $T_1^{\rm std}$, given implicitly by Eq.~(\ref{definition_T1}); we have added a label ``std'' that refers to the fact that this temperature is computed in the standard cosmological scenario, in which the universe is radiation-dominated right after the end of inflation. For $T\gtrsim \Lambda$, we obtain an equation involving $T_1^{\rm std}$ by inserting Eq.~(\ref{axion_mass_temperature0}) in Eq.~(\ref{definition_T1}) and using Eq.~(\ref{hubble_radiation}) for the Hubble expansion rate in a radiation-dominated universe,
\begin{equation}
H_{\rm rad}(T) = \bigg(\frac{8\pi^3\,g_*(T)}{90\,M_{\rm Pl}^2}\bigg)^{1/2}T^2 \approx 1.66\,\sqrt{g_*(T)}\,\frac{T^2}{M_{\rm Pl}}.
\end{equation}
In this way, we find that the parameter
\begin{equation}
x_1 \equiv \frac{T_1^{\rm std}}{\Lambda},
\end{equation}
must satisfy the expression
\begin{equation}\label{x1_equation}
\frac{x_1^6}{\left(1+\ln x_1\right)^{1.2}} = (3300\pm 1700)\,f_{a,12}^{-1}\,\sqrt{\frac{\theta_i}{\sin\theta_i\,g_*(T_1^{\rm std})}}\quad\hbox{for $x_1 \gtrsim 1$}.
\end{equation}
Eq.~(\ref{x1_equation}) defines implicitly the value of $T_1^{\rm std}$. We have checked numerically that the logarithmic term contributes a 15\% contribution to the final solution, so we neglect it because of the huge uncertainty in the parameter $a$ in Eq.~(\ref{axion_mass_new}). Neglecting the logarithmic term in Eq.~(\ref{x1_equation}) is equal to solving Eq.~(\ref{definition_T1}) with the axion mass given by  Eq.~(\ref{axion_mass_new}) instead of Eq.~(\ref{axion_mass_temperature0}). We obtain \cite{Visinelli:2009zm}
\begin{equation} \label{T_1}
T_1^{\rm std} =
\begin{cases}
\left(\frac{a\,m_a\,M_{\rm Pl}\,\Lambda_{\rm QCD}^4}{4.98\,\sqrt{g_*(T_1^{\rm std})}}\right)^{1/6} = (607\pm 27){\rm ~MeV}\,f_{a,12}^{-1/6}, & T \gtrsim \Lambda_{\rm QCD},\\
\left(\frac{m_a\,M_{\rm Pl}}{4.98\,\sqrt{g_*(T_1^{\rm std})}}\right)^{1/2} = (67 \pm 3){\rm ~MeV}\,f_{a,18}^{-1/2}, & T  \lesssim \Lambda_{\rm QCD},\\
\end{cases}
\end{equation}
In the last expression, we have used the numerical value $a = 0.017\pm 0.003$ and the values for $g_*(T)$ in Eq.~(\ref{relativistic_dof_numerical}).

\subsubsection{Axion energy density from vacuum realignment} \label{Axion energy density from vacuum realignment}

A generic scalar field $\Phi(x)$ of mass $m_\Phi$ in a quadratic potential has energy density
\begin{equation}
\rho_\Phi = \frac{1}{2}\,\left(m_\Phi^2\,\Phi^2 + \dot{\Phi}^2\right).
\end{equation}
In the case of the axion, the scalar field $a(x) = f_a \theta/N$ does not move in a quadratic potential but rather in the residual potential given by Eq.~(\ref{axion_potential}). This sinusoidal potential behaves like a quadratic potential if $\theta_i \ll 1$, but when the initial value of the misalignment angle is of order one, anharmonicities cannot be neglected, as shown in Fig.~\ref{motionplot}.

To begin with, we will consider the case in which the initial value of the misalignment angle is small, which allows the expansion
\begin{equation}\label{axion_potential_expansion}
V(\theta) \approx \frac{1}{2}\,m_a^2(T)\,\left(\frac{f_a}{N}\right)^2\,\theta^2 \quad\hbox{for $\theta_i \ll 1$}.
\end{equation}
This potential turns on at the temperature $T_1^{\rm std}$ defined by Eq.~(\ref{definition_T1}): for higher values of $T$, the angular variable $\theta$ is frozen to its initial value $\theta_i$, so that we can assume $\theta = \theta_i$ when $T = T_1^{\rm std}$. Then, the axion number density is initially
\begin{equation} \label{number_density}
n_a\left(T_1^{\rm std}\right) = \frac{1}{2} \chi\,m_a(T_1^{\rm std})\,\left(\frac{f_a}{N}\right)^2\,\langle\theta_i^2\rangle.
\end{equation}
Here, the factor $\chi$ models the temperature dependence of the axion mass around $T_1^{\rm std}$ \cite{Turner:1985si}, and depends on the number of quark flavors $N_f$ that are relativistic at temperature $T_1^{\rm std}$ through the index $c$ appearing in Eq.~(\ref{axion_mass_temperature}) as
\begin{equation}
\chi = 0.44 + 0.25\,c = 0.44+\frac{21+N_f}{24}.
\end{equation}
Here, we consider $N_f = 3$ because the axion field starts to oscillate at temperatures below the GeV, so that we take $\chi = 1.44$. The number density at the present time is found by imposing the conservation of the comoving axion number density Eq.~(\ref{conservation_comoving_number}), in the form \cite{kolb1990early}:
\begin{equation} \label{conservation}
\delta\!\left(\frac{n_a(T)}{s(T)}\right) = 0,
\end{equation}
where the entropy density is given by Eq.~(\ref{entropy_density}),
\begin{equation}\label{entropy}
s(T) = \frac{2\pi^{2}}{45}\,g_{*S}(T)\,T^3,
\end{equation} 
and $g_{*S}(T)$ denotes the entropy number of degrees of freedom at temperature $T$, see Eq.~(\ref{entropy_dof}). We approximate $g_{*S}(T) = g_*(T)$ when $T \gtrsim 4$ MeV, while the present value for $g_{*S}(T_0) = 3.91$ differs from $g_{*}(T_0) = 3.36$ \cite{kolb1990early}. For the range of interest here, for the relativistic degrees of freedom, we use the values in Eq.~(\ref{relativistic_dof_numerical}). Using Eqs.~(\ref{number_density}),~(\ref{conservation}) and~(\ref{entropy}), the present axion energy density from the misalignment mechanism $\rho_a^{\rm mis}(T_0) = m_a\,n_a(T_0)$ is
\begin{equation} \label{energydensity}
\rho_a^{\rm mis}(T_0) = \frac{1}{2}\,\chi\,m_a\,m_a(T_1^{\rm std})\,\frac{s(T_0)}{s\left(T_1^{\rm std}\right)}\,\left(\frac{f_a}{N}\right)^2\,\langle\theta_i^2\rangle.
\end{equation}
Dividing the last equation by the critical density $\rho_c = 3H^2_0
M^2_{Pl}/8\pi$ and using Eq.~(\ref{axion_mass_temperature}) for $m_a(T_1^{\rm std})$, the
cosmologically relevant ratio $\Omega^{\rm mis}_a = \rho_a/\rho_c$ is
\begin{equation} \label{energy_density_axions}
\Omega^{\rm mis}_a =  \frac{\chi}{2\,\rho_c}\,m_a^2\,\frac{s(T_0)}{s\left(T_1^{\rm std}\right)}\,\left(\frac{f_a}{N}\right)^2\,\langle\theta_i^2\rangle
\begin{cases}
a \,\left(\frac{\Lambda_{\rm QCD}}{T_1^{\rm std}}\right)^4 ,& T\gtrsim \Lambda_{\rm QCD},\\
\quad 1, & T\lesssim \Lambda_{\rm QCD}.
\end{cases}
\end{equation}
For the temperature $T_1^{\rm std}$, we substitute the expression we obtained in Eq.~(\ref{T_1}) in Eq.~(\ref{energy_density_axions}) with the numerical values $\chi = 1.44$ and $a = 0.017\pm 0.003$, obtaining
\begin{equation} \label{standarddensity}
\Omega^{\rm mis}_a h^2 =
\begin{cases}
(0.23 \pm 0.03)\,\langle\theta_i^2\rangle\,f_{a,12}^{7/6}, & f_a \lesssim \hat{f}_a,\\
(6.1\pm 0.8)\times10^{-3}\,\langle\theta_i^2\rangle\,f_{a,12}^{3/2}, & f_a \gtrsim \hat{f}_a.\\
\end{cases}
\end{equation}
The parameter $\hat{f}_a$ is defined as the energy scale at which the two expressions in Eq.~(\ref{standarddensity}) match:
\begin{equation}
\hat{f}_a = 5.4\times 10^{16}{\rm ~GeV}.
\end{equation}
The two cases in Eq.~(\ref{standarddensity}) reflect the dependence of the axion mass $m_a(T)$ on the temperature $T$ in Eq.~(\ref{axion_mass_temperature}).

\subsubsection{The role of anharmonicities} \label{The role of anharmonicities}

In the general case in which $\theta_i \sim 1$, the expression in Eq.~(\ref{standarddensity}) underestimates the actual axion energy density from the misalignment mechanism. In fact, the average value of $\theta_i^2$ over a Hubble horizon when the axion potential is treated as a pure harmonic potential like in Eq.~(\ref{axion_potential_expansion}) gives
\begin{equation} \label{average_harmonic}
\langle\theta_i^2\rangle = \frac{1}{2\pi}\,\int_{-\pi}^\pi\,d\theta_i\,\theta_i^2 = \frac{\pi^2}{3}.
\end{equation}
Instead, the potential in Eq.~(\ref{axion_potential}) is not harmonic, and the effects when the value of the angle $\theta_i$ lies close to $\pi$ become rather important. Turner \cite{Turner:1985si} found that the effect of these anharmonicities raise the value of the integral in Eq.~(\ref{average_harmonic}) by a factor 1.2-1.6.

In order to take into account these effects, we parametrize anharmonicities with a function $f(\theta_i)$ that behaves like $f(\theta_i) \to 1$ for $\theta_i\to 0$ and diverges when $\theta_i \to \pi$. This function has been considered by various authors. Turner \cite{Turner:1985si} integrated Eq.~(\ref{axion_eq_motion}) numerically and described how the anharmonicity factor can be computed, but did not give an explicit formula. Lyth \cite{Lyth:1991ub} followed the idea by Turner and performed an explicit calculation, obtaining the behavior
\begin{equation} \label{lyth_anharmonicities}
f(\theta_i) \approx \left[\ln\left(1-\frac{\theta_i}{\pi}\right)\right]^{1.175}\quad\hbox{for $\theta_i >0.9 \pi$}.
\end{equation}
Lyth also commented that his result differs from Turner's by a factor of two. The exponent $1.175$ crucially depends on the dependence of the mass on temperature, see Eq.~(\ref{axion_mass_temperature}): in fact, Lyth assumed that $m_a(T) \propto T^{-3.7}$.

Strobl and Weiler \cite{Strobl:1994wk} and Bae {\it et al.} \cite{Bae:2008ue} performed a more precise numerical analysis following the methods in Ref.~\cite{Turner:1985si}, and confirmed the result in Ref.~\cite{Lyth:1991ub} for the behavior of $f(\theta_i)$ around $\theta_i = \pi$.

Here, we look for an analytic function $f(\theta_i)$ that extends LythÕs formula in Eq.~(\ref{lyth_anharmonicities}) to values of $\theta_i < 0.9\pi$ and symmetrically to negative values of $\theta_i$. Such function must show the limiting behaviors
\begin{equation}
\lim_{\theta_i\to 0}\,f(\theta_i) = 1, \quad\hbox{and}\quad \lim_{\theta_i\to \pi}\,f(\theta_i) = +\infty.
\end{equation}
Moreover, the integral of the function must give
\begin{equation} \label{average_anharmonic}
\langle\theta_i^2\rangle = 1.4\,\frac{\pi^2}{3},
\end{equation}
where the extra factor 1.4 accounts for Turner's result in Ref.~\cite{Turner:1985si}. We parametrize the anharmonicity function as
\begin{equation} \label{anharmonicities}
f(\theta_i) = \left[\ln\frac{e}{\left[1-(\theta_i/\pi)^2\right]^\omega}\right]^\gamma,
\end{equation}
where the factor $\gamma$ is fixed by the exponent $c$ appearing in the temperature dependence of the mass $m_a(T) \propto T^{-c}$ via the formula
\begin{equation}
\gamma = \frac{c+3}{c+2} = \frac{39+N_f}{33+N_f},
\end{equation}
and $\omega$ is fixed numerically by imposing the condition in Eq.~(\ref{average_anharmonic}). When $N_f = 3$ and thus $c = 4$, we obtain
\begin{equation} \label{anharmonicity_function}
f(\theta_i) = \left[\ln\frac{e}{\left[1-(\theta_i/\pi)^2\right]^{0.25}}\right]^{7/6}.
\end{equation}
This expression for $f(\theta_i)$ is analytic on $[-\pi,\pi]$ and has the correct asymptotic behavior in Eq.~(\ref{lyth_anharmonicities}).
\newpage

\subsection{Axions from string decays} \label{Axions from string decays}

\subsubsection{The domain wall problem}\label{The domain wall problem}

Since $\theta$ has period $2\pi\,N$, the potential in Eq.~(\ref{axion_potential}) has $N$ distinct minima: when the axion field oscillates around one of these minima, the expectation value of the misalignment angle tends to zero and the CP violation in the QCD sector lies to experimentally acceptable values. However, as first noted by Sikivie \cite{Sikivie:1982qv}, when $N > 1$ there is a domain wall problem, that is, when the PQ symmetry breaks in the early universe the axion field takes different minima in causally disconnected regions, in contrast with standard cosmology \cite{Zeldovich:1974uw}. In order to avoid this problem, various models have been implemented in which $N = 1$ \cite{Lazarides:1982tw, Georgi:1982ph, Dimopoulos:1982my}. If $N > 1$, we can avoid the problem by assuming that the PQ symmetry breaks before or during inflation: a single domain with the same value of the PQ vacuum would be inflated so that the region enclosing our visible universe would lie within such a bubble. An alternative solution to the domain wall problem in the nontrivial case $N > 1$ consists on introducing a small breaking of the PQ symmetry \cite{Sikivie:1982qv}, so that the different vacua are slightly tilted and tunneling of the axion field singles out the energetically favored lowest true vacuum. If this is the case, the axion walls predominantly decay into gravitational radiation \cite{Chang:1998tb}.

\subsubsection{Parameter space of axionic strings} \label{Parameter space of axionic strings}

The spontaneous breaking of the Peccei-Quinn symmetry leads to the formation of topological defects such as axionic strings \cite{Vilenkin:1982ks}. In fact, if the PQ symmetry breaks after inflation, cold axions produced by the decay of topological strings in the early universe contribute a large fraction of the axion energy density. Instead, if the PQ symmetry breaks before or during inflation, topological defects are washed out by inflation so axions from axionic strings are not present.

The present axion energy density produced by axionic strings $\rho_a^{\rm str}(T_0)$ is proportional to the present axion energy density produced by the misalignment mechanism $\rho_a^{\rm mis}(T_0)$ \cite{Chang:1998tb, Hagmann:1990mj, Hagmann:2000ja}. The ratio $Q$ between $\rho_a^{\rm str}(T_0)$ and $\rho_a^{\rm mis}(T_0)$ can be put in the form
\begin{equation} \label{proportionality}
Q \equiv \frac{\rho_a^{\rm str}(T_0)}{\rho_a^{\rm mis}(T_0)} = \frac{\xi \,\bar{r}\,N_d^2}{\zeta}.
\end{equation}
Here, following the notation in Ref.~\cite{Hagmann:2000ja} (however, we use $\zeta$ for their parameter $\chi$ to avoid confusion with our $\chi$ in Eq.~(\ref{number_density})): $N_d$ is the number of degenerate QCD vacua, $\bar{r}$ is the factor by which the axion comoving number density increases due to string decays, averaged over all possible processes that convert strings to axions, $\xi$ is a constant factor depending on the string network model, and $\zeta$ accounts for the uncertainties in the low-energy cutoff of the radiated axion field. In the following, we add a suffix ``std'' when these parameters refer to the value in the standard cosmology. In the following, we discuss each parameter in Eq.~(\ref{proportionality}) separately in the standard cosmology, leaving the extension of previous theoretical results in generic nonstandard cosmologies to Sec.~\ref{Axions from string decays in nonstandard cosmologies}.

The values of the parameters $\bar{r}^{\rm std}$, $\zeta^{\rm std}$, $N_d$, and $\xi^{\rm std}$ have been discussed extensively both theoretically and via numerical simulations of string networks \cite{Davis:1985pt, Davis:1986xc, Harari:1987ht, Davis:1989nj, Dabholkar:1989ju, Chang:1998tb, Hagmann:1990mj, Hagmann:2000ja, Sikivie:1982qv, Battye:1993jv, Battye:1994au, Shellard:1998mi, Yamaguchi:1998gx, Bennett:1989yp, Allen:1990tv}. The number of QCD vacua is usually assumed $N_d = 1$, because a higher value $N_d > 1$ might lead to a domain wall problem \cite{Sikivie:1982qv}; here, we assume $N_d = 1$ as well. In the standard cosmology, $\zeta \sim 1$, although the theoretical uncertainty on $\zeta$ is around 50$\%$ \cite{Hagmann:2000ja}; here, we take $\zeta  =1$ without including uncertainties.

In the literature, there is still disagreement about the numerical values of $\bar{r}^{\rm std}$ and $\xi^{\rm std}$. Here, we discuss results from different authors separately.
The value of $\bar{r}$ depends on the details of the axionic string relaxation toward lower energy configurations and on the energy spectrum of the radiated axions \cite{Shellard:1998mi}. In the standard cosmology it is (see Ref.~\cite{Hagmann:2000ja})
\begin{equation} \label{bar_r}
\bar{r}^{\rm std} = \begin{cases}
\ln(t_1/\delta) \approx 70, & \hbox{for a slow-oscillating string},\\
0.8, & \hbox{for a fast-oscillating string}.
\end{cases}
\end{equation}
Here, $t_1$ is the time at which the axion field starts to oscillate and $\delta$ is the string core size \cite{Chang:1998tb}. In Eq.~(\ref{bar_r}), the first line corresponds to the string emission model in Refs.~\cite{Davis:1985pt, Davis:1986xc, Davis:1989nj, Dabholkar:1989ju, Battye:1993jv, Battye:1994au, Shellard:1998mi, Yamaguchi:1998gx}, while the second line corresponds to the model in Refs.~\cite{Chang:1998tb, Harari:1987ht, Hagmann:1990mj, Hagmann:2000ja}. In the first line in Eq.~(\ref{bar_r}), we have obtained the value $\bar{r}^{\rm std} \approx 70$ by considering the illustrative case $\delta = (10^{12} {\rm ~GeV})^{-1}$ and $T_1^{\rm std} = 1{\rm ~GeV}$: this is approximately the same value found in Refs.~\cite{Davis:1985pt, Davis:1986xc, Davis:1989nj, Dabholkar:1989ju, Battye:1993jv, Battye:1994au, Shellard:1998mi, Yamaguchi:1998gx}. We remind here that in the standard cosmology, the time $t$ relates to the temperature $T$ of the universe through the Friedmann equation $1/2t = H(T)$.
On the value of $\xi^{\rm std}$, numerical simulations for an evolving string network in Ref.~\cite{Bennett:1989yp, Allen:1990tv} obtain $\xi^{\rm std} \sim 13$, while simulations in Refs.~\cite{Hagmann:1990mj, Hagmann:2000ja, Yamaguchi:1998gx} yield $\xi^{\rm std} \sim 1$.

In this chapter, when showing numerical results, we will use the numerical values $\bar{r}^{\rm std}  = 0.8$ and $\xi^{\rm std} = 1$, corresponding to the string model studied by Harari, Hagmann, Chang, and Sikivie.
\newpage

\section{Axion isocurvature fluctuations} \label{Axion isocurvature fluctuations}

If the PQ symmetry breaks before inflation ends and if the maximum temperature after inflation is less than $f_a$, the axion field is present during inflation and is subject to quantum mechanical fluctuations. Since at such high temperatures the axion is massless, the variance of quantum fluctuations $\delta a(x)$ in the axion scalar field $a(x)$ is given by Eq.~(\ref{quantum_fluct}), which we write in the form
\begin{equation} \label{fluctuations}
\langle|\delta a(x)|^2\rangle = \bigg(\frac{H_I}{2\pi}\bigg)^2.
\end{equation}
It follows that fluctuations in the misalignment angle field $\theta(x) = a(x)/f_a$ have variance
\begin{equation} \label{standard_deviation}
\sigma^2_\theta = \bigg(\frac{H_I}{2\pi f_a}\bigg)^2.
\end{equation}
These fluctuations are over the initial misalignment angle $\theta_i$, which has a single value within one Hubble volume since it was causally connected at the onset of production.

Under the condition in Eq.~(\ref{condition_scenarioII}), axion isocurvature perturbations are present during inflation and are constrained by WMAP-7. As originally shown in Ref.~\cite{Lyth:1992tx}, this type of measurement can also set an upper limit on the inflationary scale $H_I$. Here, we revise this computation using the latest measurements in Ref.~\cite{Komatsu:2010fb}. Defining the power spectrum of axion perturbations $\Delta_a^2(k) = \langle|\delta \rho_a/\rho_a|^2\rangle$, one finds
\begin{equation} \label{axion_perturbations}
\Delta^2_a(k) = \left(\frac{H_I}{\pi\,\theta_i\,f_a}\right)^2.
\end{equation}
The axion entropy-to-curvature perturbation ratio is then
\begin{equation}
\frac{\Delta^2_a(k_0)}{\Delta^2_{\mathcal{R}}(k_0)} = \frac{H_I^2}{\pi^2 \Delta^2_{\mathcal{R}}(k_0) \theta_i^2 f^2_a} ,
\end{equation}
or, introducing the axion adiabaticity $\alpha_0(k_0)$ like in Refs.~\cite{Komatsu:2008hk, Komatsu:2010fb},
\begin{equation}
\frac{\Delta^2_a(k_0)}{\Delta^2_{\mathcal{R}}(k_0)} = \frac{\alpha_0(k_0)}{1-\alpha_0(k_0)}.
\end{equation}
The adiabaticity $\alpha_0$ is constrained by the WMAP-7 + BAO + SN data to be \cite{Komatsu:2010fb}
\begin{equation} \label{alpha}
\alpha_0 < 0.064\quad\quad \hbox{ at 95\% CL}.
\end{equation}
Using the central value for $\Delta^2_{\mathcal{R}}(k_0)$ in Eq.~(\ref{constraint_power_spectrum}), this bound is rephrased as
\begin{equation} \label{adiabaticity}
H_I < 4.05 \times 10^{-5}\,\theta_i \, f_a.
\end{equation}

\section{Parameter space of the cosmological axion} \label{Parameter space of the cosmological axion}

Depending wether the PQ symmetry breaks during inflation or after it and wether such symmetry is even restored after inflation ends, the cosmological history of the axion field greatly differs. In fact, if the axion energy scale is lower than the Gibbons-Hawking temperature $T_{\rm GH} = H_I /2\pi$ \cite{Gibbons:1977mu},
\begin{equation} \label{scenarioI}
f_a < \frac{H_I}{2\pi},
\end{equation}
the PQ scale breaks after inflation and the axion field does not participate in inflation. Furthermore, if the axion energy scale lies below the maximum temperature that the universe reaches after inflation ends due to reheating,
\begin{equation} \label{scenarioI_I}
f_a < T_{\rm MAX},
\end{equation}
the PQ symmetry is restored after inflation, to be broken again when the universe subsequently cools down in a postinflationary scenario. We remind that $T_{\rm MAX}$ is related to the reheating temperature $T_{\rm RH}$ by \cite{Chung:1998rq}
\begin{equation}
T_{\rm MAX} \sim \left(T_{\rm RH}^2\,H_I\,M_{\rm Pl}\right)^{1/4}.
\end{equation}
When either Eq.~(\ref{scenarioI}) or Eq.~(\ref{scenarioI_I}) is satisfied, the axion field exhibits specific features: different values of the misalignment angle $\theta_i(x)$ are present within one Hubble volume, giving rise to fluctuations that are adiabatic as observed in the CMB spectrum, and there are no nonadiabatic fluctuations because Eq.~(\ref{quantum_fluct}) does not apply. Moreover, topological defects like axionic strings are present because inflation could not wash them out or because the defects generated after the PQ symmetry were restored. We refer to this portion of the parameter space as Scenario I. 

On the contrary, two necessary conditions for the axion field to show nonadiabaticities have to be met: (a) the PQ symmetry breaks before inflation ends and (b) it is not restored afterwards. Condition (a) requires
\begin{equation} \label{scenario_II}
f_a > \frac{H_I}{2\pi},
\end{equation}
while condition (b) requires
\begin{equation} \label{T_max}
f_a > T_{\rm MAX},
\end{equation}
When both these conditions are met, the PQ symmetry breaks before inflation ends and it is never restored afterwards: we call this portion of the parameter space the Scenario II. Under these conditions, the salient features of the axion field consist in that nonadiabatic fluctuations are developed because the axion participates in inflation, the value of the initial misalignment field has a single value $\theta_i$ over one Hubble volume, corresponding to the inflated domain we live in, and no defects are present. Putting together the two requirements for Scenario II, we obtain that nonadiabatic axion fluctuations arise when
\begin{equation} \label{condition_scenarioII}
f_a > \max\{T_{GH},T_{\rm MAX}\}.
\end{equation}
Here, we assume $T_{\rm MAX} < H_I/2\pi$, which leads to $T_{\rm RH} \lesssim 10^{11}$ GeV. This requirement is consistent with that obtained from theories in which the inflaton decays into fermions, $T_{\rm RH} \lesssim 10^{12}\,$GeV \cite{linde1990particle}. Because of our choice, Eq.~(\ref{scenarioI}) suffices to identify Scenario I, while Eq.~(\ref{scenario_II}) describes the parameter space of Scenario II. For the reasons explained above, axionic strings in Sec.~\ref{Axions from string decays} contribute to the axion CDM energy density only in Scenario I, and the bound computed in Sec.~\ref{Axion isocurvature fluctuations} applies only in Scenario II.

\section{The axion as the CDM particle} \label{The axion as the Cold Dark Matter}

The major interest for axions in astrophysics consists in the fact that these particles may account alone for all of the observed CDM. If this is true, axions must be in highly nonthermal equilibrium and form a Bose-Einstein condensate \cite{Sikivie:2006ni, Sikivie:2009qn}. We explored the two leading mechanism that produce an axion population with zero momentum, namely the misalignment production in Sec.~\ref{Axion energy density from vacuum realignment} (see also Refs.~\cite{Preskill:1982cy, Abbott:1982af, Dine:1982ah, Stecker:1982ws, Turner:1985si}) and the axionic string decay production in Sec.~\ref{Axions from string decays} (see Refs.~\cite{Davis:1985pt, Davis:1986xc, Chang:1998tb, Harari:1987ht, Hagmann:1990mj}). The population of thermal axions discussed in Sec.~\ref{Thermal production of axions} does not constitute a fraction of the CDM, because thermal axions are still relativistic today, contributing to the HDM population.

\subsection{Axion CDM in Scenario I} \label{Axion CDM in Scenario I}

We first explore the expression for the energy density of CDM axions in Scenario I, $f_a < H_I/2\pi$. As discussed in Sec.~\ref{Parameter space of the cosmological axion}, in this region of the parameter space, there are no axion isocurvature fluctuations, but the energy density gets contributions from both the misalignment and string decay mechanisms. We have evaluated the energy density from misalignment mechanism $\rho_a^{\rm mis}$ in units of the critical density in Eq.~(\ref{standarddensity}) which, when anharmonicities are included with the function $f(\theta_i)$ in the expectation value of $\theta_i^2$, reads
\begin{equation} \label{standarddensity1}
\Omega^{\rm mis}_a h^2 =
\begin{cases}
(0.23\pm0.03)\,\langle\theta_i^2\,f(\theta_i)\rangle\,f_{a,12}^{7/6}, & f_a \lesssim \hat{f}_a,\\
(6.1\pm0.8)\times 10^{-3}\,\langle\theta_i^2\,f(\theta_i)\rangle\,f_{a,12}^{3/2}, & f_a \gtrsim \hat{f}_a,\\
\end{cases}
\end{equation}
with $\hat{f}_a =  5.4\times 10^{16}{\rm ~GeV}$. We have seen in Sec.~\ref{The role of anharmonicities} that in Scenario I, the average of $\theta_i^2$ over its possible values on a Hubble horizon gives
\begin{equation}
\langle\theta_i^2\,f(\theta_i)\rangle = 1.4\,\frac{\pi^2}{3} = 4.61.
\end{equation}
We thus use this value in Eq.~(\ref{standarddensity1}) to obtain
\begin{equation} \label{standarddensity2}
\Omega^{\rm mis}_a h^2 =
\begin{cases}
(1.06\pm 0.14)\,f_{a,12}^{7/6}, & f_a \lesssim \hat{f}_a,\\
(0.028\pm0.004)\,f_{a,12}^{3/2}, & f_a \gtrsim \hat{f}_a.\\
\end{cases}
\end{equation}
The relative contribution from axionic strings to the axion energy density has been considered in Sec.~\ref{Axions from string decays}, and it is summarized in Eq.~(\ref{proportionality}) written here as
\begin{equation}
\Omega^{\rm str}_a h^2 = \Omega^{\rm mis}_a h^2\, \frac{\xi \,\bar{r}\,N_d^2}{\zeta}.
\end{equation}
These parameters depend on the model used to describe the axionic string oscillation and the radiation spectrum of the radiated axions. Using the parameters found for the scenario of a slow-oscillating axionic string first described by Davis, Shellard, and Battye \cite{Davis:1989nj, Davis:1985pt, Davis:1986xc, Battye:1993jv, Battye:1994au} (from here on DSB Scenario) or, alternatively, the parameters computed in a fast-oscillating string scenario by Harari, Hagmann, Chang, and Sikivie \cite{Harari:1987ht, Hagmann:1990mj, Chang:1998tb, Hagmann:2000ja} (HHCS Scenario) we have
\begin{equation}
\Omega^{\rm str}_a h^2 = \begin{cases}
910\, \Omega^{\rm mis}_a h^2 &\quad  \hbox{DSB Scenario};\\
0.8\, \Omega^{\rm mis}_a h^2 &\quad  \hbox{HHCS Scenario}.\\
\end{cases}
\end{equation}
The total energy density $\Omega^{\rm tot}_a h^2 \equiv  \Omega^{\rm mis}_a h^2$ is thus
\begin{equation}\label{total_energy_density}
\Omega^{\rm tot}_a h^2 = \begin{cases}
911\, \Omega^{\rm mis}_a h^2 &\quad  \hbox{DSB Scenario};\\
1.8\, \Omega^{\rm mis}_a h^2 &\quad  \hbox{HHCS Scenario}.\\
\end{cases}
\end{equation}
Setting the total axion energy density in Eq.~(\ref{total_energy_density}) equal to the measured CDM energy density,
\begin{equation} \label{measure_CDM}
\Omega_{CDM} h^2 = 0.1126 \pm 0.0036 \quad {\rm at}\, 68\% \,{\rm CL},
\end{equation}
we compute the value of the PQ scale $f_a^{\rm std}$ for which the axion condensate accounts for the observed CDM in the universe in the standard radiation-dominated cosmology,
\begin{equation} \label{PQscale_axionCDM}
f_a^{\rm std} = \begin{cases}
(4.3\pm 0.7)\times 10^8{\rm ~GeV}, & \quad\hbox{for the DSB Scenario};\\
(8.8\pm1.5)\times 10^{10}{\rm ~GeV}, & \quad\hbox{for the HHCS Scenario}.
\end{cases}
\end{equation}
In computing $f_a^{\rm std}$, we have only considered the case $f_a^{\rm std} < \hat{f}_a$, because the PQ scale $f_a^{\rm std}$ is generically bound on Scenario I by $f_a^{\rm std} \lesssim $ and the breaking scale $\hat{f}$ lies above such bound. Thus, we do not need to consider the possibility that $f_a^{\rm std} > \hat{f}_a$ here. Notice that the two energy scales in Eq.~(\ref{PQscale_axionCDM}) differ by two orders of magnitude, because of the different contribution from axionic strings in the two scenarios. In fact, in order for the total axion energy density to equate the CDM energy density, the energy scale $f_a^{\rm std}$ and the prefactor from axionic strings are inversely proportional; since cold axions are copiously produced in the DSB Scenario, the energy scale $f_a^{\rm std}$ is much lower in the DSB Scenario than in the HHCS Scenario.
{}We also remark that the result
\begin{equation}\label{PQscale_axionCDM_HHCS}
f_a^{\rm HHCS} = (8.8\pm1.5) \times 10^{10}{\rm ~GeV}
\end{equation}
is one order of magnitude lower than the usually quoted value for the PQ energy scale, $f_a^{\rm std} \approx 10^{12} {\rm ~GeV}$. This difference by one order of magnitude comes from equating $\Omega_a^{\rm tot}$ with the precise measurement in Eq.~(\ref{measure_CDM}) for $\Omega_{\rm CDM}$, that is $\Omega_a \sim 0.1$, whereas seminal works in axion cosmology computed $f_a^{\rm std}$ from the equation $\Omega_a = 1$ that assures that axions do not overclose the universe. Also, in the computation of $\Omega_a$ we included the contribution from axionic strings decay and the careful derivation of the anharmonicities effects.

\subsection{Axion CDM in Scenario II} \label{Axion CDM in Scenario I|}

We now discuss the parameter space of CDM axions when $f_a > H_I/2\pi$, namely Scenario II. We have seen in Sec.~\ref{Parameter space of the cosmological axion} that fluctuations in the axion field arise with variance given in Eq.~(\ref{standard_deviation}); using the definition of variance, we thus write
\begin{equation}
\langle\theta_i^2\rangle = \langle\theta_i\rangle^2 + \sigma_\theta^2 = \theta_i^2 + \sigma_\theta^2.
\end{equation}
When the anharmonicity function $f(\theta_i)$ is included, we approximate $\langle\theta_i^2\,f(\theta_i)\rangle \approx \langle\theta_i^2\rangle\,f(\theta_i)$: this is a good approximation in the region of the parameter space that concern Scenario II, because in this region the term that contains an average over a Hubble volume is much smaller than the isocurvature fluctuations term, $\theta_i^2 \ll \sigma_\theta^2$, and the term $\sigma_\theta^2$ is multiplied by the anharmonicity function with no further averaging involved. In this approximation, it is then
\begin{equation}
\langle\theta_i^2\,f(\theta_i)\rangle = \left(\theta_i^2 + \sigma_\theta^2\right)\,f(\theta_i),
\end{equation}
and the total axion energy density is
\begin{equation} \label{standarddensity_ScenarioII}
\Omega^{\rm tot}_a h^2 =
\begin{cases}
(0.23\pm 0.03)\,\left(\theta_i^2 + \sigma_\theta^2\right)\,f(\theta_i)\,f_{a,12}^{7/6}, & f_a \lesssim \hat{f}_a,\\
(6.1\pm0.8)\times 10^{-3}\,\left(\theta_i^2 + \sigma_\theta^2\right)\,f(\theta_i)\,f_{a,12}^{3/2}, & f_a \gtrsim \hat{f}_a.\\
\end{cases}
\end{equation}
In Eq.~(\ref{standarddensity_ScenarioII}), the total axion energy density equates the density from the misalignment mechanism because no contributions from axionic strings are present in Scenario II. A peculiar fact about Scenario II is that, in this region of the parameter space, the PQ scale can evade the bound $f_a \lesssim 10^{12}$ GeV when the initial value of the misalignment field is small, $\theta_i \ll 1$ \cite{Linde:1987bx}.

\section{Constraining the parameter space of the axion}\label{results}

\begin{figure}[t]
\begin{center}
\includegraphics[width=13cm]{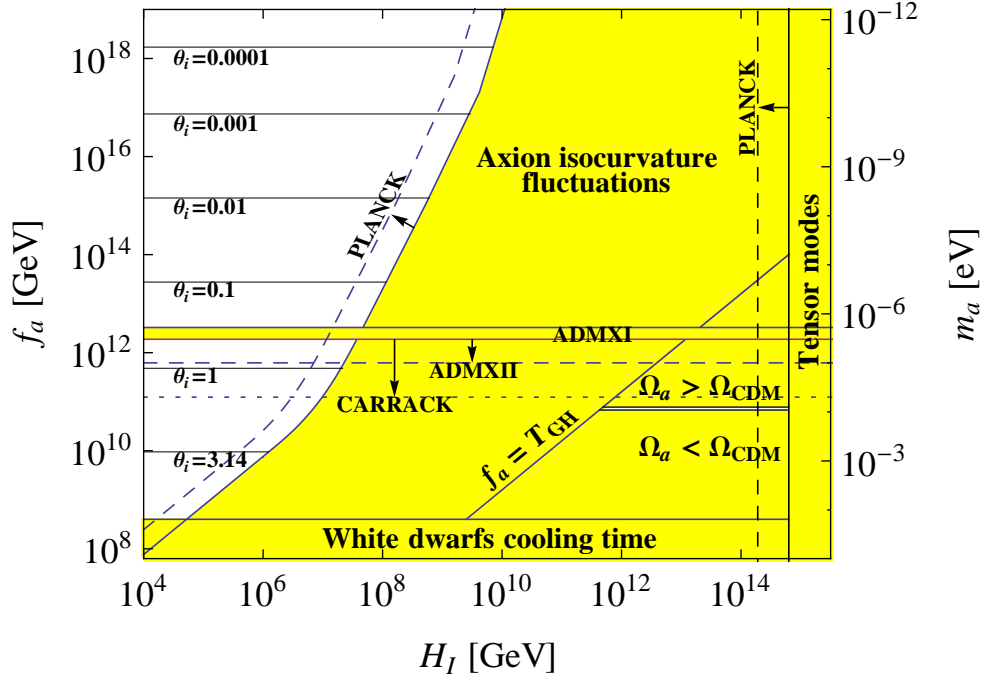}
\caption[The parameter space where the axion is 100\% CDM in the standard cosmology.]{Region of axion parameter space where the axion is 100\% of the CDM. The axion mass scale on the right corresponds to Eq.~(\ref{eq:axionmass}) with $U(1)_{PQ}$ color anomaly $N=1$. When the PQ symmetry breaks after inflation ($f_a < H_I/2\pi$), the axion is the CDM particle if $f_a =  (8.8 \pm 1.5)\times 10^{10}{\rm ~GeV}$, or $m_a=(67\pm17){\rm ~\mu eV}$, which is the narrow horizontal window shown on the right. If the axion is present during inflation ($f_a > H_I/2\pi$), axion isocurvature perturbations constrain the parameter space to the region on the top left, which is marked by the values of $\theta_i$ necessary to obtain 100\% of the CDM density. Other bounds indicated in the figure come from astrophysical observations of white dwarfs cooling times and the nonobservation of tensor modes in the Cosmic Microwave Background fluctuations. Dashed lines and arrows indicate the future reach of the PLANCK satellite and the ADMX and CARRACK microwave cavity searches.}
\label{classical}
\end{center}
\end{figure}

The axion parameter space is labeled by the axion energy scale $f_a$, the initial misalignment angle $\theta_i$, and the Hubble parameter during inflation $H_I$. Results valid within the HHCS axionic string model are shown in Fig.~\ref{classical}.

The region labeled ``Tensor modes'' shows the constraint on $H_I$ in Eq.~(\ref{HI_bound}) coming from the WMAP7 plus BAO and SN observations \cite{Komatsu:2010fb}. The newly launched PLANCK satellite will improve the actual measurement on the tensor-to-scalar ratio $r$ by at least one order of magnitude, see Ref.~\cite{Visinelli:2009zm}. If PLANCK will not detect gravitational waves and sets a new limit $r < 0.02$, $H_I$ will be bound by $H_I \lesssim 2\times 10^{14}\,$GeV. This forecast measurement is shown by a vertical dashed line labeled ``PLANCK''.

The region labeled ``White dwarfs cooling time'' is excluded because there one would have an excessively small cooling time in white dwarfs. We have used the bound in Eq.~(\ref{astrophysical_bound}) with $\cos^2\beta_H = 1$, giving the bound
\begin{equation} \label{astrophysical_bound1}
f_a > 1.3\times 10^{8} {\rm ~GeV},
\end{equation}
corresponding to an upper bound for the axion mass $m_a < 15$ meV. We remark here that this bound on the PQ scale is slightly below the constraint often used $f_a > 4 \times 10^{8} {~\rm GeV}$.

The line
\begin{equation}
f_a = H_I/2\pi
\end{equation}
divides the region in which the PQ symmetry breaks after the inflationary epoch has ended, or $f_a < H_I/2\pi$ (Scenario I), from the region where the axion field is present during inflation, $f_a > H_I/2\pi$ (Scenario II).

The region marked as ADMXI has been excluded by a direct search of axions CDM in a Sikivie-type microwave cavity detector \cite{Sikivie:1983ip} by the ADMX Phase I experiment \cite{Asztalos:2003px, Duffy:2006aa}. The window shown corresponds to $1.9\,{\rm \mu eV} < m_a < 3.3\,{\rm \mu eV}$, see Eq.~(\ref{axion_mass_bound_ADMX}), valid for the KSVZ axion model. A narrower DFSV axionic window, not shown in Fig.~\ref{classical}, has also been ruled out. The dashed line labeled ``ADMXII'' shows the forecast axionic region to be probed in the ADMX Phase II, which would search for axions with mass up to $10{\rm \mu eV}$, or $f_a = 6.2\times 10^{11}\,$GeV. The proposed CARRACK II experiment is a cavity search that will look for axions with mass in the range $(2 \div 50){\rm \mu eV}$ \cite{Yamamoto:2000si}.

For Scenario I, since $\langle\theta_i^2 f(\theta_i)\rangle = 1.4\,\pi^2/3$ and does not depend on $H_I$ or $\theta_i$, the value of $\Omega_a^{\rm mis}$ depends on $f_a$ only and it is given by Eq.~(\ref{standarddensity2}). In the HHCS model and for $f_a < \hat{f}_a$, we have obtained
\begin{equation}
\Omega_a^{\rm tot}\,h^2 = (1.90\pm 0.25)\,f_{a,12}^{7/6},\quad\hbox{for the HHCS string model}.
\end{equation}
Demanding that $\Omega_a^{\rm tot}$ accounts for the totality of the CDM, and indicating with $f_a^{\rm HHCS}$ the value of the PQ scale for which the axion is 100\% CDM in the HHCS string model, we have obtained
\begin{equation}  \label{PQ_strength}
f_a^{\rm HHCS} = (8.8\pm1.5)\times10^{10}{\rm~GeV}.
\end{equation}
The band corresponding to $f_a^{\rm HHCS}$ is drawn in Fig.~\ref{classical} as the horizontal window on the lower right. Assuming a $U(1)_{PQ}$ color anomaly $N=1$, the $\Omega_a^{\rm tot}=\Omega_{\rm CDM}$ band corresponds to an axion mass
\begin{equation} \label{axion_mass_HHCS}
m_a = (67 \pm 17) {\rm ~\mu eV}.
\end{equation}

In Scenario II, axion isocurvature fluctuations are present and lead to the bound in Eq.~(\ref{adiabaticity}). Together with the expression for $\Omega_a^{\rm mis}$ in Eq.~(\ref{standarddensity_ScenarioII}) and the condition that axions constitute 100\% of the observed CDM, the adiabaticity bound excludes the shaded region in the center of Fig~\ref{classical}. The PLANCK satellite is expected to improve the current bounds on the axion isocurvature fluctuations by at least one order of magnitude. The dashed line on the left of Fig.~\ref{classical} shows the new bound on the allowed region when Eq.~(\ref{alpha}) is replaced by $\alpha_0 < 7\times 10^{-3}$.

The leftmost boundary of this region contains two kinks and can be approximated with
\begin{equation} \label{leftmostboundary}
f_a = \begin{cases}
7.9\times 10^{9}{\rm ~GeV} \left(\frac{H_I}{10^6{\rm ~GeV}}\right) , & H_I < 9.6 \times 10^6 {\rm ~GeV} , \\
2.1\times 10^{13}{\rm ~GeV} \left(\frac{H_I}{10^8{\rm ~GeV}} \right)^{12/5} , & 9.6 \times 10^6 {\rm ~GeV} < H_I < 2.7 \times 10^9 {\rm ~GeV}, \\
1.1 \times 10^{19}{\rm ~GeV} \left( \frac{H_I}{10^{10}{\rm ~GeV}} \right)^4 , & H_I > 2.7 \times 10^9 {\rm ~GeV}.
\end{cases}
\end{equation} The upper kink occurs at $f_a = \hat{f}_a$ and is due to the change in the dependence of the axion mass on the temperature,  Eq.~(\ref{axion_mass_temperature}). The lower and smoother kink around $f_a \sim 10^{11}$GeV arises from the fact that the anharmonicity function $f(\theta_i)$ differs from one at values of $f_a$ smaller than $10^{11}$ GeV, see Eq.~(\ref{anharmonicity_function}). Notice that the simple proportionality $f_a \propto H_I$ at small $H_I \lesssim 10^7$ GeV is independent of the detailed form assumed for the function $f(\theta_i)$ near $\theta_i=\pi$, and derives in a straightforward way from Eq.~(\ref{adiabaticity}).

\begin{figure}[t]
\begin{center}
\includegraphics{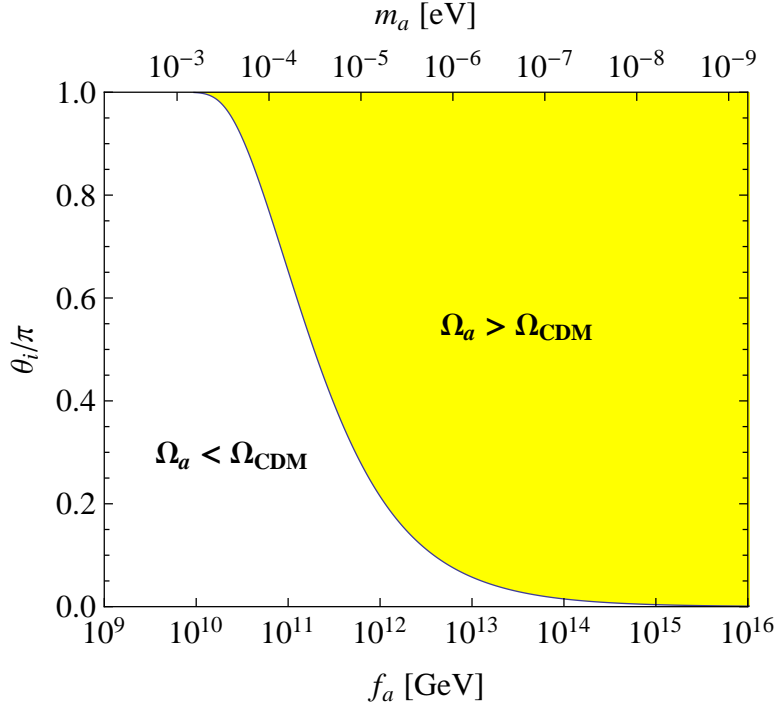}
\caption[$\theta_i$ as a function of $f_a$ in the standard cosmology.]{The misalignment angle $\theta_i$ necessary for the axion to be 100\% of the CDM in Scenario II ($f_a>H_I/2\pi$), as a function of $f_a$. Above the curve, $\Omega_a>\Omega_{CDM}$. For $f_a \gtrsim 10^{17}$ GeV, one has $\theta_i \simeq 10^{-3} (f_a/10^{17}\,{\rm GeV})^{-3/4}$; in particular, for $f_a \gtrsim 10^{19}$ GeV, the initial misalignment angle $\theta_i$ has to assume values $\theta_i \lesssim 10^{-5}$.}
\label{theta_plot}
\end{center}
\end{figure}

In the remaining region on the left of Fig.~\ref{classical}, the axion can be 100$\%$ of the CDM, provided the value of the initial misalignment angle $\theta_i$ is chosen appropriately. The $\theta_i$ contours in Fig.~\ref{classical} indicate the appropriate values of $\theta_i$ for a given $H_I$ and $f_a$. Notice that the $\theta_i$ contours are horizontal, i.e.\ independent of $H_I$, since in that region the contribution from $\sigma_{\theta}^2$ in Eq.~(\ref{standard_deviation}) is negligible compared to $\theta_i^2$. This allows us to show the full relation between $f_a$ and $\theta_i$ imposed by the constraint $\Omega_a=\Omega_{CDM}$. We find
\begin{equation} \label{const_theta}
f_{a,12} = \begin{cases}
\big(\frac{\Omega_{CDM}\,h^2}{(0.23\pm0.03)\,\theta_i^2\, f(\theta_i)}\big)^{6/7} = (0.54\pm0.07)\,\left(\theta_i^2\,f(\theta_i)\right)^{-6/7}, & f_a \lesssim \hat{f}_a {\rm ~or~} \theta_i \gtrsim 10^{-3},\\
\big(\frac{\Omega_{CDM}\,h^2}{(6.1\pm 0.8)\times 10^{-3}\,\theta_i^2\, f(\theta_i)}\big)^{2/3} = (7.0\pm0.8)\,\left(\theta_i^2\,f(\theta_i)\right)^{-2/3}, & f_a \gtrsim \hat{f}_a {\rm ~or~} \theta_i \lesssim 10^{-3}.
\end{cases}
\end{equation}
This function is illustrated in Fig.~\ref{theta_plot}.

For $f_a \gtrsim 10^{17}$ GeV ($m_a \lesssim 10^{-10}{\rm~eV}$ or $\theta_i \lesssim 10^{-3}$), one has $f(\theta_i)\simeq 1$ and Eq.~(\ref{const_theta}) simplifies to
\begin{equation}
\theta_i \simeq (0.77\pm 0.07)\times 10^{-3} \left( \frac{f_a}{10^{17}\,{\rm GeV}} \right)^{-3/4}  , \qquad {\rm for~} f_a\gtrsim10^{17}{\rm ~GeV},
\end{equation}
or
\begin{equation}
\theta_i \simeq (1.2\pm 0.1)\times 10^{-3} \left(\frac{m_a}{10^{-10}{\rm ~eV}}\right)^{3/4} , \qquad {\rm for~} m_a\lesssim 10^{-10}{\rm ~eV}.
\end{equation}
In particular, for $f_a \gtrsim 10^{19}$ GeV, the initial misalignment angle $\theta_i$ has to assume values $\theta_i \lesssim 10^{-5}$. This was also noted in Refs.~\cite{Fox:2004kb, Bae:2008ue}. These small values of $\theta_i$ may be uncomfortable in a cosmological scenario.

In the other limit of $\theta_i \simeq \pi$, the form of the function $f(\theta_i)$ assumed in Eq.~(\ref{anharmonicity_function}) gives
\begin{equation}
\pi-\theta_i \simeq \frac{\pi\,e^4}{2}\, {\rm Exp}\left[-\frac{(3.0\pm 0.4)\times10^{11}{\rm~GeV}}{f_a}\right], \qquad {\rm for~}f_a\lesssim2\times10^{10}{\rm ~GeV},
\end{equation}
So, as $\theta_i$ approaches $\pi$ from below, the corresponding $f_a$ approaches 0. This gives rise to the linear dependence of $f_a$ on $H_I$ in the lower left corner of Fig.~\ref{classical}.

\subsection{Comparison with previous work}

Other authors find different expressions for the axion energy density. Turner \cite{Turner:1985si} carefully analyzed the problem of the temperature-dependent axion mass, as reviewed in Sec.~\ref{Finite temperature effects}, and derived the parameters in Table~\ref{turner_table}. We differ from Ref.~\cite{Turner:1985si} because we use a different value of the parameter $a$, which we computed from fitting a specific function of the temperature as we discussed above. For $N_f = 3$, we confirmed the values of the parameters $b$, $c$, and $d$ in Table~\ref{turner_table}.

Hertzberg {\it et al.} \cite{Hertzberg:2008wr} use different values for $\chi$, which range from $\chi = 1$ (which they call a ``moderate'' value) to $\chi = 1/20$ (which they call a ``conservative'' value). However, in Ref.~\cite{Hertzberg:2008wr}, anharmonicities in the axion potential are not accounted for when $\theta_i$ is close to $\pi$. As shown in Sec.~\ref{The role of anharmonicities}, the anharmonicity function $f(\theta_i)$ is essential to obtain the correct behavior of the isocurvature fluctuation bound at relatively small values of $f_a$, because the initial misalignment angles that give the correct axion density are not very small. We choose $\chi = 1.44$, but most importantly we differ from \cite{Hertzberg:2008wr} in that we account for the behavior of the anharmonicity function $f(\theta_i)$. With this prescription, it turns out that the figures shown in \cite{Hertzberg:2008wr} are modified when the function $f(\theta_i)$ ceases to be of order one. This happens for $f_a \lesssim 10^{11}\,$GeV, as shown in Fig.~\ref{classical}.

Sikivie \cite{Sikivie:2006ni} studies the axion energy density in the case $T_1^{\rm std} > \Lambda$, and finds $\Omega_a = 0.15\,\theta_i^2\,(f_a/10^{12}{\rm GeV})^{7/6}$. His expression differs from ours in Eq.~(\ref{standarddensity}) because of the different numerical factors used in Ref.~\cite{Sikivie:2006ni}, namely $\chi = 1$, $b=5/12$, and $T_1^{\rm std} = 1{\rm GeV}(10^{12}{\rm GeV}/f_a)^{1/6}$.

\section{Discussion and conclusions}

We have seen that depending on the ratio $f_a/H_I$, two scenarios are possible for the axions to be 100\% of the CDM, which we called Scenario I ($f_a<H_I/2\pi$) and Scenario II ($f_a>H_I/2\pi$). In Scenario I, one needs a cosmology in which $H_I > 4.57\times10^{11}{\rm~GeV}$, and for the HHCS axionic string model, one finds a PQ scale $f_a^{\rm HHCS}=(8.8\pm1.5)\times10^{10}{\rm~GeV}$, corresponding to an axion mass $m_a= (67\pm17){\rm~\mu eV}$. In the DSB string model instead, one has $f_a^{\rm DSB} = (4.3\pm0.7)\times10^8{\rm~GeV}$, a value that is barely acceptable from the constraint on the PQ scale from astrophysical considerations, Eq.~(\ref{astrophysical_bound1}). We remind that it is the copious production of axions from axionic strings in the DSB model that leads to this extremely low value of the PQ scale. The HHCS and the DSB string scenarios can be considered as two extreme models, and a realistic measurement of the PQ scale might lead to an intermediate value between $f_a^{\rm HHCS}$ and $f_a^{\rm DSB}$. The measured value of the PQ scale would then yield information on the fraction of the total CDM axions coming from axionic strings.

In Scenario I, the Hubble scale $H_I$ is bounded from below to $H_I \gtrsim 10^{11}$GeV (leftmost edge of the $\Omega_a=\Omega_{CDM}$ horizontal window in Fig.~\ref{classical}). This window can disappear completely if the limit from tensor modes moves to the left beyond $H_I \gtrsim 10^{11}$GeV. If the limit on $r$ would become more stringent than $r \sim 10^{-8}$, Scenario I would have to be abandoned in favor of Scenario II.

We remark that for very large values of $f_a$ in Scenario II, the initial misalignment angle $\theta_i$ has to be chosen very small for the axion to be 100\% of the CDM, see Eq.~(\ref{const_theta}) and Fig.~\ref{theta_plot}. This may undermine the axion field as a dynamical solution to the strong CP problem, in that it would have to be fixed to a small value as an initial condition.

Grand Unification Theory (GUT) models that contain axions predict that $f_a$ should be of the order of the GUT scale, $\sim 10^{16}$GeV \cite{Svrcek:2006yi}. In a variety of string theory models, the PQ energy scale results in the range $10^{16}{~\rm GeV} < f_a < 10^{18}{~\rm GeV}$. From Fig.~\ref{classical}, we see that this range of $f_a$ values cannot be reconciled with axions as 100\% CDM in Scenario I, while they can be in Scenario II provided $H_I\lesssim 10^9{\rm~GeV}$.

%% file: chap6.tex
\chapter{axion CDM in nonstandard cosmologies} \label{Dark matter axions in nonstandard cosmologies}

Results presented in Chapter~\ref{Revising the axion as the cold dark matter} greatly change when the possibility that the thermal history of the universe might have differed from the standard picture is taken into account. The aim of the present chapter is to discuss the effect of such modifications for two particular modified cosmological scenarios.

\section{Motivations in considering modified cosmologies}

The standard cosmological model has been tested up to a temperature $T \sim 1\,{\rm MeV}$, or down to times as short as $\sim 1\,{\rm s}$, when Big Bang nucleosynthesis (BBN) occurred. The success of the BBN theory is due to its great precision in predicting the primordial abundance of light elements D, $^4$He, and $^7$Li. For the success of BBN, the universe must be radiation-dominated at temperatures $T \gtrsim 4 {\rm ~MeV}$ \cite{Kawasaki:1999na, Kawasaki:2000en, Hannestad:2004px, Ichikawa:2005vw, DeBernardis:2008zz}. However, due to lack of data prior to BBN, the history of the universe in the pre-BBN epoch $T \gtrsim 4$ MeV is only indirectly inferred.

In the standard cosmology, radiation has dominated the energy density of the universe before BBN since the very early time at which inflation ended. How radiation was produced at the end of inflation from a state of negligible temperature is still a topic of active research: models include the decay of the inflaton field \cite{Dolgov:1982th, Abbott:1982hn} and parametric resonance \cite{Dolgov:1989us, Traschen:1990sw, Kofman:1994rk}.

In alternative cosmological models, inflation may have ended at times close to BBN \cite{BasteroGil:2000jd, Kudo:2001ie}, or there might have been a period after inflation in which the dominant energy density was not in radiation but in some other exotic form, like the energy density of a scalar field \cite{Turner:1983he, Scherrer:1984fd}, or still more there could have been an injection of entropy into the radiation field \cite{Dine:1982ah, Steinhardt:1983ia}.
\newpage

\subsection{Cosmological probes}

A good probe of the history of the universe before $\sim 1 {\rm s}$ is a relic particle that has survived from that period. Indeed, in the standard cosmology, dark matter relics like Weakly Interacting Massive Particles (WIMPs) (see Refs.~\cite{Jungman:1995df, Bertone:2004pz, Bergstrom:2009ib} for reviews) or like the axion are produced when the universe was between $\sim 10 {\rm ns}$ and $\sim 10 {\rm ps}$ or $\sim 1 \mu{\rm s}$ old, corresponding, respectively, to the age of the universe at the WIMP freeze-out for WIMP masses between 10 GeV and 1 TeV or at the QCD phase transition $\Lambda \sim 200$ MeV. These dark matter relics are therefore excellent candidates to test the cosmological history of the universe at very early epochs \cite{Barrow:1982ei}.

Here we show that the CDM axions may work as useful probes of the pre-BBN epoch. We analyze two nonstandard pre-BBN cosmologies: the low-temperature reheating (LTR) cosmology \cite{Dine:1982ah, Steinhardt:1983ia, Turner:1983he, Scherrer:1984fd, Lazarides:1987zf, Yamamoto:1985mb, McDonald:1989jd, Moroi:1994rs, Kawasaki:1995cy, Kawasaki:1995vt, Chung:1998rq, Giudice:2000ex, Giudice:2000dp, Gelmini:2004ah, Gelmini:2006pw, Gelmini:2006pq, Drees:2006vh} and the kination cosmology \cite{Ford:1986sy, Kamionkowski:1990ni, Spokoiny:1993kt, Joyce:1996cp, Joyce:1997fc, Tashiro:2003qp, Salati:2002md, Profumo:2003hq, Rosati:2003yw, Pallis:2005hm, Pallis:2005bb, Chun:2007np, Chung:2007cn, Chung:2007vz, Gomez:2008js}. In the LTR cosmology, the expansion of the universe after inflation is driven by a massive scalar field $\Phi$ that decays and reheats the universe. This stage lasts down to a temperature $T_{\rm RH}$, after which standard radiation-dominated cosmology applies. In the kination cosmology, the energy content of the universe is dominated by the kinetic energy of a scalar field which evolves without entropy release down to a temperature $T_{\rm kin}$, after which the standard radiation-dominated cosmological scenario begins.

We find that with respect to the standard cosmology, if the Peccei-Quinn symmetry in the axion theory breaks during inflation, the allowed axion parameter space is enlarged in the LTR cosmology and restricted in the kination cosmology. Instead, if the Peccei-Quinn symmetry breaks after inflation, the mass of CDM axions  in the LTR cosmology is orders of magnitude smaller than its standard-cosmology value, and it is orders of magnitude greater in the kination cosmology.

Past work on axions in nonstandard cosmologies has examined the parameter space of hot dark matter axions in the LTR and kination cosmologies \cite{Grin:2007yg}, and the cosmological bound $\Omega_a h^2 < 1$ for cold (i.e.\ nonthermal) axions in the LTR cosmology assuming that the Peccei-Quinn symmetry breaks after inflation \cite{Dine:1982ah, Steinhardt:1983ia, Lazarides:1987zf, Yamamoto:1985mb, Kawasaki:1995vt, Giudice:2000ex}. 

Our work considers axions as 100$\%$ of the CDM, allows the PQ symmetry to break after or during inflation, and includes anharmonicities in the axion potential.
\newpage

\section{The misalignment mechanism in nonstandard cosmologies} \label{The misalignment mechanism in nonstandard cosmologies}

\subsection{Hubble rate in nonstandard cosmologies}

We review the production of cold axions by the vacuum realignment mechanism, following the conventions in Refs.~\cite{Visinelli:2009zm, Visinelli:2009kt}. The details of a cosmological axion model depend on the details of the cosmology before BBN only through the dependence of the Hubble expansion rate $H(T)$ on the temperature $T$. Different pre-BBN cosmologies differ in the choice of $H(T)$. This can be explained by noticing that Eq.~(\ref{axion_eq_motion}), which describes the evolution of the axion field in a FRW universe, depends on the specific cosmology through $H(t)$ only, the initial condition $\theta_i$ being unspecified and the axion mass being given by Eq.~(\ref{axion_mass_temperature}).

We assume a scenario in which the expansion of the universe right after inflation is described by a nonstandard cosmology, in which the Hubble rate is $H(T)$, until the universe transitions to the standard radiation-dominated cosmology. Such a transition happens at a characteristic critical temperature $T_c$ which is bound by the details on primordial nucleosynthesis to be $T_c \gtrsim 4$ MeV \cite{Kawasaki:1999na, Kawasaki:2000en, Hannestad:2004px, Ichikawa:2005vw, DeBernardis:2008zz}. When discussing specifically the LTR and kination models, we will refer to $T_c$ as $T_{\rm RH}$ or $T_{\rm kin}$, respectively. This said, the Hubble rate in this modified cosmology reads
\begin{equation} \label{H_nonstandard}
H(T) = H_{\rm rad}(T_c)\,
\begin{cases}
\sqrt{\frac{g_*(T)}{g_*(T_c)}}\,\left(T/T_c\right)^2, & \hbox{for $T < T_c$},\\
\frac{g_*(T)}{g_*(T_c)}\,\left(T/T_c\right)^\upsilon, & \hbox{for $T > T_c$}.
\end{cases}
\end{equation}
where $H_{\rm rad}(T_c)$ is the Hubble rate during the radiation-dominated epoch, Eq.~(\ref{hubble_radiation}), at temperature $T = T_c$, and $\upsilon$ specifies the modified pre-BBN cosmology. For the LTR cosmology, it is $\upsilon = 4$, and for kination cosmology it is $\upsilon = 3$.

We parametrize the time-dependence of the scale factor $a(t)$ with a generic relation,
\begin{equation} \label{a(t)}
a(t) \propto t^{\beta},
\end{equation}
where $\beta$ is a constant that depends on the details of the modified cosmology. This is a notation that has been already used consistently in the present thesis. For example, $\beta = 2/3$ for the matter-dominated and the LTR cosmologies, $\beta = 1/2$ for the radiation-dominated universe, and $\beta = 1/3$ for the kination cosmology. The Hubble rate at a given time $t$ is then
\begin{equation} \label{time_temperature_nonstd}
H(t) \equiv \frac{\dot{a}(t)}{a(t)} = \beta/t.
\end{equation}
We now see how this modification affects the value of the temperature $T_1$ at which the axion field starts to oscillate.

\subsection{Oscillation temperature in nonstandard cosmologies}

As discussed in Sec.~\ref{Computing the oscillation temperature}, coherent oscillations in the axion field begin at the temperature $T_1^{\rm NonStd}$ defined by
\begin{equation}\label{def_T1.1}
3\,H(T_1^{\rm NonStd}) = m_a(T_1^{\rm NonStd}).
\end{equation}
When inserting Eq.~(\ref{H_nonstandard}) in Eq~(\ref{def_T1.1}), we obtain a relation for $T_1^{\rm NonStd}$ in this modified cosmology. Since we have already discussed the outcome of Eq.~(\ref{def_T1.1}) for a standard radiation-dominated cosmology in Sec.~\ref{Computing the oscillation temperature}, here we only consider the case $T > T_c$. Using the expression for the axion mass in Eq.~(\ref{axion_mass_temperature}) we obtain
\begin{equation}
T_1^{\rm NonStd} = T_c\,\begin{cases}
\left[\frac{a\,m_a}{3\,H_{\rm rad}(T_c)}\,\frac{g_*(T_c)}{g_*(T^{\rm NonStd})}\,\left(\frac{\Lambda}{T_c}\right)^4\right]^{1/(\upsilon+4)}, & \hbox{for $T_1^{\rm NonStd} \lesssim \Lambda$},\\
\left(\frac{m_a\,g_*(T_c)}{3\,H_{\rm rad}(T_c)\,g_*(T_1^{\rm NonStd})}\right)^{1/\upsilon}, & \hbox{for $T_1^{\rm NonStd} \gtrsim \Lambda$}.
\end{cases}
\end{equation}

\section{Axions from string decays in nonstandard\protect\\ cosmologies} \label{Axions from string decays in nonstandard cosmologies}

We have discussed the effect of cosmic strings on the present abundance of CDM axions in Sec.~\ref{Axions from string decays}. The main equation giving the relative abundance of string axions $Q$ to the abundance of axions from the misalignment mechanism \cite{Chang:1998tb, Hagmann:1990mj, Hagmann:2000ja} is expressed in Eq.~(\ref{proportionality}),
\begin{equation} \label{proportionality1}
Q \equiv \frac{\rho_a^{\rm str}(T_0)}{\rho_a^{\rm mis}(T_0)} = \frac{\xi \,\bar{r}\,N_d^2}{\zeta}.
\end{equation}
In this section, we discuss more on the parameters appearing on the right-hand side of Eq.~(\ref{proportionality1}), and extend previous theoretical results presented in Sec.~\ref{Axions from string decays} to obtain the values of the parameters in a generic nonstandard cosmology. 

{\it Parameter} $N_d. \quad$ In the standard cosmology, it is usually assumed that $N_d = 1$, because for $N_d > 1$ a domain wall problem may arise \cite{Sikivie:1982qv}. In modified cosmologies, we take the same value $N_d = 1$ as in the standard cosmology, because $N_d$ describes a property of the axion field.

{\it Parameter} $\bar{r}. \quad $ The value of $\bar{r}$, which depends on the energy spectrum of the radiated axions \cite{Shellard:1998mi}, was expressed in the standard radiation-dominated cosmology by Eq.~(\ref{bar_r}). To extend Eq.~(\ref{bar_r}) to nonstandard cosmologies, we repeat its standard-cosmology derivation in Refs.~\cite{Hagmann:2000ja, Harari:1987ht}, but we change the relation between time and Hubble parameter to that appropriate for a nonstandard cosmology, using Eqs.~(\ref{H_nonstandard}) and~(\ref{time_temperature_nonstd}).

Harari and Sikivie \cite{Harari:1987ht} derive the axion number density from string decays $n_a^{\rm str}(t)$ from the equations
\begin{equation} \label{dn/dt}
\frac{d n_a^{\rm str}(t)}{dt} = \frac{1}{\omega(t)}\frac{d\rho_r}{dt} - 3H(t)n_a^{\rm str}(t),
\end{equation}
and
\begin{equation} \label{drho/dt}
\frac{d\rho_r}{dt} = -\frac{d\rho_s}{dt} - 2H(t)\rho_s.
\end{equation}
Here, $\rho_r$ is the energy density of the radiated axions, $\omega(t)$ is the average energy of axions radiated in string decay processes at time $t$ \cite{Hagmann:2000ja}, and $\rho_s$ is the energy density in strings, given by
\begin{equation} \label{rho_s}
\rho_s = \frac{\xi\,N_d^2}{\zeta}\,\frac{\pi\,f_a^2\ln(t/\delta)}{t^2}.
\end{equation}
From Eqs.~(\ref{dn/dt}),~(\ref{drho/dt}), and~(\ref{rho_s}) we obtain
\begin{equation}
n_a^{\rm str}(t_1) = \frac{\xi\,N_d^2\,f_a^2}{\zeta}\,\frac{2\pi\,(1-\beta)}{t_1^{3\beta}}\int_{t_{\rm PQ}}^{t_1} \frac{dt}{t^{3-3\beta}}\frac{\ln(t/\delta)}{\omega(t)},
\end{equation}
where $t_1$ is the age of the universe when its temperature is $T_1^{\rm NonStd}$, and $t_{\rm PQ} \ll t_1$ is the time at which the PQ phase transition occurs. The formula to obtain the parameter $\bar{r}$ follows from Eq.~(2.13) in Ref.~\cite{Hagmann:2000ja},
\begin{equation}
n_a^{\rm str}(t) = \frac{\xi\,\bar{r}N_d^2\,f_a^2}{\zeta\,t},
\end{equation}
which, evaluated at time $t_1$, gives
\begin{equation} \label{bar_r1}
\bar{r} = \frac{2\pi\,(1-\beta)}{t_1^{3\beta-1}}\int_{t_{\rm PQ}}^{t_1} \frac{dt}{t^{3-3\beta}}\frac{\ln(t/\delta)}{\omega(t)}.
\end{equation}

The function $\omega(t)$ depends on the model for the energy spectrum of the emitted axions. For slow-oscillating strings, Davis \cite{Davis:1985pt, Davis:1986xc} argues that the energy spectrum of the axions radiated at time $t$ is peaked around $2\pi/t$, and finds $\omega(t) = 2\pi/t$. Using this expression of $\omega(t)$ in Eq.~(\ref{bar_r1}) gives
\begin{equation} \label{bar_r_sharp}
\bar{r} = \begin{cases}
\frac{1-\beta}{3\beta-1}\ln(t_1/\delta), & \hbox{for $\beta \neq 1/3$},\\
\frac{2}{3}\ln(t_1/t_{\rm PQ})\ln(t_1/\delta), & \hbox{for $\beta = 1/3$}.
\end{cases}
\end{equation}
In particular, for the standard cosmology $\beta = 1/2$ and $\bar{r}^{\rm std} = \ln(t_1/\delta)$, as in the first line of Eq.~(\ref{bar_r}).

For fast-oscillating strings, Harari and Sikivie \cite{Harari:1987ht} argue that the energy spectrum of the radiated axions is broad, with a low-energy cutoff at energy $\pi/t_1$ and a high-energy cutoff at energy $\pi/\delta$. They find $\omega(t) = (2\pi/t)\ln(t/\delta)$. Eq.~(\ref{bar_r1}) with this expression of $\omega(t)$ leads to $\bar{r}^{\rm std} = 1$ for the standard cosmology. Numerical simulations \cite{Chang:1998tb, Hagmann:1990mj, Hagmann:2000ja} favor a slightly smaller value of $\bar{r}^{\rm std}$, namely 0.8 as quoted in Eq.~(\ref{bar_r}). Therefore, we decided to multiply Eq.~(\ref{bar_r1}) by 0.8. Hence, for the fast-oscillating strings,
\begin{equation} \label{bar_r_broad}
\bar{r} = \begin{cases}
0.8\,\frac{1-\beta}{3\beta-1}, & \hbox{for $\beta \neq 1/3$},\\
0.8\,\frac{2}{3}\,\ln(t_1/t_{\rm PQ}), & \hbox{for $\beta = 1/3$}.
\end{cases}
\end{equation}
In the particular case of the LTR cosmology ($\beta = 2/3$), Eq~(\ref{bar_r_broad}) gives $\bar{r}^{\rm LTR} = 0.27$, while in the kination cosmology ($\beta = 1/3$), Eq.~(\ref{bar_r_broad}) gives $\bar{r}^{\rm kin} = 0.53\,\ln(t_1/t_{\rm PQ})$.

In presenting our results, we use the value of $\bar{r}$ for fast-oscillating strings, Eq.~(\ref{bar_r_broad}). We discuss the alternative choice of Eq.~(\ref{bar_r_sharp}) in Sec.~\ref{Discussion}.

{\it Parameter} $\xi. \quad $ The value of $\xi^{\rm std}$ in the standard cosmology has been discussed in the literature and different authors quote different results \cite{Hagmann:1990mj, Hagmann:2000ja, Yamaguchi:1998gx,Bennett:1989yp, Allen:1990tv}. Numerical simulations for an evolving string network in Ref.~\cite{Bennett:1989yp, Allen:1990tv} yield $\xi^{\rm std} \sim 13$, while simulations in Refs.~\cite{Hagmann:1990mj, Hagmann:2000ja, Yamaguchi:1998gx} give $\xi^{\rm std} \sim 1$.

The value of $\xi$ changes in a modified cosmological scenario. The authors in Ref.~\cite{Yamaguchi:1999dy} outline a method to estimate $\xi$ in a modified cosmology in which the universe is matter-dominated from the value $\xi^{\rm std}$ for a radiation-dominated universe. Here, we generalize the results of Ref.~\cite{Yamaguchi:1999dy} to a generic cosmology with arbitrary $\beta$, following their method. We define the characteristic length $L$ for an axionic string of energy density $\rho$ and tension $\mu$ per unit length through $\rho = \mu/L^2$.  The parameter $\xi$ appears in the time dependence of the string length $L$ as $L = t/\sqrt{\xi}$. The method of Ref.~\cite{Yamaguchi:1999dy} consists in computing $\xi$ as
\begin{equation}
\label{eq:xidef}
\xi = (\gamma_0\,H\,t)^2,
\end{equation}
where $\gamma_0$ is the fixed-point value of Eq.~(14) in Ref.~\cite{Yamaguchi:1999dy},
\begin{equation} \label{yamaguchi}
\frac{d\gamma}{dt} = -\frac{H}{2}\left\{c\gamma^2 + \left[2\frac{\dot{H}}{H^2} + 3\right]\gamma - 1\right\}.
\end{equation}
Here $\gamma = (H(t) L)^{-1}$, and $c>0$ is a constant determined by the value of $\xi$ for a radiation-dominated universe.

For a generic $\beta$, we have
\begin{equation}
\frac{\dot{H}}{H^2} = -\frac{1}{\beta},
\end{equation}
and the fixed-point of Eq.~(\ref{yamaguchi}) follows from setting its right-hand side to zero,
\begin{equation} \label{gamma_0}
\gamma_0 = \frac{2-3\beta + \sqrt{(4c+9)\beta^2 -12\beta+4}}{2\beta \,c}.
\end{equation}
Then from Eq.~(\ref{eq:xidef}) and $H=\beta/t$, we find
\begin{equation} \label{xi1}
\xi = \frac{1}{4c^2}\left(2-3\beta + \sqrt{(4c+9)\beta^2 -12\beta+4}\right)^2.
\end{equation}
The constant $c$ is fixed from the requirement that $\xi = \xi^{\rm std}$ in a radiation-dominated universe, $\beta=1/2$. This gives
\begin{equation}\label{c}
c = \frac{1+2\sqrt{\xi^{\rm std}}}{4\xi^{\rm std}}.
\end{equation}
This is to be substituted in Eq.~(\ref{xi1}) to find the value of $\xi$ in the nonstandard cosmology.

For $\xi^{\rm std} = 1$, we find $c = 3/4$ and
\begin{equation} \label{xi}
\xi = \frac{4}{9}\left(2-3\beta + 2\sqrt{3\beta^2 -3\beta+1}\right)^2\quad\hbox{(for $\xi^{\rm std} = 1$)}.
\end{equation}

Then, for the LTR cosmology, in which $\beta=2/3$, we find $\xi^{\rm LTR} = 16/27 = 0.5926$; for the kination cosmology, in which $\beta = 1/3$, we find $\xi^{\rm kin} = 4(7+4\sqrt{3})/27 = 2.0634$.

The choice $\xi^{\rm std} = 13$ is discussed in Section \ref{Discussion}.

{\it Parameter} $\zeta. \quad$ In the standard cosmology, $\zeta \sim 1$ \cite{Hagmann:2000ja}. To find $\zeta$ in a generic cosmology, we use the fact that on dimensional grounds, $\zeta$ is of order $\sqrt{\xi}$ \cite{Hagmann:2000ja}, so that to a change $\Delta \xi$ there corresponds a change $\Delta \zeta/2$. However, the theoretical uncertainty on $\zeta$ is around 50$\%$ \cite{Hagmann:2000ja}, higher than the difference $\Delta \xi$ due to the change in the cosmology used. We thus consider $\zeta = 1$ in all cosmological scenarios.

\section{Axion CDM in the LTR cosmology} \label{Axion CDM in the low temperature reheating scenario}

In the LTR cosmology \cite{Dine:1982ah, Steinhardt:1983ia, Turner:1983he, Scherrer:1984fd, Lazarides:1987zf, Yamamoto:1985mb, McDonald:1989jd, Moroi:1994rs, Kawasaki:1995cy, Kawasaki:1995vt, Chung:1998rq, Giudice:2000ex, Giudice:2000dp, Gelmini:2004ah, Gelmini:2006pw, Gelmini:2006pq, Drees:2006vh}, the universe after inflation is dominated by a massive decaying scalar field $\Phi$ down to the reheating temperature $T_{\rm RH}$. The reheating temperature is defined as the temperature at which the decay width $\Gamma_\Phi$ of the scalar field $\Phi$ is equal to the Hubble expansion rate $H(T)$ \cite{Moroi:1994rs, Kawasaki:1995cy, Kawasaki:1995vt, Chung:1998rq, Giudice:2000ex, Giudice:2000dp, Gelmini:2004ah, Gelmini:2006pw, Gelmini:2006pq, Drees:2006vh},
\begin{equation}
T_{\rm RH} \equiv \left(\frac{90}{8\pi^3\,g_*(T_{\rm RH})}\right)^{1/4}\sqrt{\Gamma_{\Phi}\,M_{Pl}}.
\end{equation}
At $T>T_{\rm RH} $, the universe follows the LTR cosmology; at $T=T_{\rm RH}$, it transitions to the usual radiation-dominated era. At $T<T_{\rm RH}$, the Hubble expansion rate is given by Eq.~(\ref{hubble_radiation}) for a radiation-dominated universe; at $T>T_{\rm RH}$, the Hubble rate $H(T)$ depends on the scale factor $a^{\rm LTR}(t)$ as in a matter-dominated epoch \cite{Turner:1983he, Scherrer:1984fd, Lazarides:1987zf, Yamamoto:1985mb, McDonald:1989jd, Moroi:1994rs, Kawasaki:1995cy, Kawasaki:1995vt, Chung:1998rq},
\begin{equation}
H(T) = H_{\rm rad}(T_{\rm RH})\left(\frac{a^{\rm LTR}(T_{\rm RH})}{a^{\rm LTR}(T)}\right)^{3/2}, \quad \hbox{for $T>T_{\rm RH}$}.
\end{equation}
Here $H_{\rm rad}(T_{\rm RH})$ is in Eq.~(\ref{hubble_radiation}) for the radiation-dominated epoch. The relation between the scale factor $a^{\rm LTR}(T)$ and the temperature $T$ during the LTR epoch \cite{Lazarides:1987zf} is
\begin{equation} \label{scale_factor_LTR}
g_*(T)^{2/3}\,T^{8/3}\,a^{\rm LTR}(T) = {\rm constant}, \quad \hbox{for $T>T_{\rm RH}$},
\end{equation}
which shows that the evolution of the universe during the LTR stage is nonadiabatic because the entropy density in Eq.~(\ref{entropy_density}) does not scale with $(a^{\rm LTR})^3$. Using Eq.~(\ref{scale_factor_LTR}) and the usual relation during the radiation-dominated epoch,
\begin{equation}
g_*(T)^{1/3}\,T\,a^{\rm std}(T) = {\rm constant},\quad \hbox{for $T<T_{\rm RH}$},
\end{equation}
we find the Hubble expansion rate
\begin{equation} \label{H_LTR}
H(T) = H_{\rm rad}(T_{\rm RH})\,\begin{cases}
\sqrt{\frac{g_*(T)}{g_*(T_{\rm RH})}}\,\left(\frac{T}{T_{\rm RH}}\right)^2, & \hbox{for $T < T_{\rm RH}$},\\
\frac{g_*(T)}{g_*(T_{\rm RH})}\,\left(\frac{T}{T_{RH}}\right)^4, & \hbox{for $T > T_{\rm RH}$},\\
\end{cases}
\end{equation}
which is Eq.~(\ref{H_nonstandard}) in the case of the LTR cosmology $\upsilon = 4$, with $T_c$ replaced by $T_{\rm RH}$.

In the standard cosmology, the axion field starts to oscillate at a temperature $T_1^{\rm std}$ given by Eq.~(\ref{T_1}) with the standard radiation-dominated expansion rate $H_{\rm rad}(T)$ on the left-hand side. In the LTR cosmology, $H(T)$ differs from the standard expression at $T>T_{\rm RH}$, and the axion field may start oscillating at a different temperature $T_1^{\rm LTR}$. More precisely, if the standard temperature $T_1^{\rm std}$ is less than $T_{\rm RH}$, then the axion field starts to oscillate when the universe is radiation-dominated. Moreover, since $H(T)$ is the same in both cosmologies at $T<T_{\rm RH}$, the oscillations start at the temperature $T_1^{\rm LTR}=T_1^{\rm std}$ given in Eq.~(\ref{T_1}). In this case, the results of Sec.~\ref{Axions from the misalignment mechanism} apply.

On the other hand, if $T_1^{\rm std}$ would be larger than $T_{\rm RH}$, then the LTR temperature $T_1^{\rm LTR}$ will be smaller than $T_1^{\rm std}$. In this case, the axion field starts to oscillate when the universe is dominated by the decay of the massive scalar field $\Phi$. The temperature $T_1^{\rm LTR}$ follows from Eq.~(\ref{def_T1.1}) with $H(T)$ given by the first line of Eq.~(\ref{H_nonstandard}). Since the dependence of $H(T)$ on $T$ steepens from $T^2$ to $T^4$ as $T$ becomes greater than $T_{\rm RH}$, it follows that
\begin{equation} \label{T_1-relation-LTR}
T_1^{\rm LTR} < T_1^{\rm std}.
\end{equation}
Introducing the notation
\begin{equation}
g_{\rm RH}(T) = \frac{g_*^2(T)}{g_*(T_{\rm RH})}.
\end{equation}
and with $T_{\rm RH,MeV} = T_{\rm RH}/{\rm MeV}$, we find
\begin{equation}
T_1^{\rm LTR} = \begin{cases}
\left(a\, m_a M_{Pl} T_{\rm RH}^2 \Lambda_{\rm QCD}^{4}\sqrt{\frac{5}{4\pi^3\,g_{RH}(T_1^{\rm LTR})}}\right)^{1/8}, & \hbox{for $T_1^{\rm LTR} \gtrsim \Lambda_{\rm QCD}$},\\
\left(m_a M_{Pl} T_{\rm RH}^2 \sqrt{\frac{5}{4\pi^3\,g_{RH}(T_1^{\rm LTR})}}\right)^{1/4}, & \hbox{for $T_1^{\rm LTR} \lesssim \Lambda_{\rm QCD}$}.
\end{cases}
\end{equation}
Numerically,
\begin{equation}\label{T1LTR}
T_1^{\rm LTR} = \begin{cases}
(160\pm 2){\rm ~MeV}\,g_{RH}^{-1/16}(T_1^{\rm LTR})\,T_{\rm RH,MeV}^{1/4}\,f_{a,12}^{-1/8}, & \hbox{for $T_1^{\rm LTR} \gtrsim \Lambda_{\rm QCD}$},\\
(350\pm 8){\rm ~MeV}\,g_{RH}^{-1/8}(T_1^{\rm LTR})\,T_{\rm RH,MeV}^{1/2}\,f_{a,12}^{-1/4}, & \hbox{for $T_1^{\rm LTR} \lesssim \Lambda_{\rm QCD}$}.
\end{cases}
\end{equation}
To summarize, if $T_{\rm RH} < \Lambda_{\rm QCD}$,
\begin{equation}
T_1^{\rm LTR} = \begin{cases}
(67\pm 3){\rm ~GeV}\,g_*^{-1/4}(T_1^{\rm std})f_{a,12}^{-1/2}, & \hbox{for $T_1^{\rm std} < T_{\rm RH}$},\\
(350\pm 8){\rm ~MeV}\,g_{RH}^{-1/8}(T_1^{\rm LTR})\,T_{\rm RH,MeV}^{1/2}\,f_{a,12}^{-1/4}, & \hbox{for $T_{\rm RH} < T_1^{\rm LTR} \lesssim \Lambda_{\rm QCD}$},\\
(160\pm2){\rm ~MeV}\,g_{RH}^{-1/16}(T_1^{\rm LTR})\,T_{\rm RH,MeV}^{1/4}\,f_{a,12}^{-1/8}, & \hbox{for $\Lambda_{\rm QCD} \lesssim T_1^{\rm LTR}$};
\end{cases}
\end{equation}
if $T_{\rm RH} > \Lambda_{\rm QCD}$,
\begin{equation}
T_1^{\rm LTR} = \begin{cases}
(67\pm3){\rm ~GeV}\,g_*^{-1/4}(T_1^{\rm std})f_{a,12}^{-1/2}, & \hbox{for $T_1^{\rm std} \lesssim \Lambda_{\rm QCD}$}, \\
(607\pm 27){\rm ~MeV}\,g_*^{-1/12}(T_1^{\rm std})f_{a,12}^{-1/6}, & \hbox{for $\Lambda_{\rm QCD} \lesssim T_1^{\rm std}  < T_{\rm RH}$}, \\
(160\pm2){\rm ~MeV}\,g_{RH}^{-1/16}(T_1^{\rm LTR})\,T_{\rm RH,MeV}^{1/4}\,f_{a,12}^{-1/8}, & \hbox{for $T_{\rm RH} < T_1^{\rm std}$} .
\end{cases}
\end{equation}

Also, the present axion energy density is modified from the standard case if $T_1^{\rm std} > T_{\rm RH}$.  We examine the misalignment mechanism and the production in string decays separately.

String decays in the HHCS give the contribution to the present axion energy density expressed in Eq.~(\ref{proportionality1}),
\begin{equation} \label{strings_LTR}
\Omega_a^{\rm LTR,str} = Q^{\rm LTR}\,\Omega_a^{\rm LTR,mis} = 0.16\,\Omega_a^{\rm LTR,mis},
\end{equation}
where we have used the values $N_d = 1$, $\bar{r}^{\rm LTR} = 0.27$, $\xi^{\rm LTR} = 16/27$, and $\zeta = 1$ in Eq.~(\ref{proportionality1}), consistent with the discussion in Sec.~\ref{Axions from string decays in nonstandard cosmologies}.

For the misalignment mechanism, the axion number density at the present time can be found from considering the expression for $n_a^{\rm mis}(T_1) $ in Eq.~(\ref{number_density}), with $T_1 = T_1^{\rm LTR}$, then use the conservation of axion number in a comoving volume, $n_a(T) \propto a^{-3}(T)$. This gives
\begin{equation} \label{n(RH)}
n_a^{\rm LTR}(T_0) =
\begin{cases}
n_a(T_1^{\rm std})\,\left(\frac{a^{\rm std}(T_1^{\rm std})}{a^{\rm std}(T_0)}\right)^3 , & \hbox{for $T_1^{\rm std} < T_{\rm RH}$},\\
n_a(T_1^{\rm LTR})\,\left(\frac{a^{\rm LTR}(T_1^{\rm LTR})}{a^{\rm std}(T_0)}\right)^3 , & \hbox{for $T_1^{\rm std} > T_{\rm RH}$}.
\end{cases}
\end{equation}
Here $n_a(T_1)$ is the function given in Eq.~(\ref{number_density}). One clearly has
\begin{equation}
n_a^{\rm LTR}(T_0) = n_a^{\rm std}(T_0) \quad \hbox{for $T_1^{\rm std} < T_{\rm RH}$.}
\end{equation}

In the case $T_1^{\rm std} < T_{\rm RH}$ coherent oscillations begin in a radiation-dominated universe, and the number density of cold axions is the same as in the standard cosmology. Instead, for $T_1^{\rm std} > T_{\rm RH}$, one obtains a different axion density due to both the release of entropy and to the different function for the scale factor $a^{\rm LTR}(T)$ in this modified cosmology. To better understand the origin of the difference, it is convenient to introduce the ratio between the present density $n_a^{\rm LTR}(T_0)$ in the LTR cosmology, and the present density $n_a^{\rm std}(T_0)$ that would ensue if the cosmology were standard at temperatures $T>T_{\rm RH}$. We write, for $T_1^{\rm std} > T_{\rm RH}$,
\begin{equation}
\frac{n_a^{\rm LTR}(T_0)}{n_a^{\rm std}(T_0)} =
\frac{N^{\rm LTR}}{N^{\rm std}}\,\frac{V^{\rm LTR}}{V^{\rm std}} ,
\end{equation}
where
\begin{equation} \label{N/N_std}
\frac{N^{\rm LTR}}{N^{\rm std}} = \frac{n_a(T_1^{\rm LTR})}{n_a(T_1^{\rm std})} \left( \frac{a^{\rm std}(T_1^{\rm LTR})}{a^{\rm std}(T_1^{\rm std})} \right)^3
\end{equation}
is the standard-cosmology ratio of the comoving number of axions $N^{\rm LTR}$  at the temperature $T_1^{\rm LTR}$ to the comoving number of axions $N^{\rm std}$ at the temperature $T_1^{\rm std}$,
and
\begin{equation} \label{V/V_std}
\frac{V^{\rm LTR}}{V^{\rm std}} = \left(\frac{a^{\rm LTR}(T_1^{\rm LTR})}{a^{\rm std}(T_1^{\rm LTR})}\right)^3 ,
\end{equation}
is the ratio of the LTR-cosmology volume $V^{\rm LTR}$ to the standard-cosmology volume $V^{\rm std}$ at the temperature $T_1^{\rm LTR}$.

The ratio $N^{\rm LTR}/N^{\rm std}$ accounts for the fact that coherent oscillations in the axion field start at a different temperature in the LTR cosmology compared to the standard cosmology. The ratio $V^{\rm LTR}/V^{\rm std}$ accounts for the fact that at temperature $T_1^{\rm LTR}$ the scale factors, and so the volumes, in the LTR and in the standard cosmologies differ due to entropy production from the decay of the scalar field in the LTR case.

Using the relations between temperature and scale factor during the radiation and LTR epochs, Eqs.~(\ref{scale_factor}) and~(\ref{scale_factor_LTR}) respectively, we find
\begin{equation} \label{N/N_std1}
\frac{N^{\rm LTR}}{N^{\rm std}} = \frac{g_{*S}(T_1^{\rm std})}{g_{*S}(T_1^{\rm LTR})}\left(\frac{T_1^{\rm std}}{T_1^{\rm LTR}}\right)^7,
\end{equation}
and
\begin{equation} \label{V/V_std1}
\frac{V^{\rm LTR}}{V^{\rm std}} = \frac{g_{*S}(T_1^{\rm LTR})}{g_{*S}(T_{\rm RH})}\frac{g^2_*(T_{\rm RH})}{g^2_*(T_1^{\rm LTR})}\,\left(\frac{T_{\rm RH}}{T_1^{\rm LTR}}\right)^5.
\end{equation}

From $T_1^{\rm LTR} < T_1^{\rm std}$ (see Eq.~(\ref{T_1-relation-LTR})), we find that $N^{\rm LTR} > N^{\rm std}$. But for $T_1^{\rm LTR} > T_{\rm RH}$, $V^{\rm LTR} < V^{\rm std}$. The latter factor dominates, and $n_a^{\rm LTR}(T_0)$ is less than $n_a^{\rm std}(T_0)$.

The present axion energy density from the misalignment mechanism follows as, in units of the critical density,
\begin{equation} \label{compare}
\Omega_a^{\rm LTR,mis} = \frac{m_a n_a^{\rm LTR}(T_0)}{\rho_{\rm crit}} = \begin{cases}
\Omega_a^{\rm std,mis}, & \hbox{for $T_1^{\rm std} < T_{\rm RH}$},\\
\Omega_a^{\rm std,mis} \frac{N^{\rm LTR}}{N^{\rm std}}\,\frac{V^{\rm LTR}}{V^{\rm std}}, & \hbox{for $T_1^{\rm std} > T_{\rm RH}$}.
\end{cases}
\end{equation}
Here $\Omega_a^{\rm std,mis}$ is given in Eq.~(\ref{standarddensity1}).

The first line of Eq.~(\ref{compare}), valid for $T_1^{\rm std} < T_{\rm RH}$, is numerically equal to Eq.~(\ref{standarddensity1}). The second line of Eq.~(\ref{compare}), valid for $T_1^{\rm std} > T_{\rm RH}$, is
\begin{equation}
\Omega_a^{\rm LTR, mis} = \begin{cases}
\left(\frac{4\pi^3}{5}\right)^{3/4}\,g_*(T_0)\,g_{\rm RH}^{-1/4}(T_1^{\rm LTR})\,\frac{\chi\,m_a^{1/2}\,f_a^2\,T_{\rm RH}^2\,T_0^3}{2\,\rho_c\,a^{1/2}\,M_{\rm Pl}^{3/2}\,\Lambda_{\rm QCD}^2}\,\langle\theta_i^2\,f(\theta_i)\rangle,& {\rm for}\, f_a < \hat{f}_a(T_{\rm RH}),\\
\frac{4\pi^3}{5}\,g_*(T_0)\,\frac{\chi\,f_a^2\,T_{\rm RH}\,T_0^3}{2\,\rho_c\,M_{\rm Pl}^2}\,\langle\theta_i^2\,f(\theta_i)\rangle,& {\rm for}\, f_a > \hat{f}_a(T_{\rm RH}).
\end{cases}
\end{equation}
Numerically,
\begin{equation} \label{Omega_LTR_numerical}
\Omega_a^{\rm LTR,mis} h^2 = \begin{cases}
(2.0\pm0.2)\times 10^{-6}\,\langle\theta_i^2\,f(\theta_i)\rangle\,g_{\rm RH}^{-1/4}(T_1^{\rm LTR})\,f_{a,12}^{3/2}\,T_{\rm RH,MeV}^2, & {\rm for}\, f_a < \hat{f}_a(T_{\rm RH}),\\
(7.5\pm0.8)\times 10^{-8}\,\langle\theta_i^2\,f(\theta_i)\rangle\,f_{a,12}^2\,T_{\rm RH,MeV}, & {\rm for}\, f_a > \hat{f}_a(T_{\rm RH}),
\end{cases}
\end{equation}
where
\begin{equation} \label{PQ_scale_LTR}
\hat{f}_a(T_{\rm RH}) = (5\pm1)\times 10^{14}{\rm ~GeV}\,g_{RH}^{-1/2}(T_1^{\rm LTR})\,T_{\rm RH,MeV}^2,
\end{equation}
is the PQ scale at which the two lines in Eq.~(\ref{Omega_LTR_numerical}) match.

The present axion energy density in the LTR cosmology, for Scenario I, is given by the sum of the misalignment mechanism and the string decay contributions
\begin{equation} \label{energy_density_total_LTR}
\Omega^{\rm LTR}_a = \Omega^{\rm LTR,mis}_a + \Omega^{\rm LTR,str}_a = \begin{cases}
\Omega_a^{\rm std,mis}\,(1+Q^{\rm std}), & \hbox{for $T_1^{\rm std} < T_{\rm RH}$},\\
\Omega_a^{\rm std,mis} \frac{N^{\rm LTR}}{N^{\rm std}}\,\frac{V^{\rm LTR}}{V^{\rm std}}\,(1+Q^{\rm LTR}), & \hbox{for $T_1^{\rm std} > T_{\rm RH}$}.
\end{cases}
\end{equation}
Here, $Q^{\rm std}$ and $Q^{\rm LTR}$ are the values of the ratio $\rho_a^{\rm str}(T_0)/\rho_a^{\rm mis}(T_0)$ in Eq.~(\ref{proportionality}) in the standard and LTR cosmologies, respectively. In the HHCS model, we obtained
\begin{equation}
Q^{\rm std} = 0.8, \quad\hbox{and}\quad Q^{\rm LTR} = 0.16.
\end{equation}
Numerically, the first line of Eq.~(\ref{energy_density_total_LTR}) corresponds to the total (strings plus misalignment) energy density in axions found in the standard cosmology, Eq.~(\ref{standarddensity2}), while the second line is valid for $T_1^{\rm std} > T_{\rm RH}$ and gives
\begin{equation} \label{Omega_LTR_numerical_total}
\Omega_a^{\rm LTR} h^2 = \begin{cases}
(2.3 \pm 0.2)\times 10^{-6}\,\langle\theta_i^2\,f(\theta_i)\rangle\,g_{\rm RH}^{-1/4}(T_1^{\rm LTR})\,f_{a,12}^{3/2}\,T_{\rm RH,MeV}^2, & {\rm for}\, f_a < \hat{f}_a(T_{\rm RH}),\\
(8.7 \pm 0.9)\times 10^{-8}\,\langle\theta_i^2\,f(\theta_i)\rangle\,f_{a,12}^2\,T_{\rm RH,MeV}, & {\rm for}\, f_a > \hat{f}_a(T_{\rm RH}),
\end{cases}
\end{equation}
This equation, where axionic strings contribute to the value of $\Omega_a^{\rm LTR}$, is valid only in Scenario I ($f_a < H_I/2\pi$); otherwise, Eq.~(\ref{Omega_LTR_numerical}) suffices and yields the total axion energy density in the LTR cosmology.

\subsection{Results for the LTR cosmology} \label{Results for LTR}

We now derive the regions of axion parameter space where the axion is 100$\%$ of the CDM in the LTR cosmology. We then compare them to the standard-cosmology regions.

The CDM axion parameter space in the standard cosmology depends on the PQ energy scale $f_a$ (or alternatively the axion mass $m_a$), the initial misalignment angle $\theta_i$, and the Hubble parameter during inflation $H_I$. In the LTR cosmology, an additional parameter is included, the reheating temperature $T_{\rm RH}$.

If the PQ symmetry breaks after the end of inflation (Scenario I, $f_a<H_I/2\pi$), there is only one PQ scale $f_a$ for which the totality of the CDM is made of axions. There correspondingly is also a single value for the axion mass $m_a$. In the standard cosmology, we referred to this scale as $f_a^{\rm std}$, whose value is given in Eq.~(\ref{PQscale_axionCDM_HHCS}) for the HHCS axionic string model. Here, we repeat the same computation in the LTR cosmology: we use the equate the value for $\Omega_{\rm CDM}\,h^2$ in Eq.~(\ref{CDM}) with the expression for $\Omega_a^{\rm LTR}\,h^2$ in Eq.~(\ref{Omega_LTR_numerical_total}) setting $\langle\theta_i^2\,f(\theta_i)\rangle = 1.4\, \pi^2/3$, obtaining the PQ scale
\begin{equation} \label{naturalscale_LTR}
f_a^{\rm LTR} = (4.7 \pm 0.6)\times 10^{14}{\rm ~GeV}\,g_{RH}^{1/6}(T_1^{\rm LTR})\,T_{\rm RH,MeV}^{-4/3},
\end{equation}
or
\begin{equation} \label{naturalmass_LTR}
m_a^{\rm LTR} = (13\pm 2)\, {\rm neV}\,g_{RH}^{-1/6}(T_1^{\rm LTR})\,T_{\rm RH,MeV}^{4/3}.
\end{equation}

\begin{figure}[tb]
\begin{center}
  \includegraphics[width=13cm]{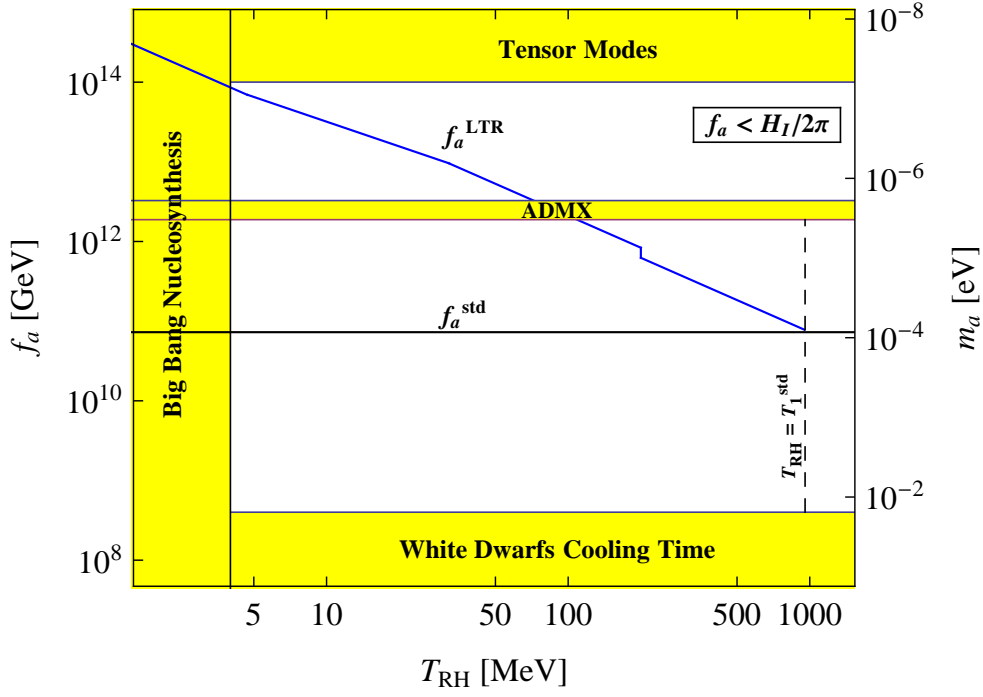}
\caption[The PQ scale $f_a^{\rm LTR}$ vs. $T_{\rm RH}$ for 100$\%$ axion CDM in the LTR cosmology]{The Peccei-Quinn scale $f_a^{\rm LTR}$ as a function of the reheating temperature $T_{\rm RH}$ for the axion to be 100$\%$ of the CDM in Scenario I ($f_a<H_I/2\pi$). Also shown are the PQ scale $f_a^{\rm std}$ in the standard cosmology, and various constraints (shaded regions).}
\label{running_Fa_LTR}
\end{center}
\end{figure}

In Fig.~\ref{running_Fa_LTR}, we plot $f_a^{\rm LTR}$ as a function of $T_{\rm RH}$. The jumps and kinks in the $f_a^{\rm LTR}$ line are due to the different values of $g_*(T_{\rm RH})$ and $g_*(T_1^{\rm LTR})$ in Eq.~(\ref{relativistic_dof_numerical}). There is also a (visually small) discontinuity between the $f_a^{\rm LTR}$ and $f_a^{\rm std}$ lines at $T_{\rm RH} = T_1^{\rm std} = T_1^{\rm LTR}$ due to different contributions from string decays. In fact, from the first line of Eq.~(\ref{standarddensity1}) and from Eq.~(\ref{energy_density_total_LTR}) we have
\begin{equation} \label{energy_density_total_LTR1}
\Omega^{\rm LTR}_a = \begin{cases}
(0.23\pm0.03)\,\langle\theta_i^2\,f(\theta_i)\rangle\,F_{a,12}^{7/6}\,(1+Q^{\rm std}), & \hbox{for $T_1^{\rm std} < T_{\rm RH}$},\\
(0.23\pm0.03)\,\langle\theta_i^2\,f(\theta_i)\rangle\,F_{a,12}^{7/6}\, \frac{N^{\rm LTR}}{N^{\rm std}}\,\frac{V^{\rm LTR}}{V^{\rm std}}\,(1+Q^{\rm LTR}), & \hbox{for $T_1^{\rm std} > T_{\rm RH}$}.
\end{cases}
\end{equation}
Equating the two lines in Eq.~(\ref{energy_density_total_LTR1}) at $T_{\rm RH} = T_1^{\rm std} = T_1^{\rm LTR}$, where $N^{\rm LTR} = N^{\rm std}$ and $V^{\rm LTR} = V^{\rm std}$, we obtain
\begin{equation} \label{condition_f_LTR}
f_a^{\rm LTR}(T_{\rm RH}\!=\!T_1^{\rm std}) = \left(\frac{1+Q^{\rm std}}{1+Q^{\rm LTR}}\right)^{6/7}\,f_a^{\rm std}.
\end{equation}
Numerically, $f_a^{\rm LTR}(T_{\rm RH}\!=\!T_1^{\rm std}) = 1.45\,f_a^{\rm std}$, which is slightly higher than $f_a^{\rm std}$.

In Fig.~\ref{running_Fa_LTR}, we also shade out the following bounds: the bound from white dwarfs cooling times in Eq.~(\ref{astrophysical_bound1}); the indirect bound on the PQ scale from the nondetection of primordial gravitational waves arising from $f_a<H_I/2\pi$ and Eq.~(\ref{HI_bound}) (region labeled ``Tensor Modes''); the bound on $T_{\rm RH} > 4$ MeV from Big Bang Nucleosynthesis; and the bound from the ADMX experiment \cite{Asztalos:2003px, Asztalos:2009yp} excluding a KSVZ axion with a mass $m_a$ between 1.9~$\,{\rm\mu eV}$ and 3.53~${\rm\mu eV}$, see Eq.~(\ref{exclusion_haloscope}). The dashed line marks the requirement that the axion starts to oscillate in the LTR cosmology, $T_{\rm RH} < T_1^{\rm std}$, with $T_1^{\rm std}$ given by Eq.~(\ref{T_1}). The ADMX bound can be rephrased as an exclusion bound for the reheating temperature $T_{\rm RH}$. Using the expression for the axion mass in Scenario I, Eq.~(\ref{naturalmass_LTR}), the ADMX result corresponds to an exclusion of the region $85{\rm ~MeV} < T_{\rm RH} < 140{\rm ~MeV}$, valid for KSVZ axions.

Depending on $T_{\rm RH}$, $f_a^{\rm LTR}$ may differ from $f_a^{\rm HHCS}$ in Eq.~(\ref{PQscale_axionCDM_HHCS}) by orders of magnitude. The maximum value of $f_a^{\rm LTR}$ is achieved for $T_{\rm RH} = 4\,{\rm MeV}$ and is, with $g_*(T_{\rm RH}) = 10.75$, $f_a^{\rm LTR} = (2.0 \pm 0.3)\times 10^{14}\,{\rm GeV}$. This value is three orders of magnitude larger than $f_a^{\rm HHCS}$ in Eq.~(\ref{PQscale_axionCDM_HHCS}). As discussed in Sec.~\ref{Discussion}, these large values of the Peccei-Quinn scale correspond to axion masses that are beyond the reach of current DM axion search experiments.

In Scenario II ($f_a>H_I/2\pi$), the parameter space is bounded by the nondetection of axion isocurvature fluctuations in the CMB spectrum, see Eq.~(\ref{adiabaticity}). For $T_{\rm RH} > T_1^{\rm std} $, the isocurvature bound has the same expression as in the standard cosmology, namely Eq.~(\ref{leftmostboundary}). For $T_{\rm RH} < T_1^{\rm std}$, we eliminate $\theta_i$ in Eq.~(\ref{adiabaticity}) by equating $\Omega_{\rm CDM}\,h^2$ with the expression for $\Omega_a^{\rm LTR}$ derived previously in Eq.~(\ref{Omega_LTR_numerical}), with no contribution from axionic strings. The LTR isocurvature bound is then, for $T_{\rm RH} < T_1^{\rm std}$ and $g_*(T_1^{\rm std}) = 61.75$,
\begin{equation} \label{leftmostboundaryLTR}
H_{I,12}=\begin{cases}
2.0\times 10^{-2}\,\left[f(\theta_i)\right]^{-1/2}\,T_{\rm RH,MeV}^{-1}\,f_{a,12}^{1/4}, & \hbox{for $8.2\times 10^{14}\,T_{\rm RH,MeV}^{-4/3}\,{\rm GeV} < f_a < \hat{f}_a(T_{\rm RH})$},\\
5.0 \times 10^{-2}\,T_{\rm RH,MeV}^{-1/2}, & \hbox{for $\hat{f}_a(T_{\rm RH}) < f_a$}.
\end{cases}
\end{equation}
Here $\hat{f}_a(T_{\rm RH}) = (5\pm1)\times 10^{14}{\rm ~GeV}\,g_{RH}^{-1/2}(T_1^{\rm LTR})\,T_{\rm RH,MeV}^2$ is given by Eq.~(\ref{PQ_scale_LTR}). The LTR isocurvature bound can be approximated by
\begin{equation} \label{leftmostboundaryLTR1}
H_{I,12} = \begin{cases}
1.3\times 10^{-4}\,f_{a,12}, & \hbox{for $f_a < 8.2\times 10^{14}\,T_{\rm RH,MeV}^{-4/3}\,{\rm GeV}$},\\
2.0\times 10^{-2}\,T_{\rm RH,MeV}^{-1}\,f_{a,12}^{1/4}, & \hbox{for $8.2\times 10^{14}\,T_{\rm RH,MeV}^{-4/3}\,{\rm GeV} < f_a < \hat{f}_a(T_{\rm RH})$},\\
5.0 \times 10^{-2}\,T_{\rm RH,MeV}^{-1/2}, & \hbox{for $f_a > \hat{f}_a(T_{\rm RH})$},
\end{cases}
\end{equation}
As in the case of the standard cosmology, there are two changes in the power-law dependence of $H_{I,12}$ on $f_{a,12}$ in the LTR cosmology, the first one being at $f_a = 8.2\times 10^{14}\,T_{\rm RH,MeV}^{-4/3}\,{\rm GeV}$ and the second one at $f_a = \hat{f}_a(T_{\rm RH})$. Notice that at large $f_a$, the isocurvature bound is independent of $f_a$.

When $T_{\rm RH} = T_1^{\rm std}$, the LTR and the standard isocurvature bounds coincide. This happens for
\begin{equation}
\begin{cases}
f_a = 4.6\times 10^{21}\,T_{\rm RH,MeV}^{-2}\,{\rm GeV}, & \hbox{for $T_1^{\rm std} < \Lambda_{\rm QCD}$},\\
f_a = 5.3 \times 10^{28}\,T_{\rm RH,MeV}^{-6}\,{\rm GeV}, & \hbox{for $T_1^{\rm std} > \Lambda_{\rm QCD}$}.
\end{cases}
\end{equation}

\begin{figure}[h!]
\begin{center}
  \includegraphics[width=12cm]{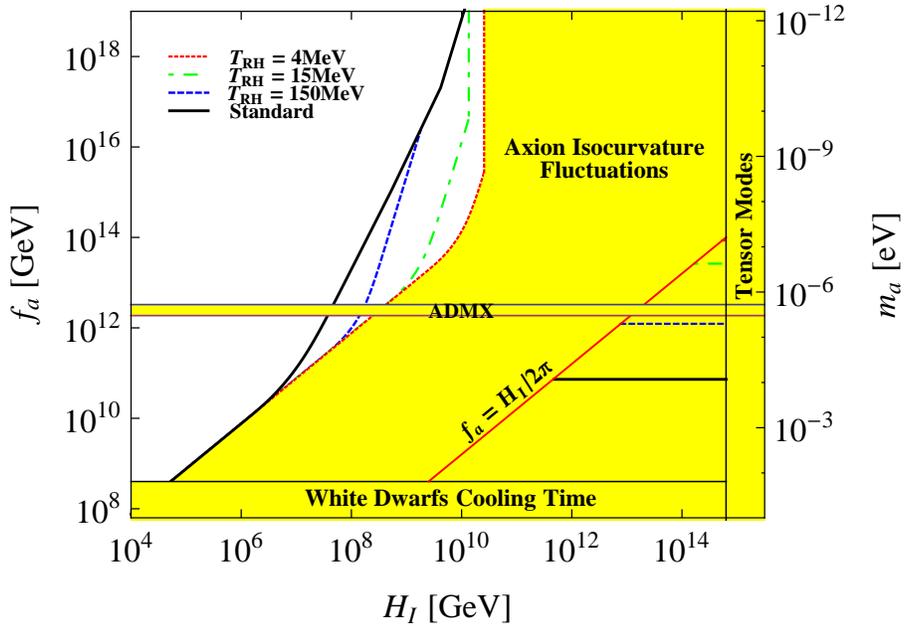}
  \caption[The parameter space where the axion is 100\% CDM in the LTR cosmology.]{In the LTR cosmology, axions are 100$\%$ of the CDM in the white region on the left (limited by a different line for each $T_{\rm RH}$) and in the narrow bands marked by horizontal lines in the lower right triangle (one line for each $T_{\rm RH}$).}
  \label{LTR}
  \end{center}
\end{figure}

Fig.~\ref{LTR} shows the regions of the parameter space $(f_a,H_I)$ where the axion is 100\% of the CDM in the LTR cosmology. The axion mass scale on the right is Eq.~(\ref{eq:axionmass}) with $N=1$. The region labeled ``Tensor Modes'' is excluded by the nonobservation of tensor modes in the CMB fluctuations, Eq.~(\ref{HI_bound}). The region labeled ``White Dwarfs Cooling Time'' is excluded from astrophysical observations of white dwarfs cooling times for KSVZ axions, Eq.~(\ref{astrophysical_bound1}) \cite{Raffelt:2006cw}.
The line $f_a = H_I/2\pi$ divides the region where the PQ symmetry breaks after inflation (Scenario I, $f_a<H_I/2\pi$) from the region where it breaks during inflation (Scenario II, $f_a>H_I/2\pi$). In the lower right region (Scenario I), the axion is the CDM particle if $f_a$ equals the value $f_a^{\rm LTR}$ given by Eq.~(\ref{naturalscale_LTR}). In the upper left region (Scenario II), we plot the isocurvature bounds to the allowed parameter space for the standard cosmology (thick line) and for $T_{\rm RH} = 4{\rm ~MeV}$ (dotted line), 15 MeV (dot-dashed line), and 150 MeV (dashed line). For a given $T_{\rm RH}$, the isocurvature bound with the LTR cosmology lies below the standard line, because the entropy dilution term $\sim (T_{\rm RH}/T_1)^5$ lowers the axion energy density. Thus, the LTR cosmology allows more CDM axion parameter space than the standard scenario.

\begin{figure}[h!]
\begin{center}
 \includegraphics[width=10cm]{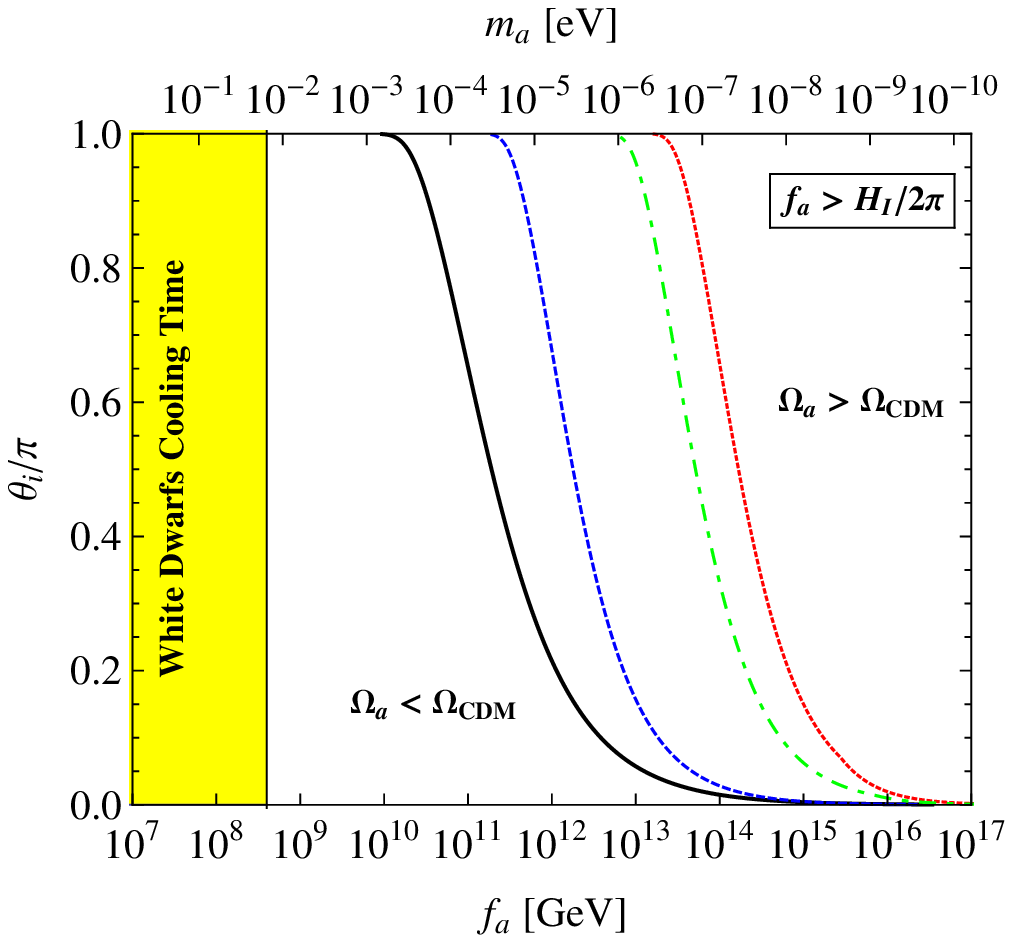}
\caption[$\theta_i$ as a function of $f_a$ in the LTR cosmology.]{The initial misalignment angle $\theta_i$ as a function of the Peccei-Quinn scale $f_a$ for the axion to be 100$\%$ of the CDM in Scenario II ($f_a>H_I/2\pi$): standard cosmology (black solid line), LTR cosmology with $T_{\rm RH} = 4$MeV (red dotted line), 15MeV (green dot-dashed line) or 150MeV (blue dashed line).}
\label{ThetaLTR}
\end{center}
\end{figure}

In the allowed region of parameter space for Scenario II, the axion can be 100$\%$ of the CDM provided the value of $\theta_i$ is chosen appropriately. In the standard cosmology, $\theta_i$ is a function of $f_a$ only \cite{Visinelli:2009zm, Hamann:2009yf}, see Eq.~(\ref{const_theta}). Instead, in the LTR cosmology, $\theta_i$ depends on both $f_a$ and $T_{\rm RH}$. Taking $g_*(T_1^{\rm LTR}) = 61.75$ we find
\begin{equation} \label{fv.theta_LTR}
f_{a,12} = \begin{cases}
\left(\frac{\Omega_{\rm CDM}h^2}{6.35\times 10^{-7}\,\theta_i^2\,f(\theta_i)T_{\rm RH,MeV}^2}\right)^{2/3}, & \hbox{for $f_a < \hat{f}_a(T_{\rm RH})\,\,{\rm or}\,\,\theta_i \gtrsim  17\,T_{\rm RH,MeV}^{-5/2}$},\\
\left(\frac{\Omega_{\rm CDM}h^2}{7.46\times 10^{-8}\,\theta_i^2\,f(\theta_i)T_{\rm RH,MeV}}\right)^{1/2}, & \hbox{for $f_a > \hat{f}_a(T_{\rm RH})\,\,{\rm or}\,\,\theta_i \lesssim 17\,T_{\rm RH,MeV}^{-5/2}$}.
\end{cases}
\end{equation}
The relation in Eq.~(\ref{fv.theta_LTR}) is plotted in Fig.~\ref{ThetaLTR} for the standard cosmology (thick line) and for $T_{\rm RH} = 4{\rm ~MeV}$ (dotted line), $15 {\rm ~MeV}$ (dot-dashed line), and $150 {\rm ~MeV}$ (dashed line). As $T_{\rm RH}$ decreases, one departs from the standard cosmology. The value of $\theta_i$ at fixed $f_a$ increases when $T_{\rm RH}$ decreases.

\newpage

\section{Axion CDM in the kination cosmology} \label{Axion CDM in the kination scenario}

We now discuss axion CDM in the kination cosmology \cite{Ford:1986sy, Kamionkowski:1990ni, Spokoiny:1993kt, Joyce:1996cp, Joyce:1997fc, Tashiro:2003qp, Salati:2002md, Profumo:2003hq, Rosati:2003yw, Pallis:2005hm, Pallis:2005bb, Chun:2007np, Chung:2007cn, Chung:2007vz, Gomez:2008js}. The Hubble parameter for this pre-BBN cosmology is (see Ref.~\cite{Chun:2007np})
\begin{equation} \label{H_kination}
H(T) = H_{\rm rad}(T_{\rm kin})\,\begin{cases}
\sqrt{\frac{g_*(T)}{g_*(T_{\rm kin})}}\,\left(\frac{T}{T_{\rm kin}}\right)^2, & {\rm for}\, T < T_{\rm kin},\\
\frac{g_*(T)}{g_*(T_{\rm kin})}\,\left(\frac{T}{T_{\rm kin}}\right)^3, & {\rm for}\, T > T_{\rm kin}.
\end{cases}
\end{equation}
Here $T_{\rm kin}$ is the temperature at which the universe transitions from kination domination to radiation domination. Entropy is conserved during the kination cosmology, so the scale factor during kination $a^{\rm kin}(T)$ follows the same temperature dependence as $a^{\rm std}(T)$,
\begin{equation} \label{scale_factor_kination}
g_{*S}^{1/3}(T)\,T\,a^{\rm kin}(T) = {\rm constant}.
\end{equation}
For this reason, here we do not make a distinction between the scale factor in the standard and in the kination cosmologies. We write $a^{\rm kin}(T) = a^{\rm std}(T) \equiv a(T)$.

If $T_1^{\rm std} < T_{\rm kin}$, coherent oscillations in the axion field start in the radiation-dominated universe, at the temperature $T_1^{\rm std}$ given in Eq.~(\ref{T_1}). On the contrary, if $T_1^{\rm std} > T_{\rm kin}$, coherent oscillations in the axion field start when the universe is in its kination stage. In this case, the temperature $T_1^{\rm kin}$ at which axion oscillations begin is given by the following expression:
\begin{equation}\label{T1kin}
T_1^{\rm kin} = \begin{cases}
2.48{\rm ~GeV}\,g_{\rm kin}^{-1/6}(T_1^{\rm kin})\,T_{\rm kin,MeV}^{1/3}\,f_{a,12}^{-1/3}, & \hbox{for $T_1^{\rm kin} \lesssim \Lambda_{\rm QCD}$},\\
331{\rm ~MeV}\,g_{\rm kin}^{-1/14}(T_1^{\rm kin})\,T_{\rm kin,MeV}^{1/7}\,f_{a,12}^{-1/7}, & \hbox{for $T_1^{\rm kin} \gtrsim \Lambda_{\rm QCD}$}.
\end{cases}
\end{equation}
Here $T_{\rm kin,MeV} = T_{\rm kin}/{\rm MeV}$.

The axion energy density in the kination cosmology has contributions from string decays and from the misalignment mechanism.

String decays give a contribution to the present axion energy density
\begin{equation} \label{strings_kination}
\Omega_a^{\rm kin,str} = Q^{\rm kin}\,\Omega_a^{\rm kin,mis} = 0.42\,\bar{r}^{\rm kin}\,\Omega_a^{\rm kin,mis},
\end{equation}
where to compute $Q^{\rm kin}$ in Eq.~(\ref{proportionality}) we used $N_d = 1$, $\xi^{\rm kin} = 2.06$, $\zeta = 4.9$, and $\bar{r}^{\rm kin}$ is given in Eq.~(\ref{bar_r_broad}) as
\begin{equation}
\bar{r}^{\rm kin} = \frac{2}{3}\,0.8\ln\left(\frac{t_1}{t_{\rm PQ}}\right) = \frac{1.6}{3}\,\ln\left(\frac{H(f_a)}{H(T_1^{\rm kin})}\right).
\end{equation}
In the last expression, we used the relation $t \propto 1/H(T)$, the fact that at the time of the PQ transition $t_{\rm PQ}$, the temperature of the universe is $T = f_a$, and the fact that at the time $t_1$, the corresponding temperature is $T_1^{\rm kin}$. Using the expression for the kination Hubble parameter in Eq.~(\ref{H_kination}) and neglecting the term $\ln(\sqrt{\frac{g_{\rm kin}(f_a)}{g_{\rm kin}(T1^{\rm kin})}})\sim 1$, we obtain
\begin{equation}
\bar{r}^{\rm kin} = 1.6\,\ln(f_a/T_1^{\rm kin}).
\end{equation}
The temperature $T_1^{\rm kin}$ is greater than $\Lambda_{\rm QCD}$ for any value of $T_{\rm RH}$ and any value of $f_a$ for which there are contributions from string decays (Scenario I, $f_a < H_I/2\pi$). Thus, using the expression for $T_1^{\rm kin}$ in the second line of Eq.~(\ref{T1kin}), we obtain
\begin{equation}
\bar{r}^{\rm kin} = 57 +\frac{16}{7}\ln f_{a,12} - \ln T_{\rm kin,MeV}.
\end{equation}
In the region of the parameters of interest for kination, $\bar{r}^{\rm kin} \sim 35$. Thus, axions from strings dominate the total axion population, the energy density $\Omega_a^{\rm kin,str}$ being one order of magnitude larger than $\Omega_a^{\rm kin,mis}$. We notice that this is opposite to what we obtained in the standard and LTR cosmologies, where the radiation of axions from axionic strings is a subdominant production mechanism for cold axions.

The contribution from the misalignment mechanism results from the conservation of the axion number in a comoving volume, $n_a(T) \propto a^{-3}(T)$. This gives
\begin{equation}
n_a^{\rm kin}(T_0) = \begin{cases}
n_a(T_1^{\rm std})\,\left(\frac{a(T_1^{\rm std})}{a(T_0)}\right)^3, & \hbox{for $ T_1^{\rm std} < T_{\rm kin}$},\\
n_a(T_1^{\rm kin})\left(\frac{a(T_1^{\rm kin})}{a(T_0)}\right)^3, & \hbox{for $ T_1^{\rm std} > T_{\rm kin}$}.
\end{cases}
\end{equation}
Here $n_a(T_1)$ is the function given in Eq.~(\ref{number_density}). One clearly has
\begin{equation}
n_a^{\rm kin}(T_0) = n_a^{\rm std}(T_0) \quad \hbox{for $T_1^{\rm std} < T_{\rm kin}$.}
\end{equation}

For $T_1^{\rm std} > T_{\rm kin}$, one obtains a different axion density. As for the LTR cosmology, we introduce the ratio between the present density $n_a^{\rm kin}(T_0)$ in the kination cosmology, and the present density $n_a^{\rm std}(T_0) = \rho_a^{\rm std}(T_0)/m_a$, see Eq.~(\ref{energydensity}), that would ensue if the cosmology were standard at temperatures $T>T_{\rm kin}$. We write, for $T_1^{\rm std} > T_{\rm kin}$,
\begin{equation}
\frac{n_a^{\rm kin}(T_0)}{n_a^{\rm std}(T_0)} =
\frac{N^{\rm kin}}{N^{\rm std}}\,\frac{V^{\rm kin}}{V^{\rm std}},
\end{equation}
where $N^{\rm kin}/N^{\rm std}$ and $V^{\rm kin}/V^{\rm std}$ are defined as follows. The ratio $N^{\rm kin}/N^{\rm std}$ is the standard-cosmology ratio of the comoving number of axions $N^{\rm kin}$  at the temperature $T_1^{\rm kin}$ to the comoving number of axions $N^{\rm std}$ at the temperature $T_1^{\rm std}$. Using Eq.~(\ref{scale_factor_kination}), we write it as
\begin{equation}
\frac{N^{\rm kin}}{N^{\rm std}} = \frac{n_a(T_1^{\rm kin})}{n_a(T_1^{\rm std})}\left(\frac{a(T_1^{\rm kin})}{a(T_1^{\rm std})}\right)^3 = \frac{n_a(T_1^{\rm kin})}{n_a(T_1^{\rm std})}\frac{g_{*S}(T_1^{\rm std})}{g_{*S}(T_1^{\rm kin})}\left(\frac{T_1^{\rm std}}{T_1^{\rm kin}}\right)^3.
\end{equation}
The ratio $V^{\rm kin}/V^{\rm std}$ is the ratio of the kination-cosmology volume $V^{\rm kin}$ to the standard-cosmology volume $V^{\rm std}$ at the temperature $T_1^{\rm kin}$,
\begin{equation}
\frac{V^{\rm kin}}{V^{\rm std}} = \left(\frac{a^{\rm kin}(T_1^{\rm kin})}{a^{\rm std}(T_1^{\rm kin})}\right)^3 = 1,
\end{equation}
The last equality follows because no significant entropy is released during the kination stage \cite{Salati:2002md}, so $a^{\rm kin}(T) = a^{\rm std}(T)$.

The present axion energy density from the misalignment mechanism, in units of the critical density, is therefore
\begin{equation} \label{compare_kin}
\Omega_a^{\rm kin,mis} = \begin{cases}
\Omega_a^{\rm std,mis}, & \hbox{for $T_1^{\rm std} < T_{\rm kin}$},\\
\Omega_a^{\rm std,mis}\,\frac{N^{\rm kin}}{N^{\rm std}}, & \hbox{for $T_1^{\rm std} > T_{\rm kin}$}.
\end{cases}
\end{equation}

Inserting numerical values, the first line of Eq.~(\ref{compare_kin}) is given by Eq.~(\ref{standarddensity}), while the second line reads
\begin{equation}\label{density_kination}
\Omega_a^{\rm kin,mis} h^2 = 1150 \,g_*^{-1/2}(T_{\rm kin})\,\langle\theta_i^2\,f(\theta_i)\rangle\,f_{a,12}\,T_{\rm kin,MeV}^{-1}.
\end{equation}
Due to the peculiar dependence of the Hubble rate with temperature in kination, $H(T) \sim T^3$, there is no distinction in Eq.~(\ref{density_kination}) between $\Omega_a^{\rm kin,mis}$ for $T_1^{\rm kin} \gtrsim \Lambda_{\rm QCD}$ and for $T_1^{\rm kin} \lesssim \Lambda_{\rm QCD}$.

Finally, the present axion energy density in the kination cosmology is given by the sum of the misalignment mechanism and the string decay contributions
\begin{equation} \label{energy_density_total_kination}
\Omega_a^{\rm kin} = \Omega_a^{\rm kin,mis} + \Omega_a^{\rm kin,str} = \begin{cases}
\Omega_a^{\rm std,mis}\,(1+Q^{\rm std}), & \hbox{for $T_1^{\rm std} < T_{\rm kin}$},\\
\Omega_a^{\rm std,mis} \frac{N^{\rm kin}}{N^{\rm std}}\,(1+Q^{\rm kin}), & \hbox{for $T_1^{\rm std} > T_{\rm kin}$}.
\end{cases}
\end{equation}
Here, $Q^{\rm std}$ and $Q^{\rm kin}$ are the values of the ratio $\rho_a^{\rm str}(T_0)/\rho_a^{\rm mis}(T_0)$ in Eq.~(\ref{proportionality}) in the standard and kination cosmologies, respectively.

\subsection{Results for kination} \label{Results for kination}

We now derive the regions of the axion parameter space where the axion is 100$\%$ of the CDM in the kination cosmology. We then compare them to the standard-cosmology regions.

The axion parameter space in kination cosmology depends on $f_a$, $H_I$, $\theta_i$, and the additional parameter $T_{\rm kin}$.

If the PQ symmetry breaks after the end of inflation (Scenario I, $f_a<H_I/2\pi$), there is only one PQ scale $f_a$ for which the totality of the CDM is made of axions. There correspondingly is also a single value of the axion mass $m_a$. In the kination cosmology, using the observed value of $\Omega_{\rm CDM} h^2$ in Eq.~(\ref{CDM}), and the expressions for $\Omega_a^{\rm kin}$ derived in this section, we find
\begin{equation} \label{naturalscale_kination}
f_a^{\rm kin} = (7.9 \pm 0.2) \times 10^6\,{\rm GeV}\,g_*^{1/2}(T_{\rm kin})\,\frac{T_{\rm kin,MeV}}{57 +\frac{16}{7}\ln f_{a,12}^{\rm kin} - \ln T_{\rm kin,MeV}},
\end{equation}
and
\begin{equation}
m_a^{\rm kin} = 739\pm22\, {\rm meV}\,g_*^{-1/2}(T_{\rm kin})\,\bar{r}^{\rm kin}\,T^{-1}_{\rm kin,MeV}.
\end{equation}
In Eq.~(\ref{naturalscale_kination}), we used the explicit expression for $\bar{r}^{\rm kin}$ derived in Sec.~\ref{Axions from string decays}.

\begin{figure}[tb]
\begin{center}
  \includegraphics[width=13cm]{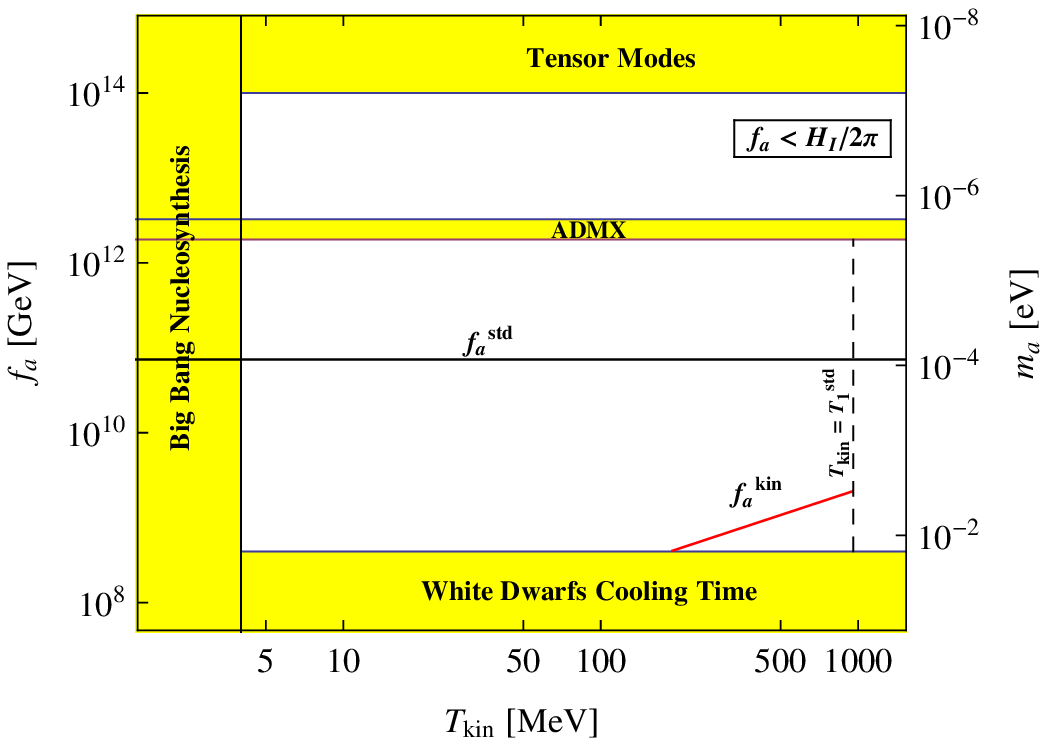}
\caption[The PQ scale $f_a^{\rm kin}$ vs. $T_{\rm kin}$ for 100$\%$ axion CDM in the kination cosmology]{The Peccei-Quinn scale $f_a^{\rm kin}$ as a function of the kination temperature $T_{\rm kin}$ for the axion to be 100$\%$ of the CDM in Scenario I ($f_a<H_I/2\pi$). Also shown are the PQ scale $f_a^{\rm std}$ in the standard cosmology, and various constraints (shaded regions).}
\label{running_Fa_kination}
\end{center}
\end{figure}

In Fig.~\ref{running_Fa_kination}, we plot $f_a^{\rm kin}$ as a function of $T_{\rm kin}$. The function $f_a^{\rm kin}$ does not present jumps, because both $g_*(T_1^{\rm kin})$ and $g_*(T_{\rm kin})$ do not change in the domain of $f_a^{\rm kin}$.

The discontinuity between the $f_a^{\rm kin}$ and $f_a^{\rm std}$ lines at $T_{\rm kin} = T_1^{\rm std} = T_1^{\rm kin}$ is due to different contributions from string decays. In fact, from Eqs.~(\ref{standarddensity}) and~(\ref{energy_density_total_kination}) we have
\begin{equation} \label{energy_density_total_kination1}
\Omega^{\rm kin}_a = \begin{cases}
1.32\,g_*^{-5/12}(T_1^{\rm std})\,\langle\theta_i^2\,f(\theta_i)\rangle\,(f_{a,12}^{\rm std})^{7/6}\,(1+Q^{\rm std}), & \hbox{for $T_1^{\rm std} < T_{\rm kin}$},\\
1.32\,g_*^{-5/12}(T_1^{\rm kin})\,\langle\theta_i^2\,f(\theta_i)\rangle\,(f_{a,12}^{\rm kin})^{7/6}\, \frac{N^{\rm kin}}{N^{\rm std}}\,\frac{V^{\rm kin}}{V^{\rm std}}\,(1+Q^{\rm kin}), & \hbox{for $T_1^{\rm std} > T_{\rm kin}$}.
\end{cases}
\end{equation}
Equating the two lines in Eq.~(\ref{energy_density_total_kination1}) at $T_{\rm kin} = T_1^{\rm std} = T_1^{\rm kin}$, where $N^{\rm kin} = N^{\rm std}$, we obtain
\begin{equation} \label{condition_f_kination}
f_a^{\rm kination}(T_{\rm kin}\!=\!T_1^{\rm std}) = f_a^{\rm std}\,\left(\frac{1+Q^{\rm std}}{1+Q^{\rm kin}}\right)^{6/7}.
\end{equation}
We find $f_a^{\rm kin}(T_{\rm kin}\!=\!T_1^{\rm std}) = 2.04\times 10^9{\rm ~GeV}$.

In Fig.~\ref{running_Fa_kination}, we also shade out the following bounds: the bound from white dwarfs cooling times in Eq.~(\ref{astrophysical_bound}); the indirect bound on $f_a$ from the nondetection of primordial gravitational waves arising from $f_a<H_I/2\pi$ and Eq.~(\ref{HI_bound}) (region labeled ``Tensor Modes''); the bound on $T_{\rm kin}$ from Big Bang Nucleosynthesis; and the bound from the ADMX experiment excluding a KSVZ axion with a mass $m_a$ between 1.9~$\,{\rm\mu eV}$ and 3.3~${\rm\mu eV}$. The dashed line marks the requirement that the axion starts to oscillate in the kination cosmology, $T_{\rm kin} < T_1^{\rm std}$, with $T_1^{\rm std}$ given by Eq.~(\ref{T_1}).

The PQ scale $f_a^{\rm kin}$ is orders of magnitude lower than the PQ scale $f_a^{\rm std}$ in the standard cosmology. The low values of $f_a^{\rm kin}$ in comparison with $f_a^{\rm std}$ is due to two different reasons. The first reason is that, since coherent oscillations of the axion field start later in the kination cosmology than in the standard cosmology, the initial comoving number of axions $N^{\rm kin}$ is higher than $N^{\rm std}$. The second reason is that the contribution from axionic strings to $\Omega_a^{\rm kin}$ in the kination cosmology is much higher than the same contribution to $\Omega_a^{\rm std}$ in the standard cosmology. Then, at a given PQ scale $f_a$, the energy density $\Omega_a^{\rm kin} > \Omega_a^{\rm std}$. A lower PQ scale is thus required in order to have the same CDM energy density $\Omega_{\rm CDM}$.

The PQ scale $f_a^{\rm kin}$ can be so small as to violate the limit from the white dwarfs cooling time in Eq.~(\ref{astrophysical_bound}). This imposes the requirement $T_{\rm kin} > 217 {\rm ~MeV}$ if axions are 100$\%$ of the CDM. This requirement is more stringent than the BBN constraint $T_{\rm kin} > 4 {\rm ~MeV}$.

In Scenario II ($f_a>H_I/2\pi$), the parameter space is bounded by the nondetection of axion isocurvature fluctuations in the CMB spectrum, Eq.~(\ref{adiabaticity}). For $T_{\rm kin} > T_1^{\rm std}$, the isocurvature bound has the same expression, Eq.~(\ref{leftmostboundary}), as in the standard cosmology. For $T_{\rm kin} < T_1^{\rm std}$, we eliminate $\theta_i$ in Eq.~(\ref{adiabaticity}) by equating $\Omega_{\rm CDM}$ with the expression for $\Omega_a^{\rm kin}$ derived in this section. The resulting kination isocurvature bound for $T_{\rm kin} < T_1^{\rm std}$ is
\begin{equation} \label{leftmostboundarykination}
H_{I,12} < 7.48\times 10^{-7}\,\sqrt{f(\theta_i)}\,f_{a,12}^{1/2}\,T_{\rm kin,MeV}^{1/2}.
\end{equation}
This bound can be approximated by
\begin{equation} \label{leftmostboundarykination1}
H_{I,12} = \begin{cases}
1.31\times 10^{-4}\,f_{a,12}, & \hbox {for $f_a < 3.26\times 10^7{\rm ~GeV}\,T_{\rm kin,MeV}$},\\
7.48\times 10^{-7}\,f_{a,12}^{1/2}\,T_{\rm kin,MeV}^{1/2}, & \hbox{for $f_a > 3.26\times 10^7{\rm ~GeV}\,T_{\rm kin,MeV}$}.
\end{cases}
\end{equation}
Contrary to the cases of standard and LTR cosmologies, in the kination cosmology, there is only one change in the power-law dependence of $H_{I,12}$ on $f_{a,12}$, namely at $f_a =  3.26\times 10^7{\rm ~GeV}\,T_{\rm kin,MeV}$. This change is due to the effects of anharmonicities.

When $T_{\rm kin} = T_1^{\rm std}$, the kination and the standard isocurvature bounds coincide. This happens for
\begin{equation}
\begin{cases}
f_a = 4.6\times 10^{21}\,T_{\rm kin,MeV}^{-2}\,{\rm GeV}, & \hbox{for $T_1^{\rm std} < \Lambda_{\rm QCD}$},\\
f_a = 5.3 \times 10^{28}\,T_{\rm kin,MeV}^{-6}\,{\rm GeV}, & \hbox{for $T_1^{\rm std} > \Lambda_{\rm QCD}$}.
\end{cases}
\end{equation}

\begin{figure}[tb]
\begin{center}
  \includegraphics[width=13cm]{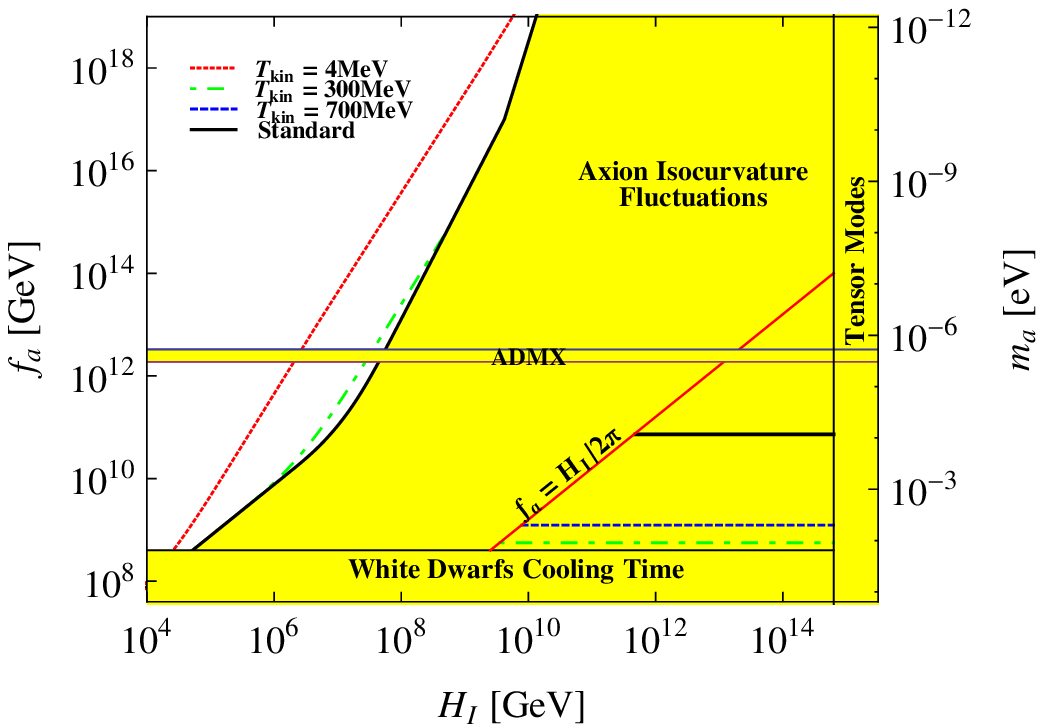}
\caption[The parameter space where the axion is 100\% CDM in the kination cosmology.]{In the kination cosmology, axions are 100$\%$ of the CDM in the white region on the left (limited by a different line for each $T_{\rm kin}$) and in the narrow bands marked by horizontal lines in the lower right triangle (one line for each $T_{\rm kin}$).}
\label{kination}
\end{center}
\end{figure}

\begin{figure}[b!]
\begin{center}
  \includegraphics[width=10cm]{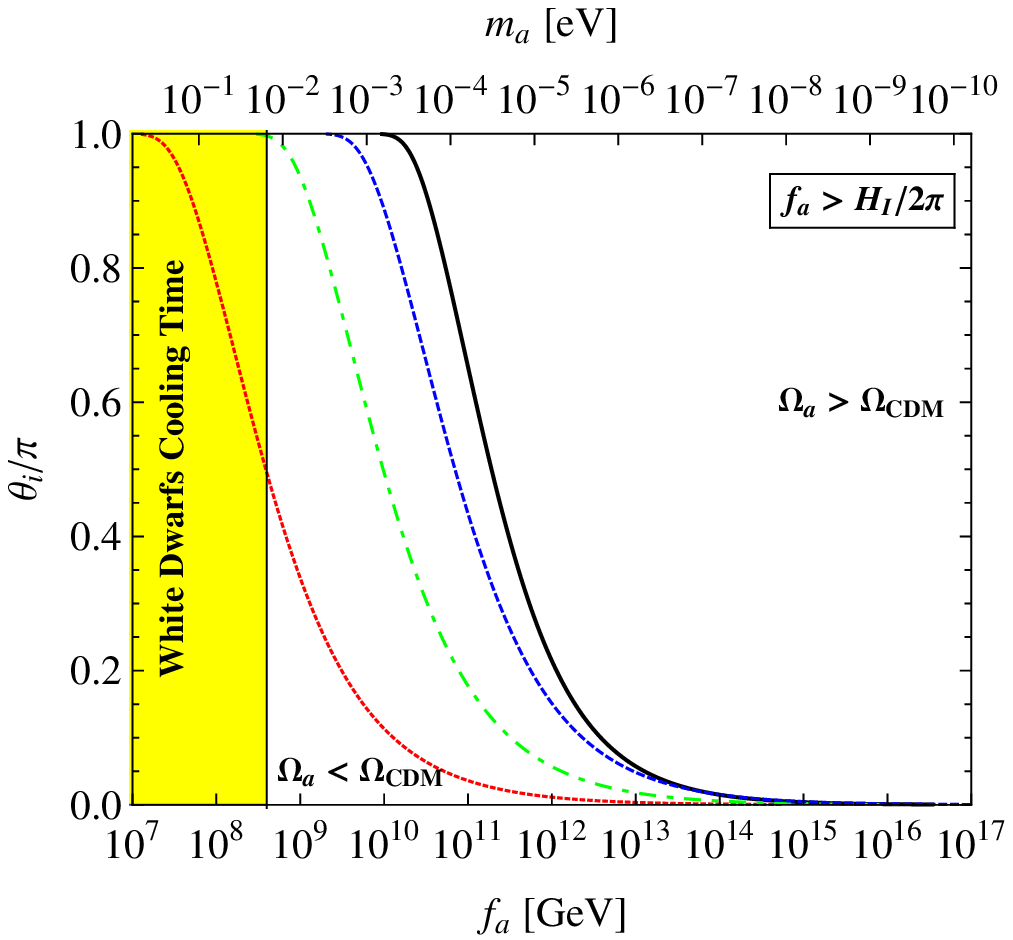}
\caption[$\theta_i$ as a function of $f_a$ in the kination cosmology.]{The initial misalignment angle $\theta_i$ as a function of the Peccei-Quinn scale $f_a$ for the axion to be 100$\%$ of the CDM in Scenario II ($f_a>H_I/2\pi$): standard cosmology (black solid line), kination cosmology with $T_{\rm kin} = 4$MeV (red dotted line), 300MeV (green dot-dashed line), or 700MeV (blue dashed line).}
\label{Thetakination}
\end{center}
\end{figure}

Fig.~\ref{kination} shows the regions of the parameter space $(f_a,H_I)$ where the axion is 100$\%$ of the CDM in the kination cosmology. The axion mass scale on the right is Eq.~(\ref{eq:axionmass}) with $N=1$. The region labeled ``Tensor Modes'' is excluded by the nonobservation of tensor modes in the CMB fluctuations, Eq.~(\ref{HI_bound}). The region labeled ``White Dwarfs Cooling Time'' is excluded from the bound in Eq.~(\ref{astrophysical_bound}). The line $f_a = H_I/2\pi$ divides the region where the PQ symmetry breaks after inflation  (Scenario I) from the region where it breaks during inflation  (Scenario II). In Scenario I, the axion is the CDM particle if $f_a$ equals the value $f_a^{\rm kin}$ in Eq.~(\ref{naturalscale_kination}). In Scenario II, we plot the isocurvature bounds to the allowed parameter space for the standard cosmology (thick line) and for $T_{\rm kin} = 4{\rm ~MeV}$ (dotted line), 300 MeV (dot-dashed line), and 700 MeV (dashed line). In the allowed region of parameter space for Scenario II, the axion is 100$\%$ CDM if
\begin{equation} \label{fv.theta_kination}
f_{a,12} = \frac{\Omega_{\rm CDM}h^2\,g_*^{1/2}(T_{\rm kin})\,T_{\rm kin,MeV}}{1150\,\theta_i^2\,f(\theta_i)}.
\end{equation}
We plot this relation between $f_a$ and $\theta_i$ in Fig.~\ref{Thetakination} for the standard cosmology (thick line) and for $T_{\rm kin} = 4{\rm ~MeV}$ (dotted line), $300 {\rm ~MeV}$ (dot-dashed line) and $700 {\rm ~MeV}$ (dashed line). As $T_{\rm kin}$ decreases, one departs from the standard cosmology. The value of $\theta_i$ at fixed $f_a$, or of $f_a$ at fixed $\theta_i$, decreases when $T_{\rm kin}$ decreases.

\section{Discussion} \label{Discussion}

\subsection{Comparison to previous work}

Axions in kination cosmology were studied only in Ref.~\cite{Grin:2007yg} and only as hot dark matter (i.e.\ thermally produced in the hot primordial soup). To the extent of our knowledge, CDM axions in kination cosmology were not examined before.

Axions in the LTR cosmology were studied before \cite{Dine:1982ah, Steinhardt:1983ia, Turner:1983he, Scherrer:1984fd, Lazarides:1987zf, Yamamoto:1985mb, Kawasaki:1995cy, Kawasaki:1995vt, Giudice:2000ex}, but only to determine the cosmological bound on the PQ scale in what we call Scenario I, namely $f_a<H_I/2\pi$, in which the Peccei-Quinn symmetry breaks after the end of inflation. Our work can also be used to set an upper bound on the PQ scale by imposing
\begin{equation}
\Omega_a\,h^2 < \Omega_{CDM}\,h^2.
\end{equation}
These bounds can be read off the figures and the equations we presented previously, all of which represent the equation $\Omega_a\,h^2 < \Omega_{CDM}\,h^2$. We remark that therefore, our work extends previous papers in that we have examined also the region $f_a>H_I/2\pi$, where the PQ symmetry breaks during inflation, have updated $\Omega_{\rm CDM}$ to the current observational value, have used an improved constraint on $T_{\rm RH}$ from Big Bang Nucleosynthesis, and have included anharmonicities in the axion potential.

Our numerical result for the highest allowed value of the PQ scale in the LTR cosmology $f_a^{\rm LTR} = 8.58\times 10^{13} {\rm ~GeV}$, obtained for $T_{\rm RH} = 4{\rm ~MeV}$, differs from previous authors. Steinhardt and Turner \cite{Steinhardt:1983ia} showed that the entropy production due to the decay of a massive scalar field raises the maximum PQ scale to $f_a^{\rm LTR} \sim 10^{18} {\rm ~GeV}$, but they were corrected by Kawasaki {\it et al.} \cite{Kawasaki:1995cy, Kawasaki:1995vt} for using the value of $T_1$ in the standard cosmology instead of the LTR cosmology. Kawasaki {\it et al.} \cite{Kawasaki:1995cy, Kawasaki:1995vt} used $T_1$ in the LTR cosmology and obtained $f_a^{\rm LTR} \sim 10^{15} {\rm ~GeV}$; however, they used $\Omega_a\,h^2 = 1$ and $T_{\rm RH} = 1{\rm ~MeV}$ instead of the current value $\Omega_{\rm CDM}h^2 = 0.1131\pm 0.0034$ and the current BBN bound $T_{\rm RH}>4{\rm ~MeV}$. Giudice {\it et al.} \cite{Giudice:2000ex} find for the maximum PQ scale in the LTR cosmology the value $f_a^{\rm LTR} \sim 10^{16} {\rm ~GeV}$, which is higher than ours for the same reasons as for Kawasaki {\it et al.}

\subsection{Effects of changing the string decay parameters}

The computation of $f_a^{\rm std}$ in Eq.~(\ref{PQ_strength}), $f_a^{\rm LTR}$ in Eq.~(\ref{naturalscale_LTR}), and $f_a^{\rm kin}$ in Eq.~(\ref{naturalscale_kination}) strongly relies on the model used to describe the axionic string evolution and the energy spectrum of emitted axions.

In Sec.~\ref{Axions from string decays}, we discussed the dependence of $\bar{r}$ and $\xi$ on the model for the axionic string oscillation and radiation spectrum. There, we showed how these quantities in a modified cosmological scenario are related to their values in the standard cosmology. For the latter, we used the values $\bar{r}^{\rm std} = 0.8$ and $\xi^{\rm std} = 1$, obtained assuming that an axionic string radiates axions in a broad energy spectrum and that the axionic string network is a global string network \cite{Sikivie:1982qv, Chang:1998tb, Yamaguchi:1998gx, Hagmann:1990mj, Hagmann:2000ja}. We have referred to this set of assumptions as the HHCS model. With the values of the axionic string parameters set by the HHCS model, the contribution from axionic string to the total axion energy density in the standard and LTR cosmologies is subdominant compared to the contribution from the misalignment mechanism, while it is dominant in the kination cosmology.

We now discuss the modification to the axion parameter space when we assume that axionic strings radiate axions in a narrow energy spectrum \cite{Davis:1985pt, Davis:1986xc, Davis:1989nj, Battye:1993jv, Battye:1994au, Shellard:1998mi}, and that the axionic strings network is a local strings network \cite{Bennett:1989yp, Allen:1990tv}. In this case, $\bar{r}^{\rm std} = 70$ and $\xi^{\rm std} = 13$. We call this set of assumptions the DSB model. With these values, the contribution from strings to the axion energy density in the standard cosmology is dominant ($N_d=1$, $\zeta = 4.9$),
\begin{equation}
\Omega_a^{\rm std,str} \sim 200\,\Omega_a^{\rm std,mis}.
\end{equation}
This affects the value of $\Omega_a^{\rm std} = \Omega_a^{\rm std,str} + \Omega_a^{\rm std,mis}$, which in the DSB model is about two hundred times higher than in the HHCS model. As a consequence, within the DSB model, the value of the PQ scale $f_a^{\rm std}$ for which $\Omega_a^{\rm std} = \Omega_{\rm CDM}$ in Scenario I is
\begin{equation}
f_a^{\rm std} = (4.3\pm 0.7) \times 10^8 {\rm ~GeV}.
\end{equation}
This value is smaller than that in Eq.~(\ref{PQ_strength}), obtained within the HHCS model. It is of the order of the astrophysical constraint from white dwarfs cooling times in Eq.~(\ref{astrophysical_bound1}).

We conclude that, depending on the model for the axionic string and its emission spectrum, $f_a^{\rm std}$ in the standard cosmology can range from the value in Eq.~(\ref{PQ_strength}) to the astrophysical bound from white dwarfs cooling times $4\times 10^{8}{\rm ~GeV}$. Correspondingly, $m_a^{\rm std}$ can range from $67\pm17{\rm ~\mu eV}$ to $15 \pm 4{\rm ~meV}$.

In nonstandard cosmologies, when going to the DSB model, we must redo the calculation of $\bar{r}$ and $\xi$ using the formulas in Sec.~\ref{Axions from string decays}. For the DSB model value $\xi^{\rm std} = 13$, Eq.~(\ref{c}) gives $c =(\sqrt{52}+1)/52 \simeq 0.158$ and Eq.~(\ref{gamma_0}) gives
\begin{equation}
\xi \sim 10\left(2-3\beta+2\sqrt{2.41\beta^2-3\beta+1}\right)^2\quad\hbox{(for $\xi^{\rm std} = 13$)}.
\end{equation}
The values of $\xi$ in the DSB model are then $\xi^{\rm LTR} = 2.8$ for the LTR cosmology ($\beta = 2/3$)and $\xi^{\rm kin} = 41$ for the kination cosmology ($\beta = 1/3$).

For the parameter $\bar{r}$, we turn to Eq.~(\ref{bar_r1}). The results from the HHCS model favor the fast-oscillating axionic strings model, which predicts $\bar{r}$ as given in Eq.~(\ref{bar_r_broad}). The DSB model points toward a slow-oscillating axionic string for which $\bar{r}$ is given by Eq.~(\ref{bar_r_sharp}). Within the DSB model, for the illustrative case $\delta = (10^{12}{\rm ~GeV})^{-1}$ and $T_{\rm RH} = T_{\rm kin} = 4\,{\rm MeV}$, we obtain $\bar{r} \simeq 70$ in the standard cosmology, $\bar{r}^{\rm LTR} \simeq 20$ in the LTR cosmology, and $\bar{r}^{\rm kin} \simeq 4300$ in the kination cosmology.

The LTR and kination axion energy densities from axionic strings in the DSB model are then, with $N_d = 1$ and $\zeta = 4.9$,
\begin{equation}
\Omega_a^{\rm LTR,str}h^2 = 11.5\,\Omega_a^{\rm LTR,mis}\quad\hbox{and}\quad\Omega_a^{\rm kin,str}h^2 = 3.6\times10^4\,\Omega_a^{\rm kin,mis}.
\end{equation}
The higher axionic string contributions in the DSB model with respect to the HHCS model sensibly lower the values of the PQ scales $f_a^{\rm LTR}$ and $f_a^{\rm kin}$ for which the axion is 100$\%$ of the CDM. We have, taking $T_{\rm RH} = T_{\rm kin} = 4\,{\rm MeV}$,
\begin{equation}
f_a^{\rm LTR} = (1.43\pm 0.04)\times 10^{13}{\rm ~GeV}\quad\hbox{and}\quad f_a^{\rm kin} = (4.3 \pm 0.1)\times 10^3{\rm ~GeV}.
\end{equation}

\subsection{Distinguishing nonstandard cosmologies observationally}

One might be able to distinguish different nonstandard cosmologies before BBN by using properties of the axion CDM particle.

Future CMB measurements in the tensor modes spectrum from the PLANCK satellite may help with this task in the following way. Suppose that the cosmology is standard and the axion is 100$\%$ CDM. A detection of tensor modes within the PLANCK range would yield the value of $H_I$, and thus that of $m_a$ \cite{Visinelli:2009zm, Hamann:2009yf}. If the axion mass is found to be different from the one derived from the PLANCK measurement, then either the axion is not the CDM particle and the cosmology is standard or the axion is the CDM particle and the cosmology is nonstandard.

PLANCK will also improve the current bounds on axion isocurvature fluctuations, or detect them. If the axion is the CDM particle, then a combination of the measurements of the axion mass and of the axion isocurvature fluctuations yields a point in the $H_I-f_a$ plane, which might be excluded in the standard cosmological scenario but not in the LTR scenario. Combining the axion searches and the PLANCK measurements thus might lead to new information on the pre-BBN history of the universe.

Outside CMB measurements, one may try to distinguish nonstandard cosmologies by measuring both the axion CDM density $\Omega_{\rm CDM}$ and the axion mass $m_a$. However, one immediately runs into the following problem.

Assume, for example, that the axion is found to be the main CDM component and the axion mass is measured at $m_a \simeq 10^{-3}{\rm ~eV}$. These facts can be ascribed to two different cosmological models. The first model involves the axion field evolving in the standard cosmology, with the dominant contribution to the total axion energy density coming from axionic strings and only a tiny fraction from the misalignment mechanism, as the DSB model would predict. The second model involves a stage of kination before BBN lasting until $T_{\rm kin} \sim 900 {\rm ~MeV}$, with the contribution from axionic strings and from the misalignment mechanism of the same order of magnitude, as in the HHCS model.

These uncertainties in the production of axion from strings decay prevent distinguishing nonstandard cosmologies with this method alone.

One may complement the measurements of $\Omega_{\rm CDM}$ and $m_a$ with a measurement of the axion CDM velocity dispersion $\delta v$. The latter allows nonstandard cosmologies to be distinguished, at least in principle. The argument proceeds as follows.

When axions start to oscillate at the temperature $T_1$, axions from vacuum realignment and axionic string decay have a momentum dispersion of the order of the Hubble scale at $T_1$ \cite{Sikivie:2006ni},
\begin{equation} \label{momentum_dispersion}
\delta p(T_1) \simeq H(T_1).
\end{equation}
The momentum dispersion scales with the scale factor as $\delta p(T) \propto 1/a(T)$. In the standard cosmology, the velocity dispersion at present is then
\begin{equation}
\delta v^{\rm std} \simeq \frac{H(T_1^{\rm std})}{m_a}\left(\frac{a(T_1^{\rm std})}{a(T_0)}\right) = \left(\frac{\rm\mu eV}{m_a}\right)^{5/6}\,1.4\times10^{-8}{\rm ~m/s}.
\end{equation}

In the kination cosmology,
\begin{equation}
\delta v^{\rm kin} \simeq \frac{H(T_1^{\rm kin})}{m_a}\left(\frac{a(T_1^{\rm kin})}{a(T_0)}\right) = T_{\rm kin,MeV}^{-5/7}\,\left(\frac{\rm\mu eV}{m_a}\right)^{5/7}\,8.5\times 10^{-7}{\rm ~m/s}.
\end{equation}
It is clear that if one has measured $m_a$, a measurement of $\delta v$ will give the value of $T_{\rm kin}$.

Similarly, in the LTR cosmology,
\begin{equation}
\delta v^{\rm LTR} \simeq \frac{H(T_1^{\rm LTR})}{m_a}\left(\frac{a(T_1^{\rm LTR})}{a(T_0)}\right) = \left(\frac{\rm\mu eV}{m_a}\right)^{5/6}\,7.6\times 10^{-9}{\rm ~m/s}.
\end{equation}

A difficulty in measuring $\delta v$ may arise from virialization of the axion population within galactic dark halos, although it has been claimed that $\delta v$ would be preserved in the phase-space evolution \cite{Sikivie:2006ni}.

\section{Conclusions} \label{Conclusions}

In this section we have examined the parameter regions in which the axion is 100\% of the CDM density in cosmologies that are nonstandard before Big Bang nucleosynthesis. We have recognized two ways in which these regions change in going from the standard cosmology to the nonstandard cases. If the Peccei-Quinn symmetry breaks after the end of inflation (Scenario I), the axion CDM regions shift to different values of the axion mass $m_a$ (or of the corresponding PQ scale $f_a$). If the PQ symmetry breaks during inflation (Scenario II), the axion CDM regions can shrink or expand according to the cosmological model.

We have considered two different nonstandard cosmologies that change the axion CDM regions in opposite directions. In the LTR cosmology, the axion CDM regions shift to lower axion masses in Scenario I and expand in Scenario II. In the kination cosmology, the axion CDM regions shift to higher axion masses in Scenario I and shrink in Scenario II.

Different axionic string models lead to different quantitative results, but the overall modifications from the standard cosmology follow the same trend.

We have also commented on the possibility to distinguish standard and nonstandard cosmologies using observable properties of the axion CDM population. We have tentatively concluded that the axion velocity dispersion may be a good indicator of the cosmology before Big Bang nucleosynthesis.

%% file: chap7.tex
\chapter{Natural warm inflation}\label{Natural warm inflation}

In this chapter, we turn our attention to the more generic family of axion-like particles. Throughout this chapter, we consider the role of the axion-like field in the inflationary paradigm; in particular, we assume that the inflaton field responsible for the primordial inflationary period of the universe is an axion-like field. We carry out our computation in the context of the warm inflation scenario.

\section{Motivations}

In Sec.~\ref{inflation}, we have discussed the reasons that led to the formulation of the inflationary theory and its embedding in the standard model of cosmology. Briefly, inflation \cite{Kazanas:1980tx, Starobinsky:1980te, Guth:1980zm, Sato:1981ds, Albrecht:1982wi, Linde:1981mu} provides a mechanism for generating the inhomogeneities observed in the CMBR \cite{Mukhanov:1981xt, Guth:1982ec, Hawking:1982cz, Starobinsky:1982ee, Bardeen:1983qw}, and explains the observed flatness, homogeneity, and the lack of relic monopoles that posed severe problems in the standard Big-Bang cosmology \cite{linde1990particle, kolb1990early}. Realistic microphysical models of inflation, in which the expansion of the universe is governed by the energy density of the inflaton field $\phi$, are complicated by the requirement  that the inflaton potential $U(\phi)$ be very flat in order to explain the anisotropies observed in the CMBR \cite{Adams:1990pn}.

Natural Inflation \cite{Freese:1990rb, Adams:1992bn, Savage:2006tr} is a viable model in which the inflaton is identified with an axion-like particle: in fact, the shift symmetry $\phi \to \phi$ + const. present in axionic theories assures a flat inflaton potential. Although Natural Inflation is well-motivated and it is consistent with the WMAP measurements \cite{Savage:2006tr}, it is not an easy task to embed this model in fundamental theories like string theory \cite{Banks:2003sx}, the main complication coming from the fact that the energy scale $f$ at which the shift symmetry spontaneously breaks must be $f>0.6~M_{\rm Pl}$, in order to agree with the constraints on the scalar spectral index $n_s$.

In this section, we show that the energy scale $f$ for axion-like particles can be as low as the Grand Unification Theory (GUT) scale $\Lambda_{\rm GUT}~\sim~10^{16}{\rm~GeV}$, if Natural Inflation is considered in the context of warm inflation (\cite{Berera:1995ie, Berera:1995wh}; see also Refs.~\cite{Hosoya:1983ke, Moss:1985wn, Lonsdale:1987kn, Yokoyama:1987an, Liddle:1988tb}); we refer to this as the Natural Warm Inflation (NWI) model.

Since axion-like particles arise in generic four-dimensional models \cite{Masso:1995tw, Masso:1997ru, Coriano:2006xh, Coriano:2007fw}, string theory compactifications \cite{Svrcek:2006yi}, and generic Kaluza-Klein theories \cite{Chang:1999si}, and possess attractive features for inflation models like a flat potential already embedded in the theory, such particles have been extensively discussed in the inflation literature \cite{Kim:2004rp, Dimopoulos:2005ac, Mohanty:2008ab, McAllister:2008hb, Kaloper:2008fb, Anber:2009ua, Kaloper:2011jz}. In particular, other attempts at lowering the Natural Inflation scale $f$ within the warm inflation scenario have been discussed in Refs.~\cite{Mohanty:2008ab, Mishra:2011vh}.

This chapter is organized as follows. We first fix our notation for the warm inflation scenario and for the axion particle physics in Sec.~\ref{The warm inflation scenario} and~\ref{Axion-like particles}, respectively. In Sec.~\ref{Warm natural inflation}, we analyze the dynamic of the inflaton field using the slow-roll conditions and number of e-folds for sufficient inflation, and we study the parameter space of our NWI model. In Sec.~\ref{Perturbations from inflation}, we discuss the bounds on the NWI model resulting from the measurements of cosmological parameters, in light of the WMAP7+BAO+SN data \cite{Komatsu:2010fb}. Finally, discussions and conclusions are drawn in Sec.~\ref{discussion}.

\section{The warm inflation scenario}\label{The warm inflation scenario}

In the warm inflation scenario, the inflaton field appreciably converts into relativistic matter (from here on referred to as ``radiation'') during the inflationary period. This mechanism is parametrized by the appearance of a dissipative term $\Gamma$ in the dynamics of the inflaton field. In the following, we reasonably assume that radiation thermalizes on a time scale much shorter than $1/\Gamma$ \cite{Berera:1995ie, Berera:1995wh}.

The energy density in radiation is given by (see Sec.~\ref{Application to the radiation-dominated universe})
\begin{equation}\label{definition_radiation}
\rho_r = \frac{\pi^2}{30}\,g_*(T)\,T^4,
\end{equation}
where $g_*(T)$ is the number of relativistic degrees of freedom of radiation at temperature $T$. Here we do not specify $g_*(T)$ in the equations, but in the figures we will always use $g_*(T) = 228.75$, corresponding to the number of relativistic degrees of freedom in the MSSM.

Warm inflation is achieved when thermal fluctuations dominate over quantum fluctuations, or \cite{Berera:1995ie, Berera:1995wh}
\begin{equation} \label{condition_warm_inflation}
H(T) < T,
\end{equation}
where $H(T)$ is the Hubble expansion rate at temperature $T$. The effectiveness at which the inflaton converts into radiation is measured by the ratio
\begin{equation} \label{effectiveness}
\mathcal{Q} = \frac{\Gamma}{3H};
\end{equation}
for $\mathcal{Q} \gg 1$ a strongly dissipative regime is achieved, while $\mathcal{Q} <1$ represents the weak regime of warm inflation. Throughout this section, we will present the general equations for warm inflation, focusing on the case $\mathcal{Q} \gg 1$ in the subsequent sections when specified.

In the following, we model the inflaton field with a scalar field $\phi = \phi(x)$ minimally coupled to the curvature and moving in a potential $U = U(\phi)$. The evolution of the inflaton field in a Friedmann-Robertson-Walker metric is described by
\begin{equation} \label{eq_motion}
\ddot{\phi} + (3H+\Gamma)\dot{\phi} + U_\phi = 0,
\end{equation}
where a dot indicates the derivation with respect to the cosmic time $t$ and $U_\phi = \partial U/\partial\phi$. Imposing the conservation of the total energy of the system, we find that the radiation energy density $\rho_r$ satisfies
\begin{equation} \label{energy_conservation_radiation}
\dot{\rho_r} + 4H\rho_r = \Gamma\,\dot{\phi}^2,
\end{equation}
with the term on the RHS of Eq.~(\ref{energy_conservation_radiation}) describing the effectiveness of conversion of the inflaton field into radiation.

The total energy density of the system at any time is $\rho_{\rm tot} = \dot{\phi}^2/2 + U(\phi, T)$, where $U(\phi,T)$ is an effective potential that accounts for temperature effects. As discussed in Ref.~\cite{Moss:2008yb}, a requirement for building a consistent model of warm inflation is that finite temperature effects on the inflaton potential must be suppressed; this makes it possible to separate the effective potential of the inflaton as $U(\phi, T) = U(\phi) + \rho_r(T) = U + \rho_r$.

The total energy density during the warm inflation period is
\begin{equation}
\rho_{\rm tot} = \frac{1}{2}\dot{\phi}^2 + U + \rho_r,
\end{equation}
and the corresponding Friedmann Eq.~(\ref{friedmann}) reads
\begin{equation} \label{friedmann_warm}
H^2 = \frac{8\pi}{3M_{\rm Pl}^2}\left(\frac{1}{2}\dot{\phi}^2+U+\rho_r\right).
\end{equation}
Inflation takes place when the potential $U$ is approximately constant and dominates over the other forms of energy, assuring the Hubble expansion rate $H$ to be constant. These conditions are generically imposed by requiring that the inflaton potential and the dissipation term satisfy slow-rolls conditions, which in the warm inflation scenario read \cite{Taylor:2000ze}
\begin{equation} \label{slow_roll}
\epsilon \ll 1+\mathcal{Q},\quad |\eta| \ll 1+\mathcal{Q},\quad |B| \ll 1+\mathcal{Q}.
\end{equation}
Here, we have introduced the slow-roll parameters
\begin{equation}\label{slow_roll_parameters}
\epsilon = \frac{1}{16\pi G}\left(\frac{U_\phi}{U}\right)^2, \quad \eta = \frac{1}{8\pi G}\,\frac{U_{\phi\phi}}{U},\quad B = \frac{1}{8\pi G}\left(\frac{\Gamma_{\phi}\,U_\phi}{\Gamma\,U}\right)
\end{equation}
Notice that these conditions reduce to the usual slow-roll requirements in the cool inflation, where $\mathcal{Q} \ll 1$.

During the slow-roll regime, the higher derivatives in Eqs.~(\ref{eq_motion}) and~(\ref{energy_conservation_radiation}) can be neglected. So, Eq.~(\ref{eq_motion}) in the slow-roll regime reads
\begin{equation} \label{eq_motion_slow_roll}
\dot{\phi} \simeq -\frac{U_\phi}{3H+\Gamma},
\end{equation}
where here and in the following we use the symbol ``$\simeq$'' for an equality that holds only in the slow-roll regime. Eqs.~(\ref{energy_conservation_radiation}) and~(\ref{friedmann_warm}) in the slow-roll regime, respectively, read
\begin{equation} \label{energy_conservation_sl}
\rho_r \simeq \frac{3\mathcal{Q}}{4}\,\dot{\phi}^2,
\end{equation}
and
\begin{equation} \label{friedmann_sl}
H^2 \simeq \frac{8\pi G}{3}\,U,
\end{equation}
from which we see that a shallow potential $U$ gives rise to a nearly constant expansion rate $H$. Notice that we have used the fact that $U \gg \rho_r$, which can be combined with the definition of $\rho_r$ in Eq.~(\ref{definition_radiation}), the requirement for a warm inflation regime in Eq.~(\ref{condition_warm_inflation}), and the Friedmann Eq.~(\ref{friedmann_sl}), to yield the constraint
\begin{equation}
U^{1/4} \ll \left(\frac{135}{32\,g_*(T)\,\pi^4}\right)^{1/4}\,M_{\rm Pl} = 5.57\,g_*^{-1/4}(T) \times 10^{18}{\rm~GeV}.
\end{equation}
This is not a stringent bound, since in most theories of grand unification the inflaton potential is related to the unification scale, $U^{1/4} \sim 10^{16}$ GeV.

\section{Axion-like particles}\label{Axion-like particles}

Axion-like particles are pseudo-scalars that arise naturally whenever an approximate global symmetry is spontaneously broken. Examples include generic four-dimensional models \cite{Masso:1995tw, Masso:1997ru, Coriano:2006xh, Coriano:2007fw}, string theory compactifications \cite{Svrcek:2006yi}, and generic Kaluza-Klein theories \cite{Chang:1999si}. Contrary to the model of the invisible axion, in which the mass $m_a$ and the energy scale $f_a$ of the invisible axion are related by Eq.~(\ref{eq:axionmass}), in a generic axion-like theory, the mass $m_\phi$ of the pseudo-scalar particle and the decay constant $f$ appearing in the model are not linked, and are treated as two independent variables. However, it is customary to introduce a scale $\Lambda$ that relates to the energy scale of the underlying theory as
\begin{equation}\label{def_lambda}
\Lambda = \sqrt{m_{\phi}\,f}.
\end{equation}
This definition mimics that appearing in the invisible axion theory, in which $\sqrt{f_a\,m_a}$ is of the order of  $\Lambda_{\rm QCD}$. In some specific models, the value of $\Lambda$ for pseudo-scalar particles can be predicted and can range up to the GUT scale $\Lambda_{\rm GUT}~\sim~10^{16}{\rm~GeV}$ or even the Planck scale $M_{\rm Pl}$. Here, we do not specify the value of $\Lambda$ and we keep $m_\phi$ and $f$ as two distinct parameters.

Axion-like particles move in the potential, see Eq.~(\ref{axion_potential}),
\begin{equation} \label{potential_axion_like}
U = U(\phi) = \Lambda^4 \left[1 + \cos\left(N\,\frac{\phi}{f}\right)\right],
\end{equation}
with $N$ integer. It is well known that axion-like particles serve as suitable candidates for the inflaton field because the potential $U(\phi)$ is naturally flat whenever $\phi/f \ll \pi/N$. Moreover, as we will show in Sec.~\ref{Scalar power spectrum},  the self-interaction term derived from $U(\phi)$,
\begin{equation}\label{self_interaction}
\lambda_\phi = \frac{m_\phi^2}{24f^2},
\end{equation}
fulfills the requirement  from observations \cite{Adams:1990pn}, $\lambda_\phi \lesssim 10^{-8}$.

We assume that axion-like particles couple to radiation with a term similar to the axion-photon coupling in Sec.~\ref{Coupling of axions to photons}, with coupling constant $g_F$. The decay rate of axion-like particles into radiation in this model is, see Sec.~\ref{Lifetime of the axion},
\begin{equation}\label{decay_constant}
\Gamma =g_F^2\,\frac{m_\phi^3}{f^2}.
\end{equation}
This dissipative term does not depend on the axion-like field $\phi$ and we have $\Gamma_\phi = B =0$.

\section{Warm natural inflation} \label{Warm natural inflation}

From now on, we identify the axion-like particle $\phi$ with the inflaton, using the flat potential in Eq.~(\ref{potential_axion_like}) to describe the dynamics for the inflaton field in the early universe. Here, we will take $N = 1$ in Eq.~(\ref{potential_axion_like}), so that the potential has a unique minimum at $\phi = \pi f$; the inflaton field relaxes towards this minimum in its evolution. In the following, we indicate the inflaton with $\phi$, bearing in mind that we have assumed that $\phi$ is also a pseudo-scalar particle for which the formulas in Sec.~\ref{Axion-like particles} apply.

Furthermore, from here on, we are interested only in the strongly dissipative regime of warm inflation, $Q \gg 1$. In this limit, the only parameters in the theory are the inflaton mass $m_\phi$, the decay constant $f$, and the dissipation term $\Gamma$. In Table~7.1, we have collected the most important quantities of the theory in terms of these parameters

\begin{table}[h!]
\caption{Expressions for some derived quantities in the NWI theory, valid during slow-roll and $\mathcal{Q} \gg1$.}
\begin{center}
\begin{tabular}{ll}
{\bf Quantity} \hspace{4em} & {\bf Equation(s) used} \hspace{12em}\\
\hline\\
$U(\phi) = m_\phi^2\,f^2\,(1+\cos\phi/f),$ & Eq.~(\ref{potential_axion_like});\\
$U_\phi(\phi) = -m_\phi^2\,f\,\sin\phi/f,$ & derived from Eq.~(\ref{potential_axion_like});\\
$H \simeq \frac{m_\phi\,f}{M_{\rm Pl}}\,\sqrt{\frac{8\pi}{3}\,(1+\cos\phi/f)},$ & Eqs.~(\ref{friedmann_sl}) and~(\ref{potential_axion_like});\\
$\dot{\phi} \simeq \frac{m_\phi^2\,f}{\Gamma}\,\sin\phi/f,$ & Eqs.~(\ref{eq_motion_slow_roll}),~(\ref{friedmann_sl}) and~(\ref{potential_axion_like});\\
$\mathcal{Q} \simeq \frac{\Gamma\,M_{\rm Pl}}{m_\phi\,f}\,\sqrt{\frac{1}{24\pi(1+\cos\phi/f)}},$ & Eqs.~(\ref{potential_axion_like}) and~(\ref{effectiveness_sl});\\
$\rho_r \simeq \sqrt{\frac{3}{128\pi}}\frac{m_\phi^3\,f\,M_{\rm Pl}}{\Gamma}\,\frac{\sin^2\phi/f}{\sqrt{1+\cos\phi/f}},$ & Eqs.~(\ref{eq_motion_slow_roll}),~(\ref{energy_conservation_sl}),~(\ref{friedmann_sl}) and~(\ref{potential_axion_like});\\
$T^4 \simeq \sqrt{\frac{675}{32\pi^5}}\frac{m_\phi^3\,f\,M_{\rm Pl}}{\Gamma\,g_*(T)}\,\frac{\sin^2\phi/f}{\sqrt{1+\cos\phi/f}},$ & Eqs.~(\ref{definition_radiation}),~(\ref{eq_motion_slow_roll}),~(\ref{energy_conservation_sl}),~(\ref{friedmann_sl}) and~(\ref{potential_axion_like}).\\
$\epsilon = \frac{1}{16\pi\,G\,f^2}\,\frac{\sin^2\phi/f}{(1+\cos\phi/f)^2},$ & Eqs.~(\ref{slow_roll_parameters}) and~(\ref{potential_axion_like}).\\
$\eta \simeq  -\frac{1}{8\pi\,G\,f^2}\,\frac{\cos\phi/f}{1+\cos\phi/f},$ & Eqs.~(\ref{slow_roll_parameters}) and~(\ref{potential_axion_like}).\\
\\
\hline\\
\end{tabular}
\label{table1.1}
\end{center}
\end{table}
We now examine the constraints on the NWI model coming from the slow-roll conditions in Eq.~(\ref{slow_roll}), the requirement for sufficient inflation, and the WMAP measurements on the power spectrum of density and tensor perturbations.

\subsection{Slow-roll conditions}

During warm inflation, the values of the slow-roll parameters $\epsilon$, $|\eta|$ must be smaller than $1+\mathcal{Q}$, see Eq.~(\ref{slow_roll}); inflation ends when one of these two conditions is violated. In the following, we only consider the case in which $\epsilon \ll 1+ \mathcal{Q} \approx \mathcal{Q}$ is violated in the strongly dissipating regime. Writing the parameter $\mathcal{Q}$ during slow-roll as
\begin{equation} \label{effectiveness_sl}
\mathcal{Q} \simeq \frac{\Gamma}{\sqrt{24\pi G\,U}},
\end{equation}
the slow-roll condition $\epsilon \ll \mathcal{Q}$ reads
\begin{equation}
\epsilon = \frac{1}{16\pi G}\left(\frac{U_\phi}{U}\right)^2 \ll \frac{\Gamma}{\sqrt{24\pi G\,U}},
\end{equation}
or
\begin{equation} \label{slow_roll_epsilon}
\sqrt{\frac{3}{32\pi}}\,\frac{M_{\rm Pl}\,U_{\phi}^2}{\Gamma\,U^{3/2}} \ll 1.
\end{equation}
Using the expression for the potential $U(\phi)$ in Eq.~(\ref{potential_axion_like}), this slow-roll condition gives
\begin{equation} \label{slow_roll_epsilon1}
\frac{\sin^2\phi/f}{(1 + \cos\phi/f)^{3/2}} \ll \sqrt{\frac{32\pi}{3}}\frac{\Gamma\,f}{M_{\rm Pl}\,m_{\phi}}.
\end{equation}
Since the combination on the RHS of Eq.~(\ref{slow_roll_epsilon1}) appears frequently, we define
\begin{equation} \label{def_alpha}
\alpha \equiv \sqrt{\frac{32\pi}{3}}\frac{\Gamma\,f}{M_{\rm Pl}\,m_{\phi}} = 4.74\,\Gamma_{12}\,f_{16}\,m_{\phi \,9}^{-1}.
\end{equation}

The slow-roll regime ends when the field $\phi$ reaches a value $\phi_f$ for which the condition $\epsilon \ll \mathcal{Q}$ is no longer satisfied,
\begin{equation} \label{solution_slow_roll}
\cos(\phi_f/f) = \frac{2+\alpha^2 - \sqrt{\alpha^4 +8\alpha^2}}{2}.
\end{equation}
We use Eq.~(\ref{solution_slow_roll}) as a definition of $\phi_f$.

From Eq.~(\ref{slow_roll_epsilon1}), we see that it is always $\alpha >1$. When $\alpha \gg 1$, a Taylor expansion of Eq.~(\ref{solution_slow_roll}) yields
\begin{equation}  \label{definition_phi2}
\frac{\phi_f}{f}= \pi - \frac{\sqrt{8}}{\alpha}.
\end{equation}
Since observations favor a large value of $\alpha > 10$ (see below), here we use Eq.~(\ref{definition_phi2}) to obtain $\phi_f$.

\subsection{Number of E-folds}

The number of e-folds is defined as
\begin{equation} \label{number_efoldings}
N_e \equiv \ln(a_2/a_1) = \int_{t_1}^{t_2} H dt,
\end{equation}
where $a_1$ and $a_2$ are the values of the scale factor $a(t)$ appearing in the Friedmann metric when inflation begins and ends, respectively. Sufficient inflation requires
\begin{equation}
N_e > 60.
\end{equation}
During the inflationary stage, the value of the inflaton field decreases from the initial value $\phi_i$ to the value at the end of inflation $\phi_f$ defined via Eq.~(\ref{solution_slow_roll}). We now derive a relation between $\phi_i$ and $\phi_f$, using the definition of $N_e$ above. In the case of a slow-rolling of the inflaton and in the strongly dissipative regime, Eq.~(\ref{number_efoldings}) reads
\begin{equation} \label{number_efoldings1}
N_e \simeq -\int_{\phi_i}^{\phi_f}\frac{H\,\Gamma}{U_\phi}d\phi,
\end{equation}
where we used Eq.~(\ref{eq_motion_slow_roll}) with $\Gamma \gg 3H$. Using the axion-like potential in Eq.~(\ref{potential_axion_like}) and the Friedmann equation, we obtain
\begin{equation}
N_e = \sqrt{\frac{8\pi}{3}}\,\frac{\Gamma\,f}{m_\phi\,M_{\rm Pl}}\,\int_{\phi_i/f}^{\phi_f/f}dx\,\frac{\sqrt{1+\cos x}}{\sin x} = \frac{\alpha}{\sqrt{2}}\,\ln\,\tan \frac{x}{4}\,\bigg|_{\phi_i/f}^{\phi_f/f},
\end{equation}
or
\begin{equation} \label{relation_phi1_phi2}
\tan\frac{\phi_f}{4f}= \tan\frac{\phi_i}{4f}\,{\rm Exp}\left[\frac{\sqrt{2}\,N_e}{\alpha}\right].
\end{equation}
In general, we use Eqs.~(\ref{definition_phi2}) and~(\ref{relation_phi1_phi2}) to obtain the value of $\phi_i/f$, 
\begin{equation}\label{definition_phi1}
\frac{\phi_i}{f} = 4\arctan \left[\frac{1-\tan\frac{1}{\sqrt{2}\alpha}}{1+\tan\frac{1}{\sqrt{2}\alpha}}\,{\rm Exp}\left(-\frac{\sqrt{2}N_e}{\alpha}\right)\right]
\end{equation}
The appearance of trigonometric functions is due to the shape of the potential $U(\phi)$, that differs from a pure quadratic one and contains a cosine function itself.

\begin{figure}[b!]
\begin{center}
\includegraphics[width=13cm]{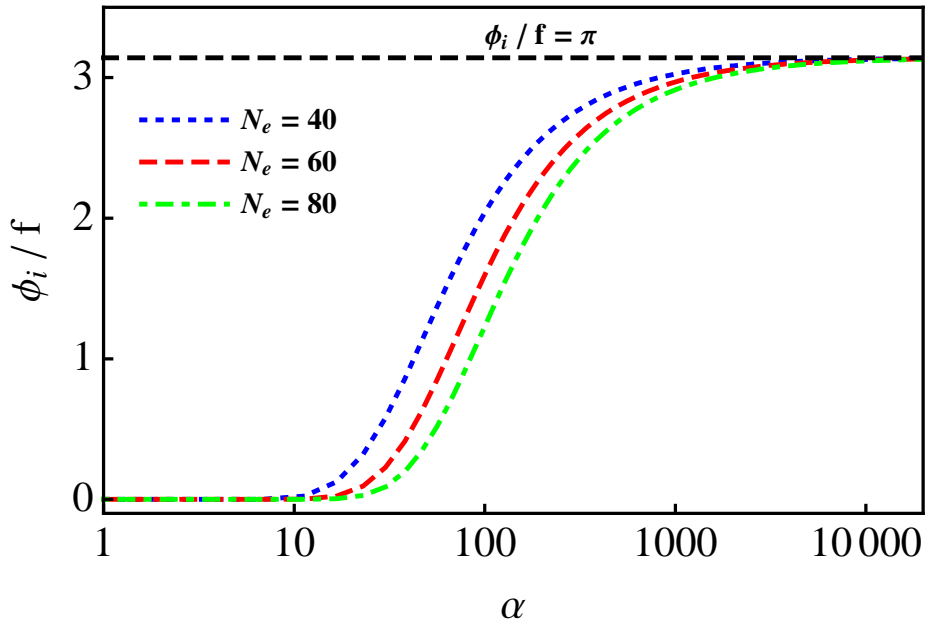}
\caption[$\phi_i$ as a function of $\alpha$ for different values of the number of e-folds $N_e$.]{The initial value of the angle $\phi_i/f$ as a function of $\alpha = \sqrt{32\pi/3} \,\Gamma f/m_\phi M_{\rm Pl}$ for different values of the number of e-folds $N_e$. Blue dotted: $N_e=40$; Red dashed: $N_e=60$; Green dot-dashed: $N_e=80$. Also shown is the line $\phi_i = \pi f$ (Black dashed line).}
\label{phi1_f}
\end{center}
\end{figure}

In Figure~\ref{phi1_f}, we show the value of $\phi_i/f$ given in  Eq.~(\ref{definition_phi1}) as a function of $\alpha$ for different values of the number of e-folds $N_f$. Blue dotted is $N_e = 40$, red dashed is $N_e = 60$, and green dot-dashed is $N_e = 80$. Also outlined is the value $\phi_f/f = \pi$ as a black dashed line. Considering the three parameters $m_\phi$, $f$, and $\Gamma$ from which $\alpha$ depends, moving towards greater values of $\alpha$ corresponds to a decrease in $m_\phi$ or to an increase in $f$ or $\Gamma$, once the other two parameters have been fixed.

\begin{figure}[t!]
\begin{center}
\includegraphics[width=13cm]{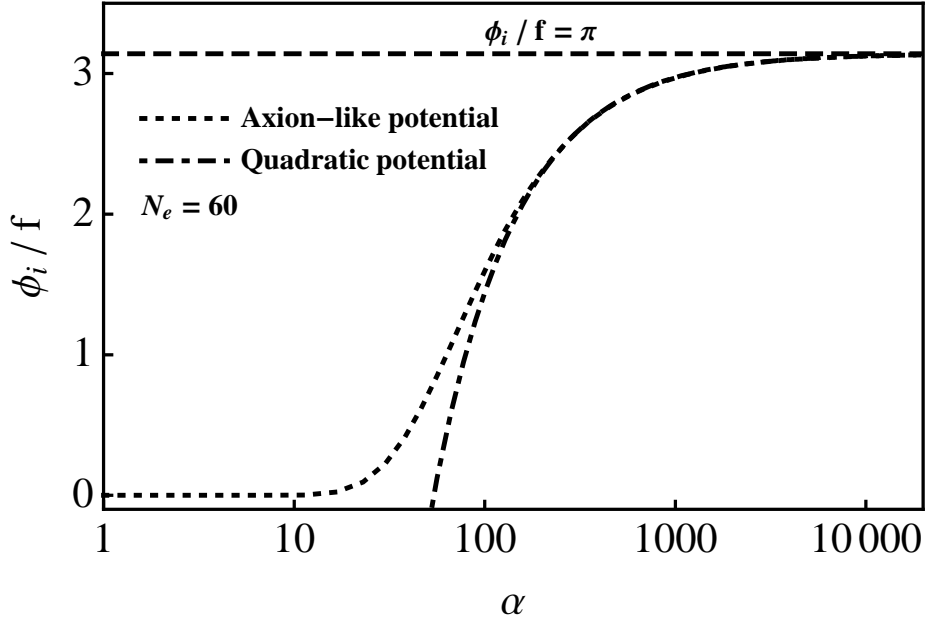}
\caption[$\phi_i$ vs. $\alpha$ for the axion and the quadratic potentials.]{The initial value of the angle $\phi_i/f$ as a function of $\alpha = \sqrt{32\pi/3} \,\Gamma f/m_\phi M_{\rm Pl}$ for the case of the axion-like potential in Eq.~(\ref{potential_axion_like}) (black dotted line) and for the quadratic potential in Eq.~(\ref{quadratic_potential}) that approximates Eq.~(\ref{potential_axion_like}) around $\phi_i \sim \pi\,f$ (black dot-dashed line). For these lines, we fixed $N_e = 60$. Also shown is the horizontal line corresponding to $\phi_f/f = \pi$.}
\label{phi1_f2}
\end{center}
\end{figure}

For larger values of $\alpha$, the initial angle $\phi_i/f$ approaches $\pi$ and the axion-like potential is not distinguishable from a pure quadratic one. This fact is shown in Fig.~\ref{phi1_f2}, where we compare the value of the angle $\phi_i/f$ with $N_e = 60$ (black dashed line), together with the value of $\phi_i$ if the inflaton potential were a pure quadratic one (black dot-dashed line). The quadratic potential used for the plot in Fig.~\ref{phi1_f2}, in place of $U(\phi)$ in Eq.~(\ref{potential_axion_like}), is
\begin{equation} \label{quadratic_potential}
U_{\rm quad}(\phi) = \frac{\Lambda^4}{2}\,\left(\pi-\frac{\phi}{f}\right)^2.
\end{equation}
For $\alpha \gtrsim N_e$, Eq.~(\ref{definition_phi1}) is approximated by 
\begin{equation} \label{phi1_approx}
\phi_i \approx \pi - \sqrt{8}\,N_e/\alpha,
\end{equation}
and the dynamics of the field can no longer discern between the two potentials $U(\phi)$ and $U_{\rm quad}(\phi)$.

We now discuss the constraints on the parameter space of the NWI theory, based on the inequalities used only.

\subsection{Parameter space of the NWI} \label{parameter_space}

We have derived our results in the strongly dissipating regime $\Gamma \gg 3H$: this constrains the parameters as
\begin{equation} \label{constraint_G}
\alpha \gg 16\pi\,\left(\frac{f}{M_{\rm Pl}}\right)^2\,\sqrt{1+\cos\phi_i}.
\end{equation}
This constraint is not severe for values of $f$ around the GUT scale, which is the region of interest here. In fact, for $f\sim 10^{16}$ GeV a numerical solution of Eq.~(\ref{constraint_G}) is
\begin{equation}\label{cnstr1}
\alpha \gg 4.8\times 10^{-5}\,f_{16}^2,
\end{equation}
independently of $N_e$. This bound is much more loose than the requirement of flatness of $U(\phi)$, which yielded $\alpha > 1$. When $f \sim M_{\rm Pl}$, the constraint in Eq.~(\ref{cnstr1}) starts depending mildly on the number of e-folds: for example, setting $f = M_{\rm Pl}$ in Eq.~(\ref{constraint_G}) gives $\alpha~\gg~20 N_e^{0.28}$.

\section{Perturbations from inflation} \label{Perturbations from inflation}

In this section, we specialize the tools described in Sec.~\ref{Fluctuations during inflation} to the NWI model. In particular, we will consider the scalar spectrum $\Delta^2_{\mathcal{R}}(k)$, see Eq.~(\ref{curvature_perturbations}), the scalar spectral index $n_s(k)$, see Eq.~(\ref{derivative_power_spectrum}), and the tensor perturbations spectrum $\Delta^2_{\mathcal{T}}(k)$, see Eq.~(\ref{tensor_perturbations}), using results from the warm inflation literature to set constraints on the NWI model.

\subsection{Scalar power spectrum} \label{Scalar power spectrum}

The spectrum of the adiabatic density perturbations generated by inflation is specified by the power spectrum $\Delta^2_{\mathcal{R}}(k)$, given in Eq.~(\ref{curvature_perturbations}), and a scalar spectral index $n_s$, given in Eq.~(\ref{derivative_power_spectrum}). In both warm and cool inflation models, the scalar power spectrum is generically expressed by Eq.~(\ref{scalar_spectrum}),
\begin{equation}\label{scalar_spectrum1}
\Delta^2_{\mathcal{R}}(k) = \left(\frac{H}{\dot{\phi}}\right)^2\,\langle|\delta \phi|^2\rangle,
\end{equation}
with $\dot{\phi} \simeq -U_\phi/(3H+\Gamma)$. In warm inflation, the spectrum of fluctuations in the inflaton field $\langle\delta \phi\rangle$ has been computed in Ref.~\cite{BasteroGil:2004tg},
\begin{equation} \label{thermal_fluctuations}
\langle\delta \phi \rangle_{\rm thermal} = \left(\frac{\Gamma\,H\,T^2}{(4\pi)^3}\right)^{1/4}.
\end{equation}
Using Eqs.~(\ref{scalar_spectrum1}) and~(\ref{thermal_fluctuations}) for the variance of fluctuations, we obtain the scalar power spectrum in the strongly dissipative regime of warm inflation as
\begin{equation} \label{power_spectrum_1}
\Delta^2_{\mathcal{R},{\rm warm}}(k_0) = \frac{1}{(4\pi)^{3/2}}\,\frac{H^{5/2}\,\Gamma^{1/2}T}{\dot{\phi}^2},
\end{equation}
which is the same result as in Refs.~\cite{Taylor:2000ze, Moss:2008yb}, with an extra factor $1/(2\pi)^2$ that accounts for our normalization of the power spectrum. For comparison, the power spectrum in the usual cool inflation is given in Eq.~(\ref{power_spectrum_cool}); density perturbations are larger in warm inflation by a factor
\begin{equation}\label{def_xi}
\kappa \equiv \frac{\Delta^2_{\mathcal{R},{\rm warm}}(k_0)}{\Delta^2_{\mathcal{R},{\rm cool}}(k_0)} = \left(\frac{\pi}{4}\,\frac{\Gamma\,T^2}{H^3}\right)^{1/4} \approx 10^9\,\left(\frac{\Gamma_{12}}{m_{\phi\,9}^3\,f_{16}^5}\right)^{1/4}.
\end{equation}

\begin{figure}[h!]
\begin{center}
\includegraphics[width=11cm]{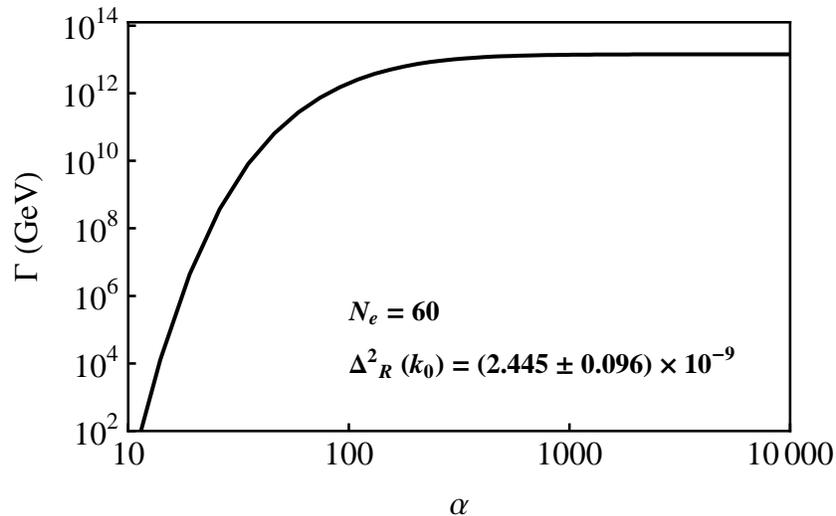}
\caption[The dissipation term $\Gamma$ as a function of $\alpha$.]{The dissipation term $\Gamma$ as a function of $\alpha$, given by Eq.~(\ref{constraint_power_spectrum}) with $\Delta^2_{\mathcal{R}}(k_0) = (2.445 \pm 0.096) \times 10^{-9}$. We used Eq.~(\ref{power_spectrum2}) for the analytic expression of the scalar power spectrum.}
\label{plotGamma}
\end{center}
\end{figure}

From now on, scalar perturbations are considered only in the warm inflation scenario, so we suppress the index ``warm'' in Eq.~(\ref{power_spectrum_1}). Using the expressions for $H$ and $T$ in Table~7.1 we write Eq.~(\ref{power_spectrum_1}) as
\begin{equation} \label{power_spectrum2}
\Delta^2_{\mathcal{R}}(k_0) \simeq 3.7\times 10^{-13}\,\left(\frac{g_*(T)}{228.75}\right)^{-1/4}\,\alpha^{3/4}\,\Gamma_{12}^{3/2}\,\left(\frac{(1+\cos\phi_i/f)^{3/4}}{\sin\phi_i/f}\right)^{3/2}.
\end{equation}
In deriving this last expression, we have used the fact that the largest density perturbations are produced when $\phi = \phi_i$ \cite{Freese:1990rb}. Eq.~(\ref{power_spectrum2}) defines the power spectrum in terms of $\alpha$, $\Gamma$, and $N_e$, the latter variable appearing implicitly in the definition of $\phi_i$.  We equate the expression for the power spectrum in Eq.~(\ref{power_spectrum2}) to the measured value from WMAP in Eq.~(\ref{constraint_power_spectrum}) to obtain a relation between $\Gamma$ and $\alpha$, as shown in Fig.~\ref{plotGamma} for case $N_e=60$. We have checked that different values of $N_e$ do not modify the curves sensibly.

From Fig.~\ref{plotGamma}, the dissipation $\Gamma$ reaches a maximum value $\Gamma_{\rm max}$ at large values of $\alpha$, found analytically by equating Eqs.~(\ref{constraint_power_spectrum}) and~(\ref{power_spectrum2}) in the limit $\alpha \gg 1$,
\begin{equation}\label{gamma_max}
\Gamma_{\rm max} = \frac{3.52\pm 0.09}{\sqrt{N_e}}\,\times 10^{13}{\rm ~GeV}\,\left(\frac{g_*(T)}{228.75}\right)^{1/6}.
\end{equation}

\begin{figure}[b!]
\begin{center}
\includegraphics[width=11cm]{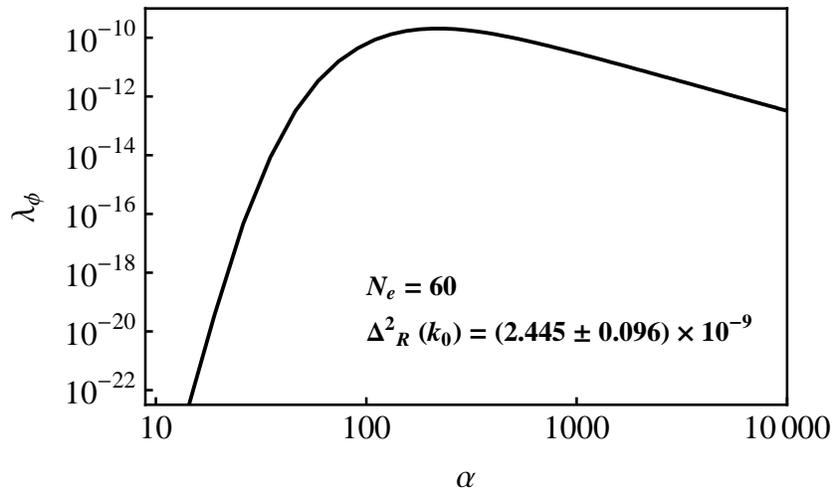}
\caption[The inflaton quartic self-coupling $\lambda_\phi$ as a function of $\alpha$.]{The inflaton quartic self-coupling $\lambda_\phi$ as a function of the parameter $\alpha$, Eq.~(\ref{power_spectrum3}). The parameters $\Delta^2_{\mathcal{R}}(k_0)$ and $N_e$ are fixed as indicated in the figure.}
\label{plotLambda}
\end{center}
\end{figure}

To study the strength of the quartic self-interaction in the NWI model, we use Eqs.~(\ref{self_interaction}) and~(\ref{def_alpha}) to eliminate $\Gamma$ in Eq.~(\ref{power_spectrum2}) and obtain a relation between $\lambda_\phi$ and $\alpha$,
\begin{equation} \label{power_spectrum3}
\Delta^2_{\mathcal{R}}(k_0) \simeq 6.89\times 10^{-14}\,\left(\frac{g_*(T)}{228.75}\right)^{-1/4}\,\alpha^{9/4}\,\lambda_\phi^{3/4}\,\left(\frac{(1+\cos\phi_i/f)^{3/4}}{\sin\phi_i/f}\right)^{3/2}.
\end{equation}
Using the measured value for $\Delta^2_{\mathcal{R}}(k_0)$ in Eq.~(\ref{constraint_power_spectrum}), we obtain the dependence of the quartic self-interaction $\lambda_\phi$ on $\alpha$, as shown in Fig.~\ref{plotLambda}. We see from Fig.~\ref{plotLambda} that the self-interaction is $\lambda_\phi \lesssim 10^{-10}$ for any $\alpha$, and the NWI model can easily satisfy the constraint for the quartic self-interaction term  $\lambda_\phi \lesssim 10^{-8}$ obtained in Ref.~\cite{Adams:1990pn}. When $\alpha \gg N_e$, Eq.~(\ref{power_spectrum3}) with the measured value of the scalar power spectrum is approximated by
\begin{equation}
\lambda_\phi = 1.2 \times 10^{-10}\,\left(\frac{g_*(T)}{228.75}\right)^{1/3}\,\left(\frac{\alpha}{10^4}\right)^{-5}\,\left(\frac{N_e}{60}\right)^2.
\end{equation}

\newpage

\subsection{Scalar spectral index}

The scalar spectral index $n_s$ describes the mild dependence of the scalar power spectrum on the wavenumber $k$, as in Eq.~(\ref{curvature_perturbations}). Here, we do not consider the spectral tilt $\tau$ in Eq.~(\ref{spectral_tilt}), because this quantity depends on higher orders in the slow-roll parameters. With this approximation, the scalar spectral index in the warm inflation scenario results in \cite{Hall:2003zp}
\begin{equation}
n_s -1 = \frac{1}{\mathcal{Q}}\left(-\frac{9}{4}\epsilon+\frac{3}{2}\eta-\frac{9}{4}\,B\right).
\end{equation}
Using the expressions for $\epsilon$ and $\eta$ in Eq.~(\ref{slow_roll_parameters}), and with $B = 0$, we find
\begin{equation}
n_s -1 = \frac{3}{16\pi\,G\,U^2\,\mathcal{Q}}\left(U_{\phi\phi}\,U-\frac{3}{4}U_\phi^2\right),
\end{equation}
or, using $U(\phi)$ in Eq.~(\ref{potential_axion_like}) and its derivatives at $\phi = \phi_i$, together with the values of $\mathcal{Q}$, $\epsilon$ and $\eta$ in Table~7.1,
\begin{equation} \label{ns_eq}
n_s  = 1 -\frac{3}{8}\sqrt{\frac{3}{2\pi}}\,\frac{m_\phi\,M_{\rm Pl}}{\Gamma\,f}\frac{3+\cos\phi_i/f}{\sqrt{1+\cos\phi_i/f}} = 1 -\frac{1.50}{\alpha}\frac{3+\cos\phi_i/f}{\sqrt{1+\cos\phi_i/f}}.
\end{equation}

\begin{figure}[h!]
\begin{center}
  \includegraphics[width=11cm]{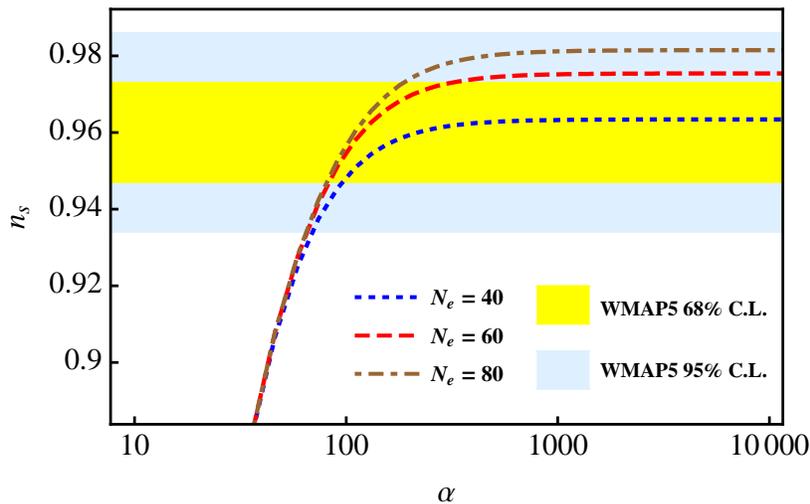}
\caption[The scalar spectral index $n_s$ as a function of $\alpha$, for different values of $N_e$.]{The scalar spectral index $n_s$, Eq.~(\ref{ns_eq}), as a function of $\alpha$, for different values of the number of e-folds. Blue dotted: $N_e=40$; Red dashed: $N_e=60$; Brown dot-dashed: $N_e=80$. Also shown are the C.L. regions from the combined WMAP+BAO+SN measurement, Eq.~(\ref{ns_exp}). Yellow: 68\%; violet: 95\%.}
\label{plot_ns}
\end{center}
\end{figure}

Figure~\ref{plot_ns} shows the dependence of the spectral index on $\alpha$ as in Eq.~(\ref{ns_eq}), for different values of the number of e-folds $N_e$.

 We have compared the values of $n_s$ with the  combined WMAP +SN+BAO measurement \cite{Komatsu:2010fb},
\begin{equation} \label{ns_exp}
n_s = 0.960 \pm 0.013 \quad\hbox{at 68\% C.L.},
\end{equation}
so that the shaded yellow region in Fig.~\ref{plot_ns} corresponds to the 68\% C.L., while the light blue region is the corresponding 95\% C.L. region. We see that small values of $N_e \approx 50$ and large values of $\alpha > O(100)$ are favored, although values of $N_e = 60-70$ can be accommodated in the 68\% C.L. region for $100\lesssim \alpha\lesssim 300$ or in the 95\%C.L. region for all values of $\alpha$. For large values of $\alpha$ the scalar spectral index $n_s$ is essentially independent of $\alpha$, while for $\alpha \lesssim N_e$ it is the dependence on $N_e$ that vanishes. This behavior can be summed up by considering Eq.~(\ref{ns_eq}) in these two limits,
\begin{equation} \label{ns_approx}
1 - n_s = \begin{cases}
4.2/\alpha & \hbox{$\alpha \lesssim N_e$},\\
1.5/N_e & \hbox{$\alpha\gtrsim N_e$}.
\end{cases}
\end{equation}
This type of dependence of $n_s$ on either $N_e$ or $\alpha$ has been also noticed in models of Natural Inflation set in the usual cool inflation scenario, see for example Eq.~(12) in Ref.~\cite{Savage:2006tr}. In the case in Ref.~\cite{Savage:2006tr}, in which there is no dissipation term $\Gamma$, and the inflaton mass $m_\phi$ is adjusted so that $\Lambda\approx 10^{16}$ GeV, the result analogous to Eq.~(\ref{ns_approx}) is expressed in terms of $f$ instead of $\alpha$.

\subsection{Tensor power spectrum}

Since warm inflation considers thermal fluctuations instead of quantum fluctuations to generate scalar perturbations, it is only density fluctuations that modify in this scenario while tensor perturbations show the same spectrum as in the usual cool inflation \cite{Moss:2008yb}. Defining the tensor power spectrum as
\begin{equation} \label{tensor_perturbations1}
\Delta^2_{\mathcal{T}}(k) \equiv \frac{k^3\,P_{\mathcal{T}}(k)}{2\pi^2} = \Delta_{\mathcal{T}}^2(k_0)\,\left(\frac{k}{k_0}\right)^{n_T},
\end{equation}
where the tensor spectral index $n_T$ is assumed to be independent of $k$, because current measurement cannot constraint its scale dependence.
WMAP does not constrain $\Delta_{\mathcal{T}}^2(k_0)$ directly, but rather the tensor-to-scalar ratio
\begin{equation}
r \equiv \frac{\Delta_{\mathcal{T}}^2(k_0)}{\Delta_{\mathcal{R}}^2(k_0)}.
\end{equation}
which qualitatively measures the amplitude of gravitational waves per density fluctuations. The WMAP+BAO+SN measurement constrains the tensor-to-scalar ratio as \cite{Komatsu:2010fb}
\begin{equation}
r < 0.22 \quad \hbox{at 95$\%$ C.L.}.
\end{equation}

Since in warm inflation the scalar power spectrum is enhanced by the quantity $\kappa$ in Eq.~(\ref{def_xi}) with respect to the value in cool inflation, the tensor-to-scalar ratio and thus gravitational waves are reduced in the NWI model by the same amount $\kappa$. Using Eq.~(\ref{power_spectrum_1}) for the scalar power spectrum and the expression for the tensor power spectrum,
\begin{equation}
\Delta_{\mathcal{T}}^2(k_0) = \frac{16\,H^2}{\pi M_{\rm pl}^2},
\end{equation}
the tensor-to-scalar ratio in warm inflation is
\begin{equation}\label{r_measure}
r = \frac{128\sqrt{\pi}}{M_{\rm Pl}^2}\,\frac{\dot{\phi}^2}{\sqrt{H\Gamma}\,T}.
\end{equation}
With the values in Table~7.1, we find
$$r \simeq 128\,\left(\frac{\pi^7}{150}\right)^{1/8}\,g_*^{1/4}(T)\,\left(\frac{f^5\,m_{\phi}^{11}}{\Gamma^9\,M_{\rm Pl}^7}\right)^{1/4}\,\frac{(\sin\phi_i/f)^{3/2}}{(1+\cos\phi_i/f)^{1/8}} = $$
\begin{equation}
= 1.62 \times 10^{-13}\,\left(\frac{g_*(T)}{228.75}\right)^{1/4}\,\left(\frac{f_{16}^5\,m_{\phi\,9}^{11}}{\Gamma_{12}^9}\right)^{1/4}\,\frac{(\sin\phi_i/f)^{3/2}}{(1+\cos\phi_i/f)^{1/8}}.
\end{equation}
In the limit $\alpha \gg N_e$, with $\phi_i \approx \pi-\sqrt{8}N_e/\alpha$, we obtain
\begin{equation}
r = 4.63 \times 10^{-14}\,N_e^{5/4}\left(\frac{g_*(T)}{228.75}\right)^{1/4}\,\left(\frac{m_{\phi\,9}}{\Gamma_{{\rm max},12}}\right)^{4},\quad\hbox{for $\alpha \gg N_e$}.
\end{equation}
Using the WMAP measure of the tensor-to-scalar ratio in Eq.~(\ref{r_measure}), we constrain the inflaton mass in the NWI model to
\begin{equation} \label{bound_r}
m_\phi < 2.7\times 10^{12}{\rm ~GeV}\,\left(\frac{r}{0.22}\right)^{1/4}\,\left(\frac{N_e}{60}\right)^{-13/16}\,\left(\frac{g_*(T)}{228.75}\right)^{5/48},
\end{equation}
where we used the expression for $\Gamma_{\rm max}$ in Eq.~(\ref{gamma_max}). Future measurements will not substantially improve the bound in Eq.~(\ref{bound_r}), because of the power $1/4$ raising $r$. As an example, the forecast PLANCK measurement will constrain the tensor-to-scalar ratio by one order of magnitude with respect to WMAP, $r \lesssim 0.01$, thus the bound in Eq.~(\ref{bound_r}) will approximately lower by a factor of two.

\begin{figure}[tb]
\begin{center}
\includegraphics[width=11cm]{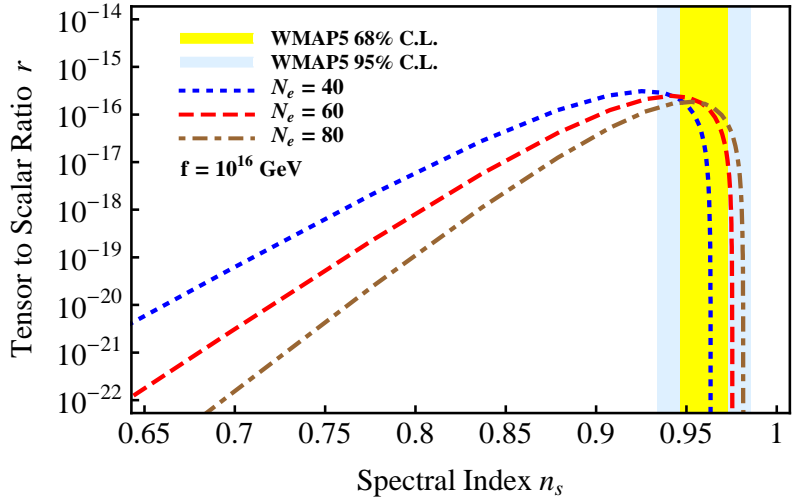}
\caption[Constraints on the NWI model in the $r$-$n_s$ plane, for $f=10^{16}$ GeV.]{Prediction from the NWI model and WMAP constraints in the $r$-$n_s$ plane for $f=10^{16}$ GeV. Blue dotted line: $N_e=40$; Red dashed line: $N_e=60$; Brown dot-dashed line: $N_e=80$. The yellow and violet regions are the parameter spaces allowed by WMAP+BAO+SN at 68\% and 95\% C.L., respectively.}
\label{plot16}
\end{center}
\end{figure}

Results for the case $f = 10^{16}{\rm ~GeV}$ are summarized in Fig.~\ref{plot16}, where we show the constrains on the NWI model from the WMAP data in the $r$-$n_s$ plane. The WMAP data are better in agreement with a low number of e-folds $N_e = 40-50$, although values up to $N_e = 80$ can be accommodated within the 95\% C.L. region.

We conclude that a value of the axion decay constant of the order of the GUT scale $f~\sim~\Lambda_{\rm GUT}$ can be easily embedded in the NWI model, making it possible to construct microscopic theories of warm inflation with axion-like particles at the GUT scale. For this value of the axion decay constant, the expected value of $r$ and thus the amount of gravitational waves that are produced with this type of inflation is extremely low. If gravitational waves are found with $r \sim 10^{-14}$ or above, this model has to be abandoned and a $\Lambda_{\rm GUT}$ valued axion decay constant $f$ is no longer viable.

\section{Discussion and conclusions}\label{discussion}

The parameter space of the inflaton field in the NWI model is broader than that in the usual Natural Inflation, because of the presence of the extra quantity $\Gamma$ describing dissipation and of the enlarged slow-roll conditions in Eq.~(\ref{slow_roll}).

The NWI model allows us to lower the value of the axion decay constant $f$ from the Planck scale resulting in Natural Inflation to the GUT scale; a ratio $f/M_{\rm Pl} \approx 10^{-3}$ helps overcoming some difficulties encountered in Natural Inflation model-building.

A decay constant of the order of the Planck scale is still possible, since $f$ is not bound from above. In the case $f \sim M_{\rm Pl}$, Natural Inflation and NWI can be distinguished observationally by a measurement of the tensor-to-scalar ratio $r$, see Fig.~\ref{plot16}. This difference in the value of the ratio $r$ comes from the fact that in the NWI model, the amplitude of gravitational waves is suppressed by the factor $\kappa$ in Eq~(\ref{def_xi}) with respect to the standard cool inflation.

Measurements of the scalar spectral index $n_s$ favor larger values of $\alpha \gtrsim N_e$, as shown in Fig.~\ref{plot_ns}. In this region, the value of $\phi_i$ in Eq.~(\ref{definition_phi1}) can be approximated with $\phi_i \approx \pi - \sqrt{8}N_e/\alpha$, see Eq.~(\ref{phi1_approx}), and the axion-like potential in Eq.~(\ref{potential_axion_like}) cannot be distinguished from a pure quadratic one through the dynamic of the inflaton. In this condition, one would need to obtain the value of the self-interaction $\lambda_\phi$ independently, for example from considering density fluctuations in the CMBR; Natural Inflation and NWI models predict a precise ratio between the height of the potential and the self-interaction coupling constant, $U/\lambda_\phi = 12\,f^2$, whereas other inflaton models predict different values of this ratio.

The possibility that the axion-like energy scale $f$ is of the order of the GUT scale when warm inflation is considered has been already taken into account in Ref.~\cite{Mohanty:2008ab}. However, we differ from Ref.~\cite{Mohanty:2008ab} in various points: for example, we do not consider the temperature as an independent variable, because in warm inflation the radiation bath is thermalized and temperature that is expressed is linked to the radiation energy density $\rho_r$ as in Eq.~(\ref{definition_radiation}). Moreover, here we have included a detailed analysis of the CMBR observables $r$, $\Delta^2_{\mathcal{R}}(k)$ and $n_s$ that was missing in Ref.~\cite{Mohanty:2008ab}.

We have presented a model in which Natural Inflation takes place within the warm inflation scenario: in such a model, the decay constant $f$ no longer ties to the Planck scale as in the usual Natural Inflation model, but it can be as low as the GUT scale, $f \sim 10^{16}$ GeV. We have shown the viability of the NWI model and its agreement with current astrophysical data.